\documentclass{aa}

\usepackage{graphicx}

\usepackage[varg]{txfonts}

\usepackage{natbib}
\bibpunct{(}{)}{;}{a}{}{,} 

\begin{document}

   \title{Amplification and generation of turbulence during self-gravitating collapse}

   \author{Patrick Hennebelle \inst{\ref{inst1}} }

   \institute{AIM, CEA, CNRS, Université Paris-Saclay, Université Paris Diderot, Sorbonne Paris Cité, F-91191 Gif-sur-Yvette, France,
      \label{inst1} }



\abstract
{The formation of astrophysical structures, such as stars, compact objects but also galaxies,  entail an
enhancement of densities by many orders of magnitude which occurs through gravitational collapse.}
{The role played by turbulence during this process is important. Turbulence generates density fluctuations, 
exerts a support against gravity and possibly delivers angular momentum. How turbulence exactly behave during 
the collapse and get amplified remains a matter of investigation.}
{Spherical averaging of the fluid equations is carried out, leading to 1D fluid equations that describe the evolution of mean 
quantities in particular the mean radial velocity as well as the mean radial and transverse turbulent velocities. These equations
differ from the ones usually employed in the literature. We then perform a series of 3D numerical simulations of collapsing clouds
for a wide range of  thermal and turbulent supports with two polytropic equation of state, $P \propto \rho^\Gamma$, with $\Gamma=1$ and 1.25. For each
 3D simulations we perform a series of 1D simulations  using the spherically averaged equations and with the same initial conditions.}
{By performing a detailed comparison between 3D and 1D simulations, we can analyse in great details the observed behaviours.
Altogether we find that the two approaches agree remarkably well demonstrating the validity of the inferred equations
although when  turbulence is initially strong, major deviations from spherical geometry certainly preclude quantitative 
comparisons. The detailed comparisons lead us to an estimate of the turbulent dissipation parameter that when the turbulence 
is initially low, is found to be  in good 
agreement with previous estimate of non self-gravitating supersonic turbulence.  When turbulence is initially dynamically important
larger values of the dissipation appear necessary for the 1D simulations to match the 3D ones. We find that the behaviour of turbulence 
depends on the cloud thermal support. If it is high, initial turbulence is amplified as proposed earlier in the literature. However if 
 thermal support is low, turbulence is also generated by the development of local non-axisymmetric gravitational instabilities 
reaching values several times larger and in equipartition with gravitational energy.}
{The inferred 1D equations offer an easy way to estimate the level reached by turbulence during gravitational collapse. Depending 
on the cloud thermal support, turbulence is either amplified or locally generated. }


   \maketitle


\section{Introduction}

Gravitational collapse is a common and  major process, which takes place in our Universe. Indeed 
all self-gravitating objects such as stars, white dwarfs, neutron stars or black holes emerge 
after a phase during which the density of the matter they contain has increased by 
several orders of magnitude. Other objects such as galaxies, protostellar dense core as well 
as proto-stellar cluster clumps and stellar clusters, although less extreme, are also several 
orders of magnitude denser than the background from which they have been assembled by gravity. 
For many of these objects, turbulence is believed to play a role before or even during the collapse. 
As a matter of fact, turbulence deeply affects gas dynamics by various effects which  depend of the situation.
For instance in the context of star formation, the density fluctuations induced by turbulence
within a collapsing core have been proposed to induce fragmentation \citep[e.g.][]{bate2003,goodwin04a,goodwin2004,leeh2018a,leeh2018b,h2019}
 while in the context of supernova progenitors,  density fluctuation generation  has been advocated 
to be a source of shock distortion which  can trigger the explosion \citep[e.g.][]{foglizzo2001,couch2013,muller2015}.
Turbulence within a collapsing prestellar core  can also help transporting away magnetic flux 
\citep[e.g.][]{santos2012,seifried2012,joos2013} but may also help generating magnetic field in 
primordial clouds through turbulent dynamo \citep{schleicher2010,federrath2011,schober2012}.
Turbulence can also strongly influence the angular momentum evolution in a 
complex manner. Indeed while generally speaking turbulence can help transporting angular momentum 
outwards \citep[e.g.][]{pringle1981,balbus1999}, it can also 
bring or even generate angular momentum by triggering axisymmetry breaking \citep{misugi2019,verliat2020}.

Understanding how turbulence behaves during gravitational collapse is therefore of 
primary importance for a broad class of astrophysical problems. It is also worth mentioning that 
induced spherical contraction is also most relevant in the context of inertial fusion for instance. 
Here again turbulence is believed to play an important role \citep{davidovits2016,viciconte2018}. 

To tackle the amplification of turbulence in a compressed medium, 
\citet{robertson2012} have performed 3D numerical simulations in which the density enhancement
is imposed by adding to the fluid equations analytical source terms which describe an homologous contraction \citep[see also][]{mandal2020}. 
They also propose an analytical 
model to describe the behaviour of the turbulent component, which entails a source term induced by the 
contraction itself and a classical dissipation. By comparing the simulation results and the model, they show that it is 
able to capture well the fluid behaviour. In the context of the formation of massive stars and the collapse of
turbulent cores, \citet{murray2015} have performed 1D simulations in which the turbulence is modelled using 
a generalised form of the  equation proposed by \citet{robertson2012} namely 
\begin{eqnarray}
\partial _t V_T + V_r \partial _r V_T + {V_T V_r \over r} = - \nu {V_T^2 \over r}.
\label{eq_norm}
\end{eqnarray}
Performing asymptotic analysis as well as 1D numerical simulations of a spherical collapse with a
turbulence described by Eq.~(\ref{eq_norm}), they predict typical behaviour of a turbulent 
collapsing cloud, finding for instance density profiles close but slightly shallower than 
the classical $\rho \propto r^{-2}$ inferred in isothermal collapse \citep[e.g.][]{Shu77}.
Under the assumption that the turbulent velocity is proportional to the radial ones, \citet{xu2020} 
obtain self-similar solutions close to the solutions studied by \citep{Shu77}.

Recently, \citet{guerrero2020} have performed 3D collapse calculations of an unstable prestellar dense 
core. They find that the turbulence, which develops during the collapse tends to
 reach values which are in near virial equilibrium with gravity. 
They postulate that a local equilibrium between gravitational instability and turbulence is established. 
 Similar conclusion has also been reached by \citet{mocz2017} where a series of driven turbulence 
calculations with various magnetisations are presented. Their Fig.~5  unambiguously shows that 
within collapsing cores, the 
turbulent pressure is almost proportional to the gravitational energy. 
In the context of primordial minihalos, similar studies have also been conducted by \citet{higashi2021} 
who also found that turbulence is getting amplified by gravitational contraction. An analytical model based on 
the growth of gravitational instability is also provided.

All these studies strongly suggest that turbulence is definitely amplified during gravitational 
collapse. However  precisely understanding the mechanism at play remains a challenge because 
of the great non-linearity of the process. Moreover so far only few cases have been explored
while there is a great diversity of initial conditions.
The purpose of the paper is twofold. First we infer a set of 1D  equations in spherical geometry
which can be seen as a revised version of the equations used for instance by \citet{murray2015}
and \citet{xu2020}. They are inferred by performing rigorous spherical averaging of the 
3D fluid equations. Second we perform a set of 3D simulations in which we extensively vary the 
initial conditions but also consider two polytropic equations of state namely $\Gamma=1$ and 
$\Gamma=1.25$. Importantly we perform detailed comparisons between the 1D and 3D simulations 
which allows us to identify the need to add as a source of turbulence the development of 
local Jeans instabilities when the collapsing cloud is highly unstable. 
Altogether, the two sets of simulations agree very well, showing that the 1D approach can be 
used to predict the level of turbulence in collapsing objects.

The plan of the paper is as follows
\begin{description}
\item[-] In the second part of the paper, we obtain a set of two equations to describe the evolution of turbulence 
during gravitational collapse. These equations are close but not identical to the previously
proposed equation of \citet{robertson2012}. 
\item[-] In the third part, we describe the 1D code that 
we developed to solve the 1D collapse in spherical geometry taking into account turbulence and
we describe the 3D simulations that we performed as well. This includes their setup and how we proceed
to compare the 1D and 3D simulations. 
\item[-] In the fourth part, we perform a comparison between 
the 3D  and 1D simulations by comparing the time evolution of the central masses. This leads us to an estimate
of the dissipation parameter. 
\item[-] In the fifth part, a detailed comparison between 1D and 3D simulations is carried out and
we find that in the proposed equations to describe turbulence, a source 
term associated to the development of the local Jeans instability is indeed needed to reproduce the 3D simulations. 
\item[-] The sixth part is dedicated to a discussion on the nature of the velocity fluctations.
\end{description}
The seventh part concludes the paper.

\section{Formalism for turbulence  in one-dimensional spherical geometry}
\label{1Dequation}
We start with a derivation of the one-dimensional equations appropriate to 
describe a spherical collapsing turbulent cloud.

\subsection{Standard equations and spherical mean}
While it is clear that in the presence of turbulence the various fields are not spherically symmetric, one can 
nevertheless consider their spherical mean:

\begin{eqnarray}
\label{meansphe}
\bar{\rho} = \int \int d\Omega \rho,
\bar{V}_r = \int \int d\Omega {\rho \over \bar{\rho} } V_r,
\end{eqnarray}
where $d \Omega= \sin \theta d \theta d \phi$. 
Similar procedures are employed when inferring subgrid schemes for instance \citep{schmidt2006}.

We introduce the Reynolds decomposition
\begin{eqnarray}
\label{rey_U}
V_r = \bar{V}_r + v_r, 
\end{eqnarray}
implying that $\bar{v}_r=0$.

To describe turbulence, we also need to consider the orthoradial and azimuthal 
velocity components  $V_\theta$ and  $V _\phi$. 
Below, we do not need to distinguish 
between $V_\theta$ and $V_\phi$ and simply consider $v_t = \sqrt{V_\theta^2+V_\phi^2}$, the 
transverse velocity component.
Our goal is to obtain equations for $\bar{v_r^2}$ and $\bar{v_t^2}$ separately
as  they behave differently  during the collapse of a spherical cloud.

The continuity equation in spherical geometry is
\begin{eqnarray}
\label{consmat}
\partial _t \rho + {1 \over r^2} \partial_r (r^2 \rho V_r) + {1 \over r \sin \theta} \partial _\theta ( \sin \theta 
V_\theta) + {1 \over r \sin \theta} \partial _\phi V_\phi  =0.
\end{eqnarray}
With this expression, it is easy to see that when performing the spherical mean as given by Eq.~(\ref{meansphe}), all terms
involving angular derivatives cancel out
and thus for the sake of simplicity, we 
do not explicitly write the angular dependent terms below. 
The other relevant fluid equations are then

\begin{eqnarray}
\label{consmom}
 \partial _t V_r  + V_r \partial _r V_r - {v_t^2 \over r} = - {1 \over \rho} \partial_r \left( c_s^2 \rho \right) +  g_r + \nu \partial _r \sigma_{r r}, 
\end{eqnarray}
where $c_s$ is the sound speed, $g_r$ the gravitational acceleration and $\sigma_{rr}$ one of the component of the stress tensor.

\begin{eqnarray}
\label{constrans}
 \partial _t {v_t^2 \over 2}  +  V_r \partial _r {v_t^2 \over 2} + {v_t^2 V_r \over r} =  \nu (V_\theta \partial_r \sigma_{r \theta}+V_\phi \partial_r \sigma_{r \phi}),
\end{eqnarray}
To get this last equation, we simply multiplied the standard orthoradial and azimuthal moment equations
by respectively $V_\theta$ and $V_\phi$ and add them.

Finally, the Poisson equation is
\begin{eqnarray}
\label{poisson}
{1 \over r^2 } \partial_r \left( r^2 g_r \right) = - 4 \pi G \rho.
\end{eqnarray}

\subsection{Spherically averaged equations}
As explained above, we want to obtain equations on the quantities $\bar{\rho}$, $\bar{V}_r$ and $\bar{v_T^2}$, taking the 
spherical average of Eq.~(\ref{consmat}), we obtain
\begin{eqnarray}
\label{consmat_mean}
\partial _t \bar{\rho} + {1 \over r^2} \partial_r (r^2 \bar{\rho} \bar{V}_r)=0,
\end{eqnarray}

Next, to get a radial mean momentum equation, we first combine  Eqs.~(\ref{consmat}) and~(\ref{consmom}) to get 
the conservative form of the momentum equation
\begin{eqnarray}
\label{consmom2}
 \partial _t (\rho V_r)  + {1 \over r^2} \partial _r (r^2 \rho V_r^2)  - {\rho v_t^2 \over r} = - \partial_r \left( c_s^2 \rho \right) +  \rho g_r + \nu \rho \partial _r \sigma_{r r}, 
\end{eqnarray}
then with the Reynolds decomposition $V_r = \bar{V}_r + v_r$ and after performing the spherical average, we got
\begin{eqnarray}
\nonumber
 \partial _t ( \bar{\rho} \bar{V}_r)  + {1 \over r^2} \partial _r (r^2 \bar{\rho} \bar{V}_r^2)
 + {1 \over r^2} \partial _r (r^2 \bar{\rho} \bar{v_r^2})
  - {\bar{\rho} \bar{v_t^2} \over r} = \\
-  \partial_r \left( c_s^2 \bar{\rho} \right) +  \overline{ \rho g_r } + \nu \overline{ \rho \partial _r \sigma_{r r}}, 
\label{consmom_mean2}
\end{eqnarray}
that can also be written as 
\begin{eqnarray}
\nonumber
 \partial _t  \bar{V}_r  + \bar{V}_r \partial _r  \bar{V}_r
   + {2  \bar{v_r^2} -  \bar{v_t^2} \over r} = \\
-  {1 \over \bar{\rho} } \partial_r \left( (c_s^2 + \bar{v_r^2}) \bar{\rho} \right) +  
{ \overline{ \rho g_r }    \over \bar{\rho} }  + { \nu \overline{ \rho \partial _r \sigma_{r r}}  \over \bar{\rho} } , 
\label{consmom_mean3}
\end{eqnarray}

Next, we seek for an equation of $\bar{v_r^2}$. For this purpose, we add together the product of Eq.~(\ref{consmom2})
by $v_r$ and the product of Eq.~(\ref{consmom}) by $V_r= \bar{V}_r + v_r$ to which we subtract 
the product of Eq.~(\ref{consmom_mean3}) by $\bar{V}_r$. Finally, we take the spherical mean  as defined 
by Eq.~(\ref{meansphe}) and we obtain the following equation 

\begin{eqnarray}
\nonumber
\partial_t \left( { \bar{\rho} \bar{v_r^2} \over 2} \right) + {1 \over r^2} \partial _r \left(r^2 {\bar{\rho}  \bar{v_r^2} \over 2 } \bar{V}_r
\right) +  \bar{\rho} \bar{v_r^2} \partial_r \bar{V}_r + 
{1 \over r^2} \partial _r \left( r^2 {\bar{\rho} \bar{v_r^3} \over 2} \right) \\
 - {\bar{\rho} \overline{v_t^2 v_r } \over r} = 
- \overline{v_r \partial_r (c_s^2 \rho)} + \overline{v_r \rho g_r} + \nu \overline{v_r \rho \sigma_{r r}}.
\label{turb_rad}
\end{eqnarray}

To get an equation on $\bar{v_t^2}$, we combine Eq.~(\ref{constrans}) with Eq.~(\ref{consmat}) and we take the 
spherical mean, this leads to:
\begin{eqnarray}
\nonumber
 \partial _t \left( { \bar{\rho} \bar{v_t^2} \over 2} \right)
 +   {1 \over r^2} \partial _r \left( r^2 {\bar{\rho} \bar{v_t^2} \over 2} \bar{V}_r \right)
 + {\bar{\rho} \bar{v_t^2} \bar{V}_r \over r}
 \\
+   {1 \over r^2} \partial _r \left( r^2 {\bar{\rho} \overline{v_t^2 v_r} \over 2} \right)
   + {\bar{\rho} \overline{v_t^2 v_r}  \over r} 
= 
 \nu \overline{ (v_\theta \partial_r \sigma_{r \theta}+v_\phi \partial_r \sigma_{r \phi})} ,
\label{turb_trans}
\end{eqnarray}

\subsection{Turbulent-kinetic-energy models}
Equations~(\ref{turb_rad}) and~(\ref{turb_trans}) are very similar to the turbulent kinetic energy  
equations obtained in the context of incompressible flow as shown  for instance in Eq.~5.132 of 
\citet{pope2000}.

 Apart for the Lagrangian derivative, which in 
the compressible spherically symmetric case is replaced by the second term of the left-hand side, 
we have one source term that involves $\bar{v_r^2}$ and $\bar{V}_r$ (third term of the left-hand side)
and one term that involves $v_r^3$ or $v_t^2 v_r$. It is generally assumed \citep[see chapter 10 of][]{pope2000}
that these terms together with the pressure ones, leads to an effective turbulent diffusivity, i.e. 
$\propto \nabla. (\kappa_T \nabla \bar{\rho} \bar{v^2})$. 
On the other-hand, the last terms of the right-hand side, represent dissipation that following 
\citet{robertson2012} is assumed to be $\propto \rho v_T^2/ \tau_{diss}$, where $\tau_{diss}$ is the 
dissipation time. We proceed here with the same assumptions which lead 
to
\begin{eqnarray}
\nonumber
\partial_t \left( { \bar{\rho} \bar{v_r^2} \over 2} \right) + {1 \over r^2} \partial _r \left(r^2 {\bar{\rho}  \bar{v_r^2} \over 2 } \bar{V}_r
\right) +  \bar{\rho} \bar{v_r^2} \partial_r \bar{V}_r  \\
+   {1 \over r^2} \partial _r \left( r^3  \eta_{diff} \overline{v_r^2}^{1/2}  \partial _r (\bar{\rho}  \overline{v_r^2})  \right)
= 
- \eta_{diss} { \bar{\rho} \bar{v_r^2} \over \tau_{diss}}
\label{turb_rad2}
\end{eqnarray}

\begin{eqnarray}
\nonumber
 \partial _t \left( { \bar{\rho} \bar{v_t^2} \over 2} \right)
 +   {1 \over r^2} \partial _r \left( r^2 {\bar{\rho} \bar{v_t^2} \over 2} \bar{V}_r \right)
 + {\bar{\rho} \bar{v_t^2} \bar{V}_r \over r} \\
+   {1 \over r^2} \partial _r \left( r^3  \eta_{diff} \overline{v_t^2}^{1/2}  \partial _r (\bar{\rho}  \overline{v_t^2})  \right)
= 
  - \eta_{diss} { \bar{\rho} \bar{v_t^2} \over \tau_{diss}}
\label{turb_trans2}
\end{eqnarray}
where we have assumed that $\kappa _T = \eta_{diff} r \overline{v_T^2}^{1/2}$ and $\eta_{diff}$ is a dimensionless 
number on the order of a few. 
Note that to write the dissipation and diffusion terms, it is necessary to estimate the typical dissipation time 
and diffusion coefficient and it is generally assumed that the former is simply given by the local crossing time
$\tau_{diss} =r /  v_{t,r} $ while the latter is given by $ r v_{t,r}$.
While these expressions are reasonable, estimating what the relevant spatial scale, i.e. the right $r$ value that should 
be employed is not a simple task and this is further discussed in Section~\ref{1Dcode}.

Equations~(\ref{turb_rad2}) and (\ref{turb_trans2}) are the ones  used in this work (but see the discussion on the generation 
of turbulence  in Section~\ref{geneturb}). 
They are similar to the equation proposed
by \citet{robertson2012} except for the source terms  and the turbulent diffusion one.
Note however that importantly enough, Eqs.~(\ref{turb_rad2}) and (\ref{turb_trans2}) present source terms namely
$\rho v_r ^2 \partial _r V_r$ and $\rho v_t ^2 V_r / r$ that are not identical. While the second one solely depends on 
the sign of $V_r$, the first may be either positive or negative even if $V_r$ remains negative. 
In particular, during the first phase of the collapse, i.e. before the formation 
of the central singularity, the amplitude of $V_r$ increases with $r$ (in the cloud inner part), leading to an amplification 
of $v_{r}$ while in the second phase, the amplitude of $V_r$ decreases with $r$. This suggests, as will be confirmed 
later, that the 2 components behave differently.

 Apart from the difference on the turbulent velocity equations, 
 there is another important difference with the equations used for instance by 
\citet{murray2015} and \citet{xu2020} which employed the equation proposed by \citet{robertson2012}
together with a momentum equation that assumed a turbulent pressure $\rho v_T^2$, where $v_T^2 = v_{r}^2 + v_t^2$.
 From 
Eq.~(\ref{consmom_mean3}), it is seen that  the support provided by turbulence is not identical to theirs if the 
turbulence is not anisotropic as in this case $2 v_r^2 - v_t^2 \ne 0$.
Even if turbulence is fully isotropic, we note that the turbulent pressure they use 
is $\rho v_T^2$, while our analysis suggests that it should be  $\rho v_{r}^2 = \rho v_T^2 / 3$.
Generally speaking, it is worth stressing that the support exerted by the transverse component is 
similar to a centrifugal support apart for the fact that unlike the centrifugal force, it is isotropic. 
The radial component of the turbulent velocity exerts  a support that contains a
pressure-like term and a geometric contribution $2 v_r^2/r$.

Finally, let us note that at this stage we have two unknowns $\eta_{diss}$ and $\eta _{diff}$. 
Anticipating our results below, we find that $\eta _{diss} \simeq 0.25$ while we 
find no evidence that significant turbulent diffusivity should be accounted for.

\subsection{Turbulent amplification by local gravitational instability}
\label{geneturb}

Equations.~(\ref{turb_rad2}) and (\ref{turb_trans2}) describe 
the amplification of turbulence which results from the gravitational compression.
It does not take into the possibility to have turbulence being generated by 
local gravitational instability as discussed for instance in \citet{guerrero2020}.
As will be seen below this indeed appears to be needed and to 
 take into account a new source of 
turbulence in Eqs.~(\ref{turb_rad2}) and (\ref{turb_trans2}) which
 accounts for the development of local gravitational instabilities on top 
of the global collapse, we need to estimate the growth rate as a function of 
time and radius. Generally speaking, this is a complicated problem because the cloud
is dynamically evolving. We therefore follow a phenomenological approach. 
First we compute the Jeans \citep{jeans1902} growth rate as a function of radius
\begin{eqnarray}
\omega(r) = \sqrt{4 \pi G \rho(r) - c_s^2 ({2 \pi / 2 r})^2}.
\end{eqnarray}
Second, we find the maximum $\omega_{max}$ which occurs at a radius 
$r_{max}$. 
We assume that as a mode grows, it amplifies the velocity within all points 
of radius, $r < r_{max}$. Thus 
\begin{eqnarray}
\tau_{jeans} = 1/ \omega_{max} {\rm \; if \;} r < r_{max}, \\
\nonumber
\tau_{jeans} = 1/ \omega(r) {\rm \; if \;} r > r_{max}.
\end{eqnarray}

This leads to rewrite Eqs.~(\ref{turb_rad2}) and (\ref{turb_trans2}) as
\begin{eqnarray}
\label{turb_rad2_jeans}
\partial_t \left( { \bar{\rho} \bar{v_r^2} \over 2} \right) + {1 \over r^2} \partial _r \left(r^2 {\bar{\rho}  \bar{v_r^2} \over 2 } \bar{V}_r
\right) +  \bar{\rho} \bar{v_r^2} \partial_r \bar{V}_r  \\
= - \eta_{diss} { \bar{\rho} \bar{v_r^2}^{3/2} \over r} +  { \bar{\rho} \bar{v_r^2} \over 2} {1 \over  \tau _{jeans} },
\nonumber
\end{eqnarray}
and
\begin{eqnarray}
\label{turb_trans2_jeans}
 \partial _t \left( { \bar{\rho} \bar{v_t^2} \over 2} \right)
 +   {1 \over r^2} \partial _r \left( r^2 {\bar{\rho} \bar{v_t^2} \over 2} \bar{V}_r \right)
 + {\bar{\rho} \bar{v_t^2} \bar{V}_r \over r}
\\
=  - \eta_{diss} { \bar{\rho} \bar{v_t^2}^{3/2} \over r}  +  { \bar{\rho} \bar{v_t^2} \over 2 } {1 \over \tau _{jeans}}.
\nonumber
\end{eqnarray}

\subsection{A Lagrangian perspective}
Equation~(\ref{turb_trans2_jeans}) can be recasted in a different form. By multiplying it by $r^2$
and defining $J_T =r v_t$, we obtain
\begin{eqnarray}
\label{turb_angmom}
 \partial _t \left(  \bar{\rho} \bar{J_T ^2} \right)
 +   {1 \over r^2} \partial _r \left( r^2 \bar{\rho} \bar{J_T ^2} \bar{V}_r \right)
=  - 2 \eta_{diss} { \bar{\rho} \bar{J_T ^2}^{3/2} \over r^2}  +   { \bar{\rho} \bar{J_T ^2}  \over \tau _{jeans}}.
\end{eqnarray}
If we consider a shell of radius $r$ and thickness $dr$ that we follow in a Lagrangian way, 
we get that $dm = 4 \pi \rho r^2 dr$ is conserved and therefore 

\begin{eqnarray}
\label{turb_angmom_lag}
 d _t \left(  \bar{J_T ^2} \right)
=  - 2 \eta_{diss} { \bar{J_T ^2}^{3/2} \over r^2}  +   \sqrt{4 \pi G \rho} \bar{J_T ^2}.
\end{eqnarray}
where we have assumed for the sake of simplicity that $\tau_{jeans} = \sqrt{4 \pi G \rho}$. 

Two possible asymptotic regimes are worth discussing.

Assuming stationarity, we thus get 
\begin{eqnarray}
\label{turb_stat}
  \sqrt{ \bar{J_T ^2} }   =  { \sqrt{ \pi G \rho} r^2 \over    \eta_{diss} }.
\end{eqnarray}
while in a situation where gravitational instability does not develop
we get
\begin{eqnarray}
\label{turb_decay}
\sqrt{ {\bar J_T^2} } =  \left( J_{T,0} ^{-1} + \eta_{diss} \int dt r^{-2}  \right) ^{-1}.
\end{eqnarray}
that is to say $J_T$ is continuously decaying. 

\subsection{Isotropic turbulent equation}

If for the sake of simplicity we  assume that $v_r$ and $v_t$ are on average equal, although we see 
from Eqs.~(\ref{turb_rad}) and (\ref{turb_trans}) that in principle the spherical geometry introduced 
an isotropy, one can summing up Eqs.~(\ref{turb_rad}) and (\ref{turb_trans}), replacing the third order terms by a
turbulent diffusion term and the viscous ones by $\eta \rho V_T^3/r$ we obtain 
\begin{eqnarray}
\nonumber
 \partial _t \left( { \bar{\rho} \bar{v_T^2} \over 2} \right)
 +   {1 \over r^2} \partial _r \left( r^2 {\bar{\rho} \bar{v_T^2} \over 2} \bar{V}_r \right)
 + {\bar{\rho} {2 \over 3} \bar{ v_T^2} \bar{V}_r \over r} \\
\nonumber
+  \bar{\rho} {1 \over 3} \bar{v_T^2} \partial_r \bar{V}_r 
+   {1 \over r^2} \partial _r \left( r^3  \eta_{diff} \overline{v_T^2}^{1/2}  \partial _r (\bar{\rho}  \overline{v_T^2})  \right)
= \\
- \eta_{diss} { \bar{\rho} \overline{v_T^2}^{3/2} \over r}.
\label{turb_isotrop}
\end{eqnarray}

\setlength{\unitlength}{1cm}
\begin{figure*}
\begin{picture} (0,12)
\put(0,9){\includegraphics[width=6cm]{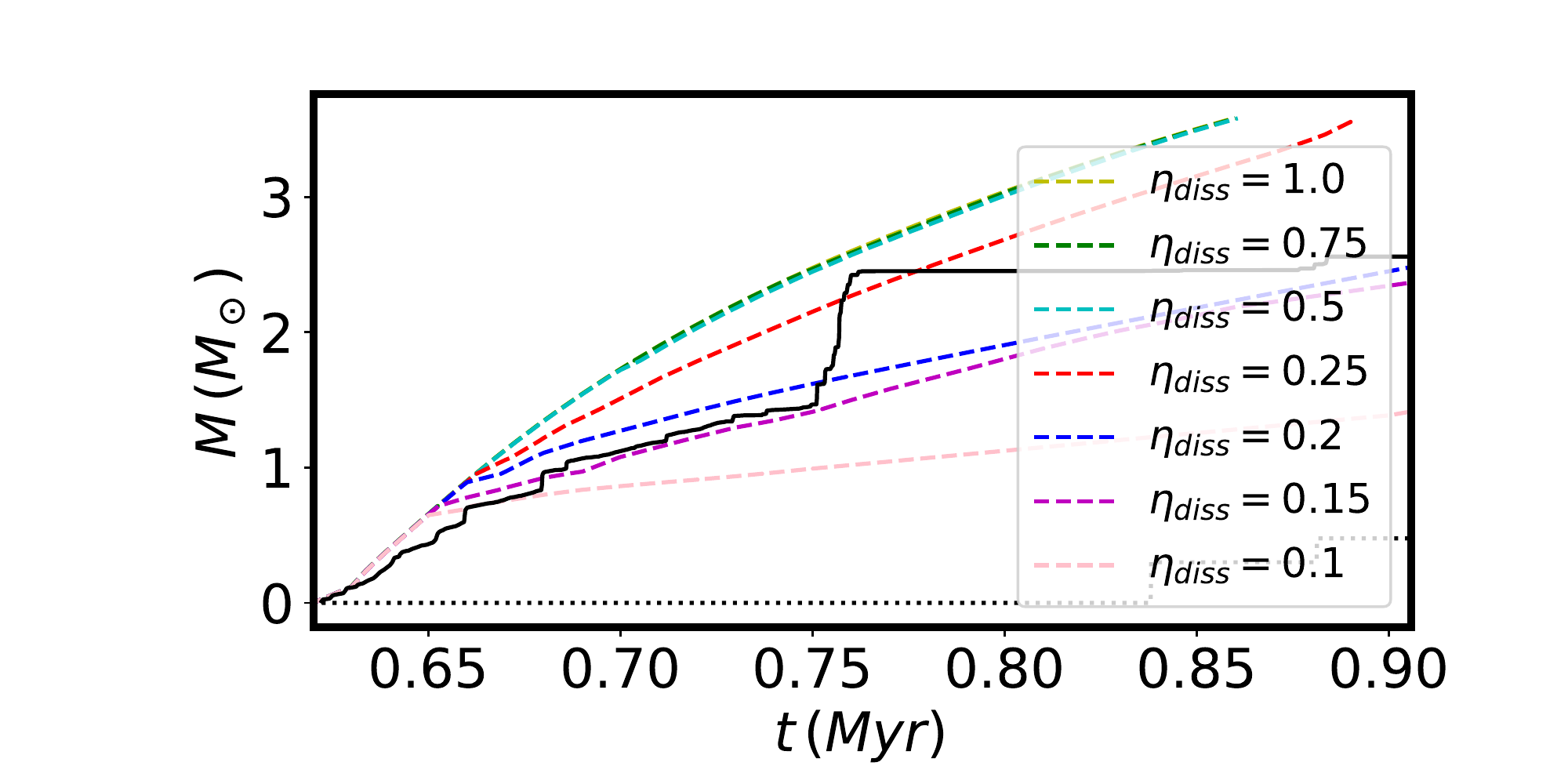}}
\put(1.2,11.2){$A0.5M0.04$}
\put(6,9){\includegraphics[width=6cm]{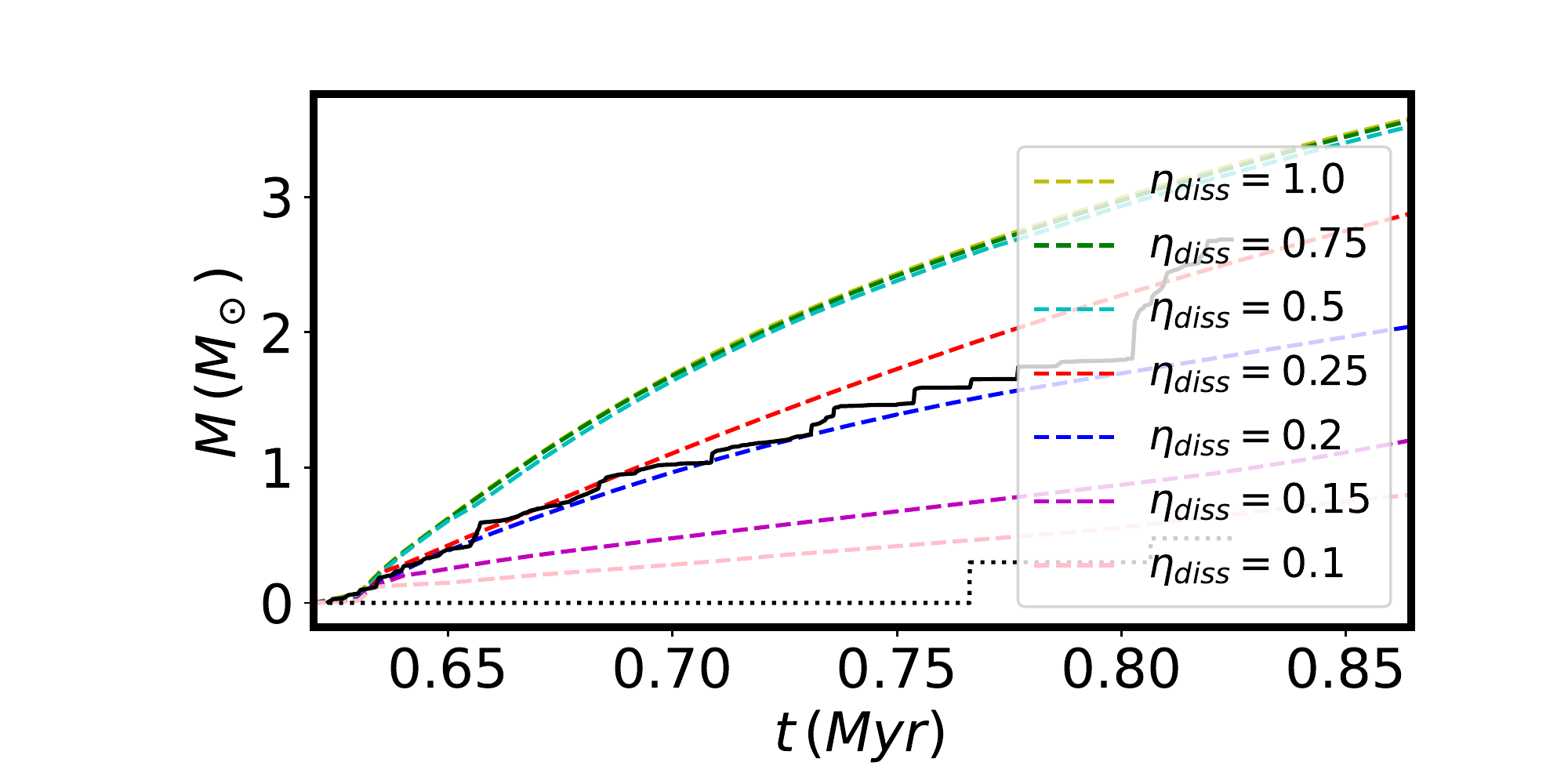}}
\put(7.2,11.2){$A0.5M0.1$}
\put(12,9){\includegraphics[width=6cm]{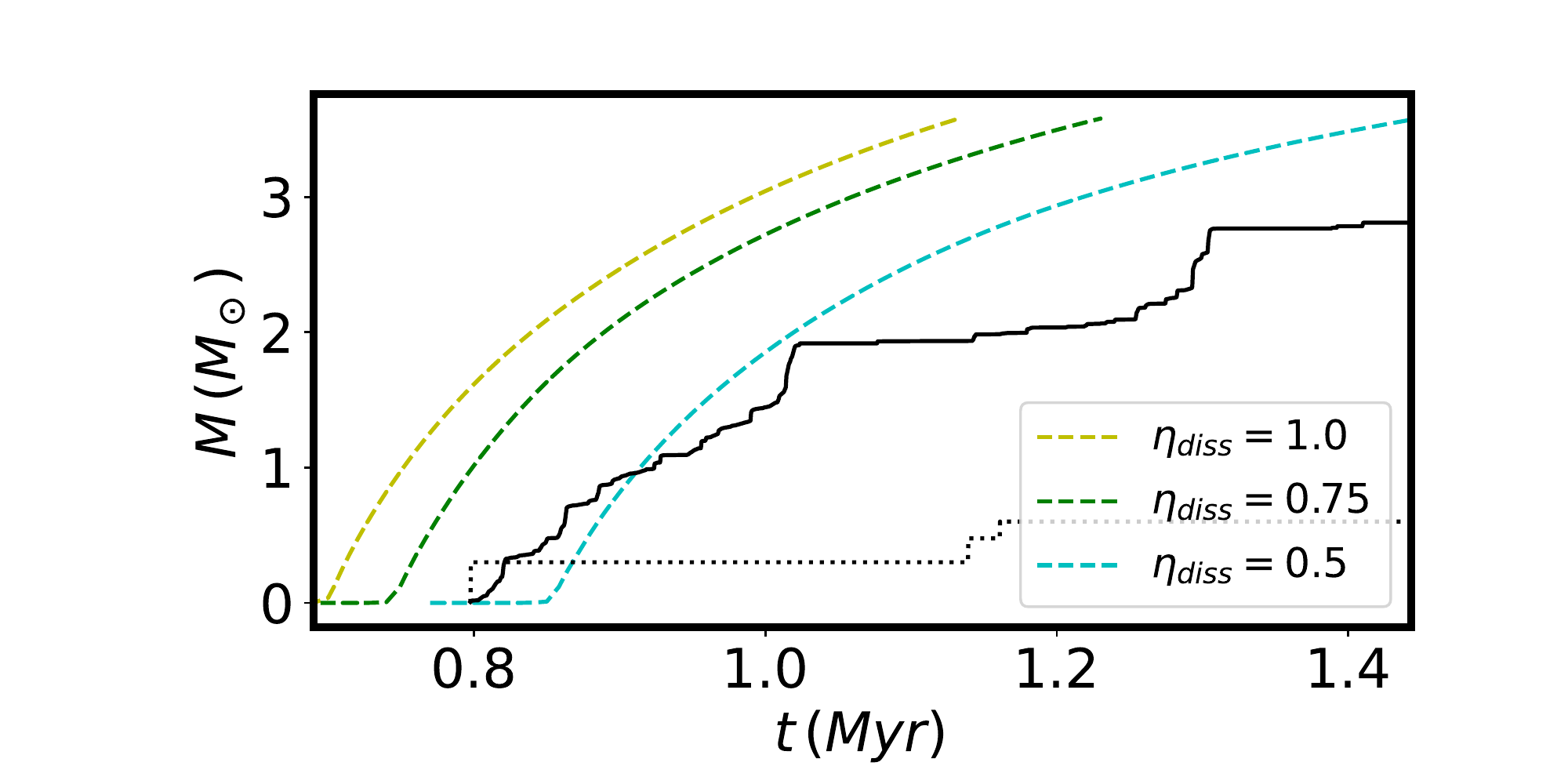}}
\put(13.2,11.2){$A0.5M1$}
\put(6,6){\includegraphics[width=6cm]{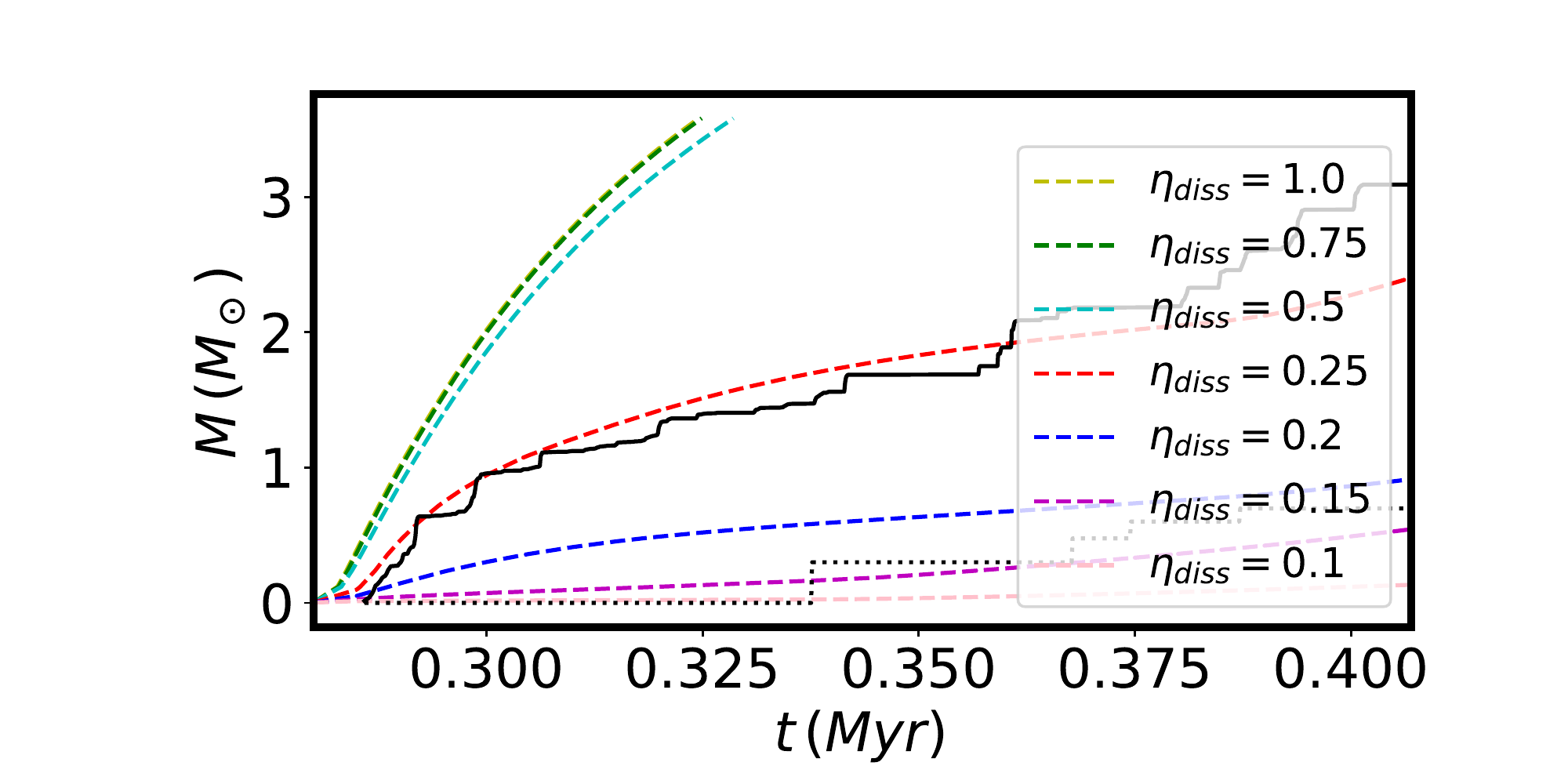}}
\put(7.2,8.2){$A0.3M0.1$}
\put(6,3){\includegraphics[width=6cm]{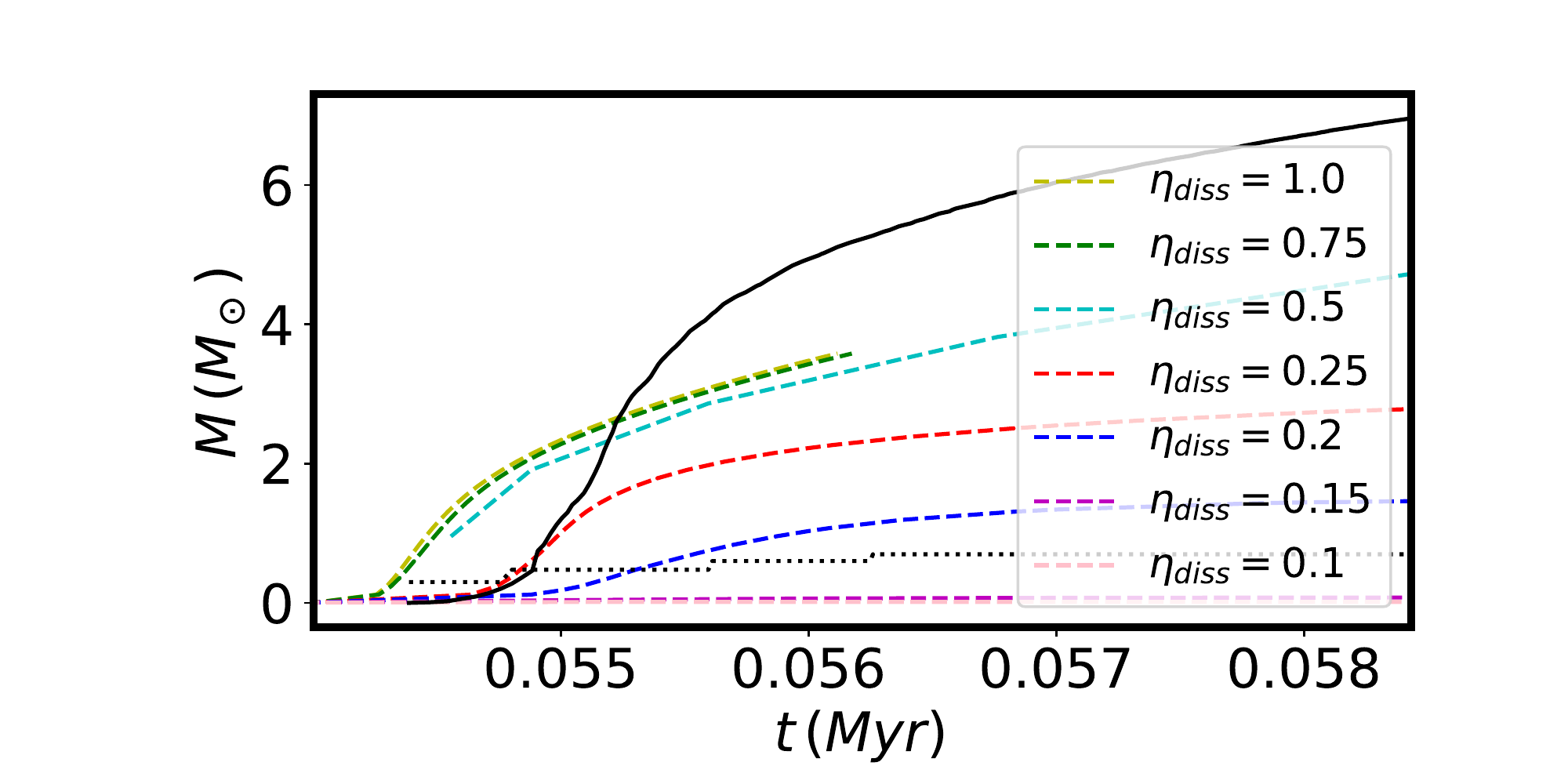}}
\put(7.2,5.2){$A0.1M0.1$}
\put(12,3){\includegraphics[width=6cm]{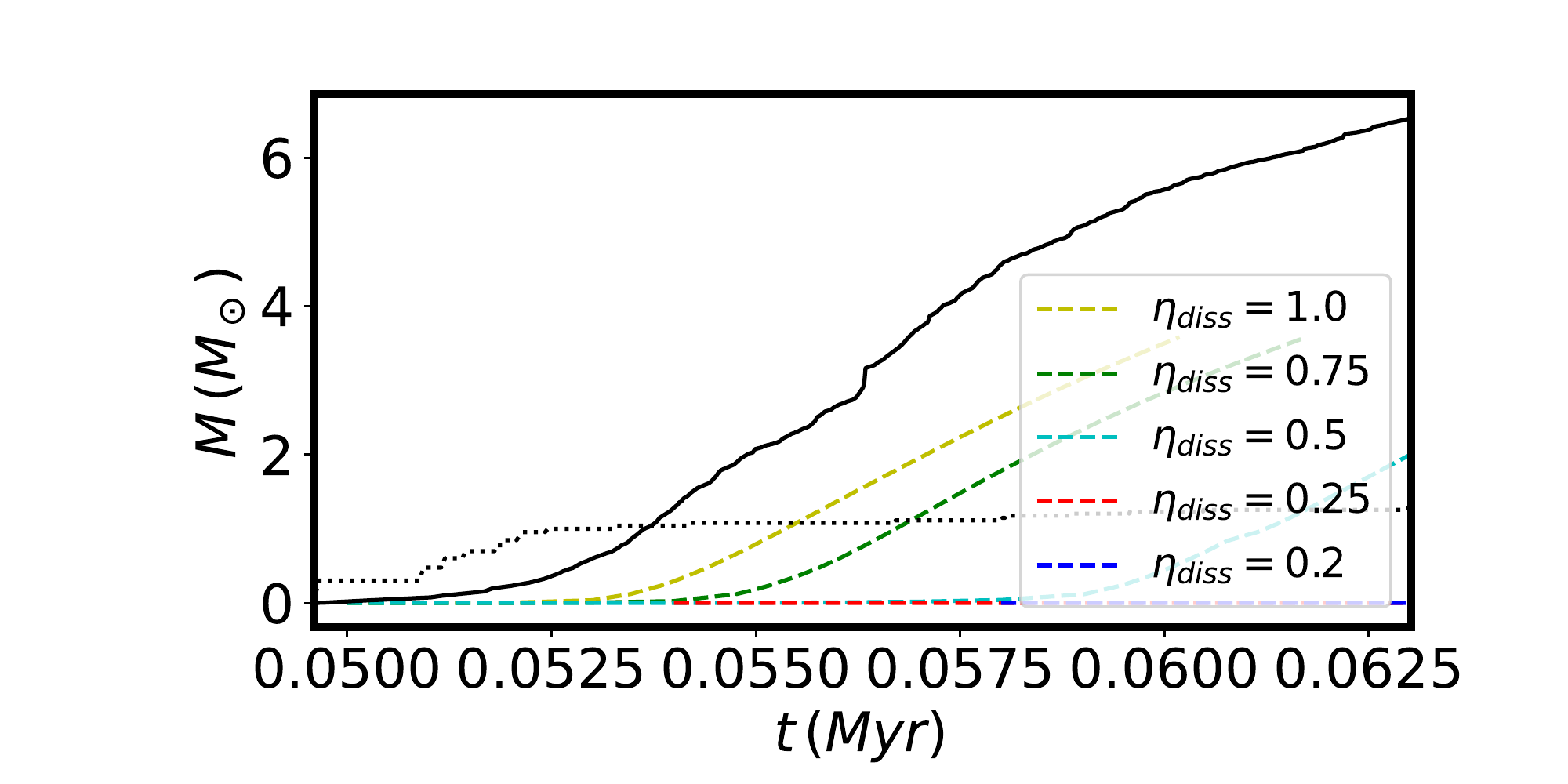}}
\put(13.2,5.2){$A0.1M1$}
\put(6,0){\includegraphics[width=6cm]{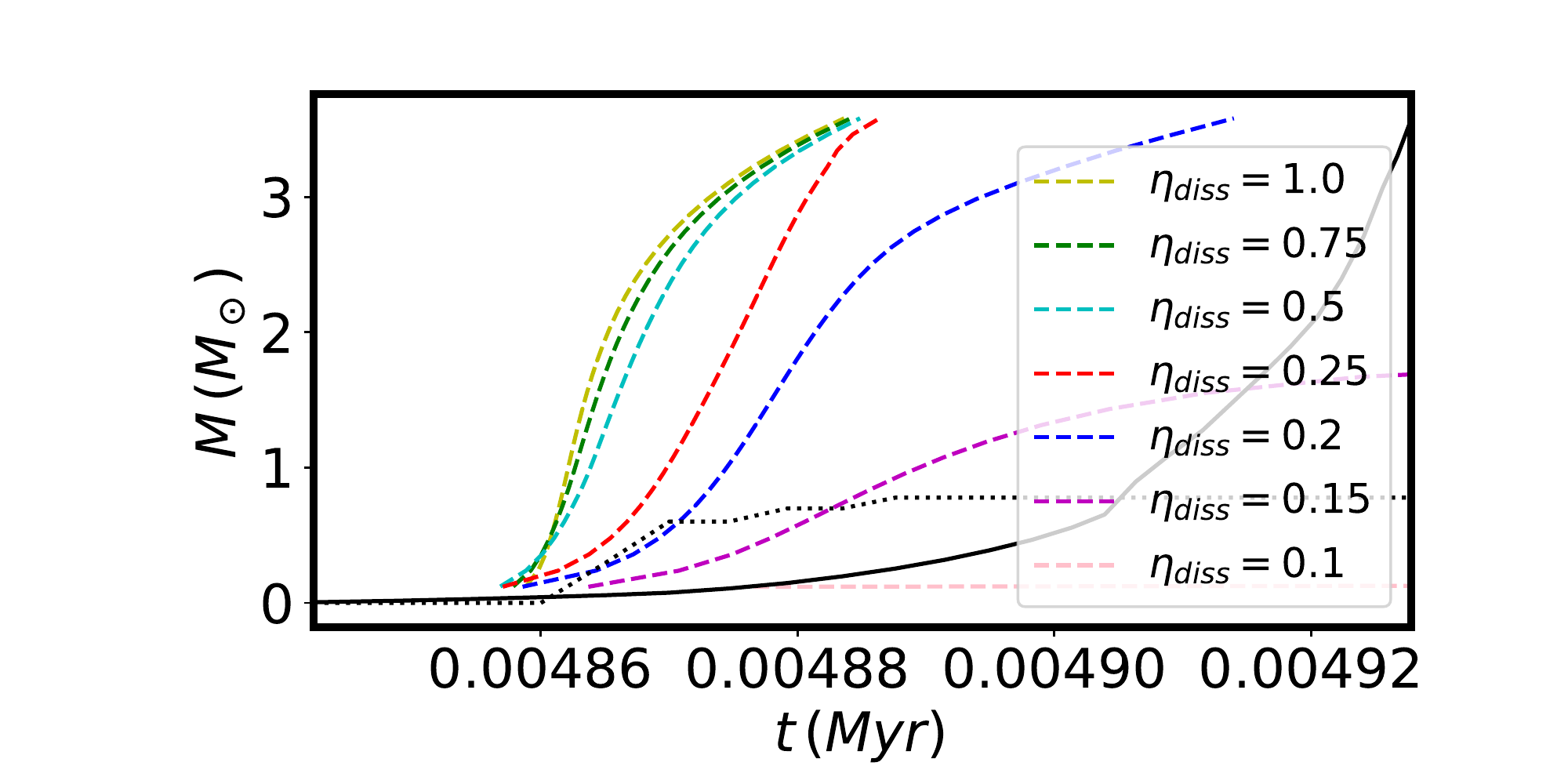}}
\put(7.2,2.2){$A0.02M0.1$}
\put(12,0){\includegraphics[width=6cm]{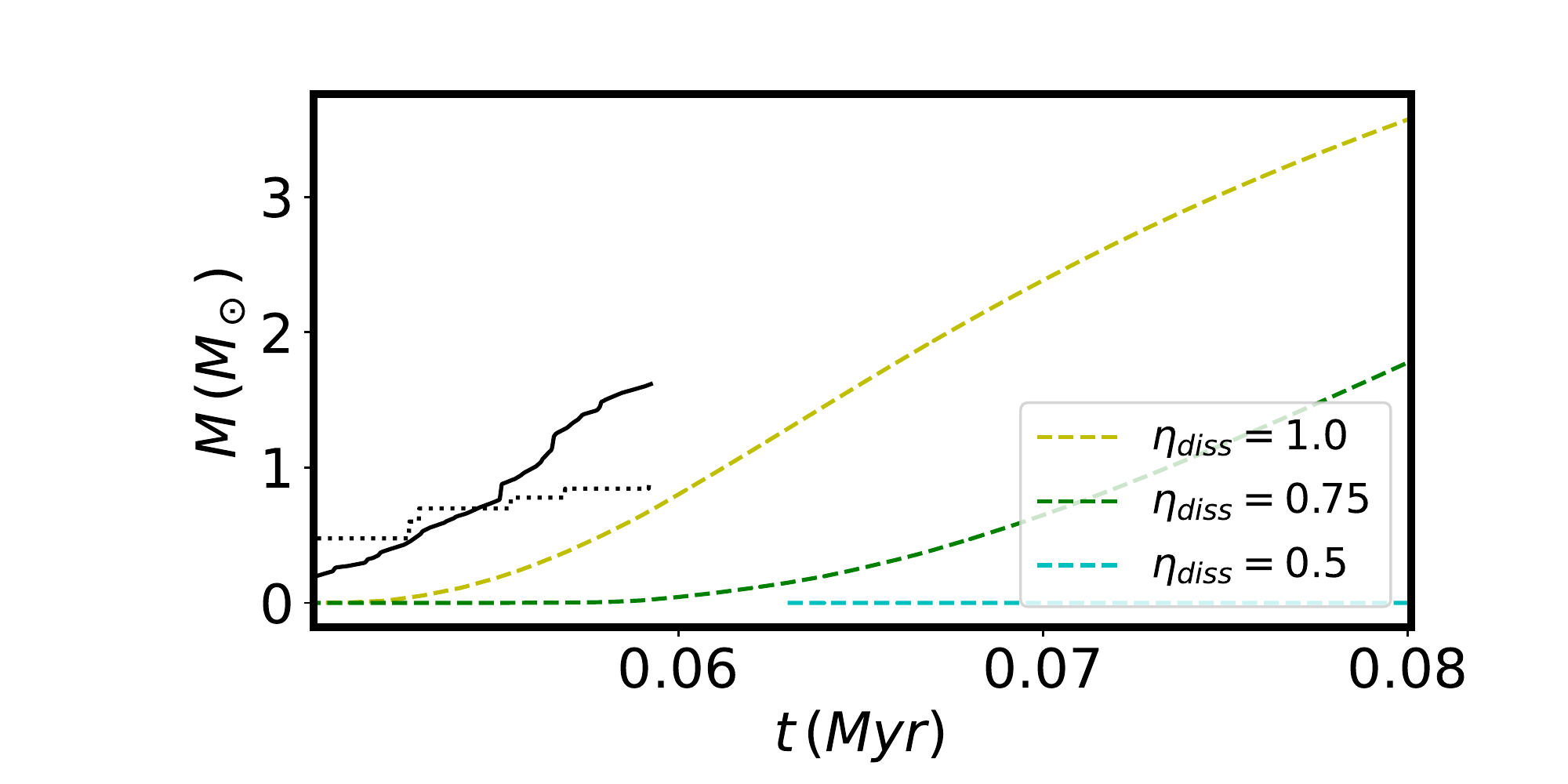}}
\put(13.2,2.2){$A0.1M3$}
\end{picture}
\caption{Accreted mass as a function of time for the various 3D simulations performed and labelled 
in the panel (solid lines) and a series of 1D simulations having the same initial conditions that 
the 3D ones performed with several values of $\eta_{diss}$. The dotted lines is the log of 
the number of sink 
particles that form in the 3D simulations. 
}
\label{accret_comp}
\end{figure*}
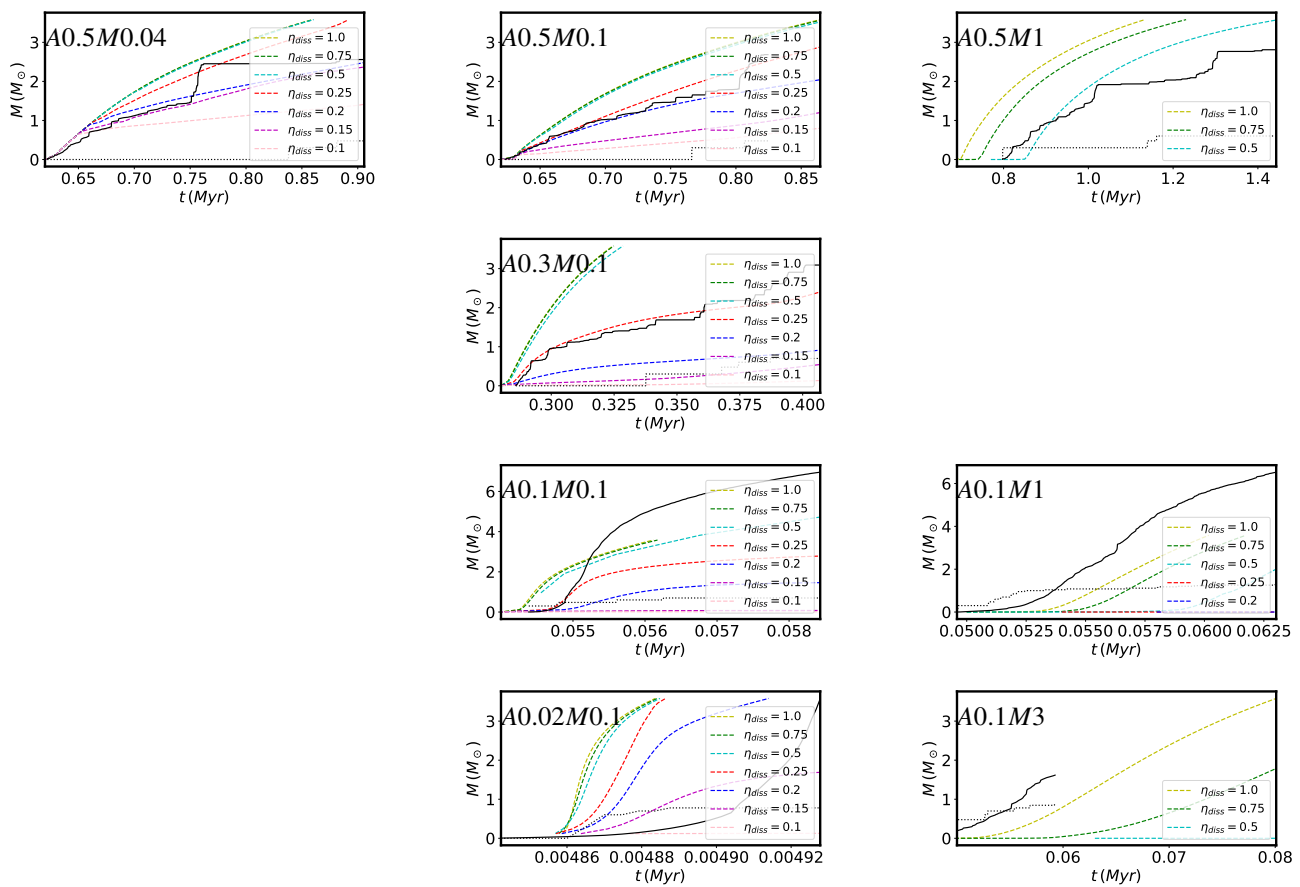

\section{Codes and setup}
Since our goal is to understand the behaviour of turbulence during the gravitational 
collapse, we perform a set of 1D and 3D simulations that we closely compare in 
Section~\ref{results}.

\subsection{1D simulations}
\label{1Dcode}
We have developed a 1D spherical code that solves the set of equations obtained in Section~\ref{1Dequation}, namely 
Eqs.~(\ref{consmat_mean}),~(\ref{consmom_mean2}),~(\ref{turb_rad2_jeans}),~(\ref{turb_trans2_jeans}) together with the Poisson 
equation. 

The code uses a logarithmic spatial grid and employes vanishing gradient boundary conditions both 
at the inner and outer boundaries. A finite volume, Godunov method is employed with a second order Muscl-Hancock scheme.
The flux at the cell interfaces are computed using an HLL solver which in Cartesian geometry leads to exact conservation 
of mass, energy and momentum. 
Note that Eqs.~(\ref{consmom_mean2}),~(\ref{turb_rad2_jeans}),~(\ref{turb_trans2_jeans}) entail source terms which cannot 
be expressed in conservative form. 

The code has been widely tested using exact homologous collapse solutions which are easy to infer. As the comparisons
performed below with the 3D simulations, indeed, constitute another set of extensive tests, we do not detail further
them here.

In Eqs.~(\ref{turb_rad2}) and (\ref{turb_trans2}) the dissipation term, is proportional to $1/r$. While such a dependence 
is meaningful once the  collapse is fully developed, it is not appropriate before this is the case. In particular, 
if initially we consider a uniform density, there is no reason why the energy dissipation should 
be faster in the cloud inner part. To estimate the cloud local dynamical scale, we simply use the density
field and write  $L _{diss} =  \sqrt{\rho_0 / \rho}$, where $\rho_0$ is the initial cloud density. 

The spherical mesh extends from $3 \, 10^{-3} \times r_c$ to $5 \times r_c$, where $r_c$ is the cloud radius. 
The initial conditions consist in a uniform density medium inside $r_c$ and one hundredth this value outside 
the cloud.  The radial velocity is initially zero while the turbulent ones $v_t$ and $v_{t,r}$ are 
proportional to $r^{1/3}$ as it should for a turbulent flow. Their mean values are adjusted 
in such a way that the mean Mach number inside the cloud is equal to the desired value. We also impose 
$v_{t}^2=2 v_{r}^2$ initially.

\subsection{3D simulations}
The 3D simulations are performed using the adaptive mesh refinement code Ramses \citep{Teyssier02,Fromang06}. 
Initially, the computational box, which has  a total size equal to $8 \times r_c$, is described by a grid base
of $64^3$ points (corresponding to level 6 in Ramses). 
Since we use at least 40 cells per Jeans length, the cloud has initially a resolution that corresponds
to level 8, leading to about 64 cells to describe the cloud diameter initially. As time proceeds, 
more amr levels are being added until level 14 is reached. Finally once the density reaches a value 
of 10$^{11}$ cm$^{-3}$ a Lagrangian sink particle is being added \citep{Bleuler14}. 

The initial density distribution is strictly identical to the one of the 1D simulation, i.e. a spherical 
uniform density cloud 100 times denser than the surrounding medium. One fundamental difference is however with the 
initial turbulence. The velocity field is initialised using a stochastic field that employed random 
phases and presents a powerspectrum with a slope equal to $-5/3$. This allows to mimic the expected 
statistical properties of a turbulent velocity field.

\subsection{Runs performed}
To understand the behaviour of turbulence within a collapsing cloud, we perform 
a series of runs and vary 3 parameters, the mean Mach number, $\mathcal{M}$, the 
initial ratio of thermal over gravitational energy, $\alpha$ and the effective adiabatic index.
Table~\ref{table_param_num} summarizes the various runs performed. 
Since all runs are either isothermal or employed and effective polytropic exponent of 1.25, they do not present 
any characteristic spatial scale and can be freely rescaled once the value of $\alpha$ is determined.
To fix ideas, we choose a mass, $M_c$, equal to 10 $M_\odot$ and a radius as specified in Table~\ref{table_param_num}.
This is typical of dense prestellar cores \citep[e.g.][]{ward2007} but we stress that theses results are not 
restricted to these specific choices and could certainly be applied to objects as different than 
a collapsing star or  a massive star forming clump.
For the isothermal runs the gas temperature is kept fixed to 10K while for the run with an effective 
barotropic equation of state, we simply have $T = 10 \, K \, (n/n_0)^{0.25}$, where $n_0$ is the initial
cloud density.

      \begin{table}
         \begin{center}
            \begin{tabular}{lcccccccc}
               \hline\hline
               Name & $r_c$ (pc) & $\alpha$  & $\mathcal{M}$   & $\Gamma$    \\
               \hline
               $A0.5M0.044$ &  0.239 & 0.5 & 0.044 & 1 \\
               $A0.5M0.1$ &  0.239 & 0.5 & 0.1 & 1 \\
               $A0.5M01$ &  0.239 & 0.5 & 1 & 1 \\
               $A0.5M0.3$ &  0.1435 & 0.3 & 0.1 & 1 \\
               $A0.1M0.1$ & 0.0478 & 0.1 & 0.1 & 1 \\
               $A0.1M1$ &  0.0478 & 0.1 & 1 & 1 \\
               $A0.1M3$ &  0.0478 & 0.1 & 3 & 1 \\
               $A0.02M0.1$ &  0.00957 & 0.02 & 0.1 & 1 \\
               $A0.02M10$ &  0.00957 & 0.02 & 10 & 1 \\
               $A0.5gam1.25M0.1$ & 0.239 & 0.5 & 0.1 & 1.25 \\
            \end{tabular}
         \end{center}
         \caption{Summary of the runs performed. $r_c$ is the initial cloud radius, $\alpha$ the initial thermal over gravitational energy ratio, 
$\mathcal{M}$ is the mean cloud Mach number while $\Gamma$ is the effective adiabatic index. 
}
\label{table_param_num}
      \end{table}

As will be seen below, the thermal over gravitational energy ratio, $\alpha$, is a key parameter to interpret the results. 
Runs $A0.5gam1.25M0.1$ and  $A0.5M0.1$ have initially $\alpha=0.5$ and are therefore 
near thermal equilibrium, which is generally found for values slightly larger than this.
This helps keeping the 3D runs reasonably spherical which is mandatory to perform comparisons with the 
1D simulations. For the same reason, both runs have a low Mach number equal to 0.1 which makes turbulence of 
marginal dynamical significance during most of the collapse.   
 The major difference 
between the 2 runs is obviously that as the collapse proceeds, the thermal support rapidly drops  for run 
 $A0.5M0.1$. Indeed it is well known that 
\begin{eqnarray}
\alpha = { {3 \over 2} k_B T {M_c \over m_p} \over {3 G M_c^2 \over 5 r_c} } \propto r_c^{4 - 3 \Gamma}.
\end{eqnarray}
For $\Gamma=1.25$, we have $\alpha \propto r_c^{0.25}$, while for $\Gamma=1$, $\alpha \propto r_c$.
Note that run $A0.5gam1.25M0.1$ would broadly correspond to a collapsing star on the verge 
to form a neutron star or a black hole \citep[e.g.][]{janka2007}  while run $A0.5M0.1$ is typical 
of a low mass quiescent prestellar dense core \citep{ward2007}.

Runs $A0.3M0.1$,  $A0.1M0.1$ and  $A0.02M0.1$  also have a Mach number of 0.1 initially but 
they  their initial thermal over gravitational energy $\alpha$  are respectively  0.3, 0.1 and 0.03.
Together with run $A0.5M0.1$, this allows us to explore widely different physical regime. 

 Finally runs $A0.1M1$, $A0.1M3$ and $A0.02M10$  have $\alpha=0.1$ and $\alpha=0.02$ but present Mach numbers that are 
respectively equal to 1, 3 and 10. 
These initially conditions are more representative of high mass 
cores \citep[e.g.][]{ward2007} or high mass star forming clumps \citep[e.g.][]{elia2017}.
This is complemented by run $A05M1$ which presents high thermal and turbulent support and 
is more representative of lower mass cores. 
For these 3 runs the turbulence is truly dynamically significant. The drawback for the present study is that, 
this induced major departure from the spherical symmetry making the comparison with the 1D simulations more qualitative.

\subsection{Choice of timesteps and displayed quantities}
By performing close comparisons between the 2 sets of simulations, we expect to 
accurately test the validity of Eqs.~(\ref{turb_rad2}) and (\ref{turb_trans2}).
However performing such 
detailed comparisons is not completely straightforward because the collapse time 
decreases as $1/\sqrt{\rho}$ and therefore small differences, due to
the intrinsic differences between a 1D configuration and a 3D one, which 
is not fully spherical,
lead to significant  shift in time once the collapse is more advanced.
For this reason we have chosen in order to perform comparisons, to select timesteps which have nearly identical
central densities before the central sink forms, and which have nearly identical  sink masses after it forms.
Practically many timesteps of the 1D calculations are stored and the closest timestep 
from the 3D ones are chosen. 
We usually display 3 timesteps before the pivotal stage, i.e. the instant where a singularity forms which is 
signed by the appearance of the sink particle and 2 time steps after, selecting a sink mass of few 0.1  and $\simeq$2 $M_\odot$
respectivelly.

For the 1D simulations, the displayed quantities are directly the computed ones, i.e. the density, the radial 
velocity, the transverse velocity and the root mean square radial velocity fluctuations.
For the 3D simulations, all quantities are computed in concentric shells as the mass weighted mean value. 
For instance we have
\begin{eqnarray}
v_t^2 =  { \int \rho \left( {\bf V} \times {{\bf r} \over r} \right)^2  d \Omega r^2 dr  \over \int \rho d \Omega r^2 dr},
\end{eqnarray}
\begin{eqnarray}
v_{r}^2 =  { \int \rho \left( {\bf V} . {{\bf r} \over r}  - V_r \right)^2  d \Omega r^2 dr  \over \int \rho d \Omega r^2 dr}.
\end{eqnarray}
The shell center is chosen to be the most massive sink particles.

\setlength{\unitlength}{1cm}
\begin{figure*}
\begin{picture} (0,8)
\put(7,4){\includegraphics[width=8cm]{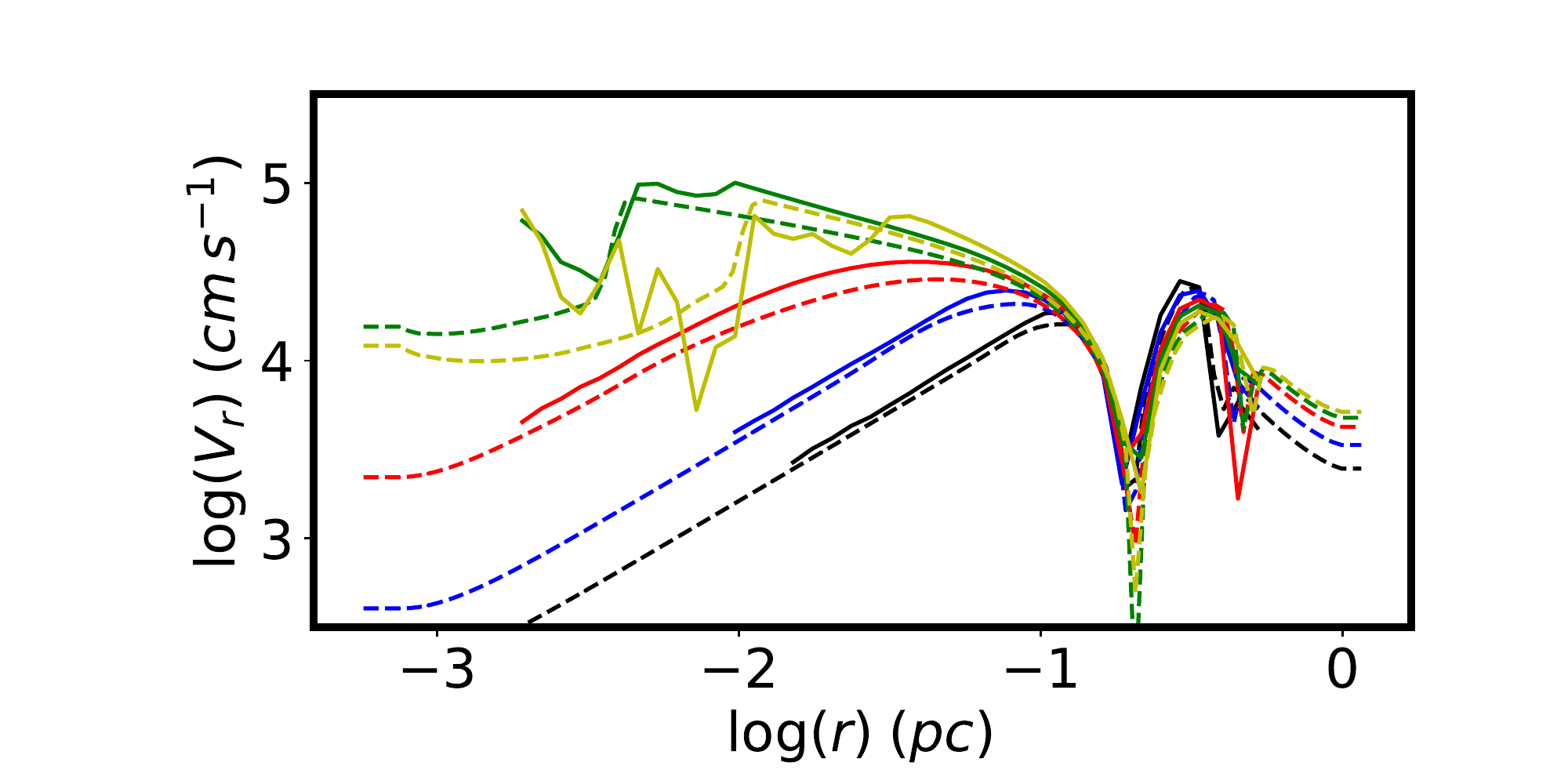}}
\put(0,4){\includegraphics[width=8cm]{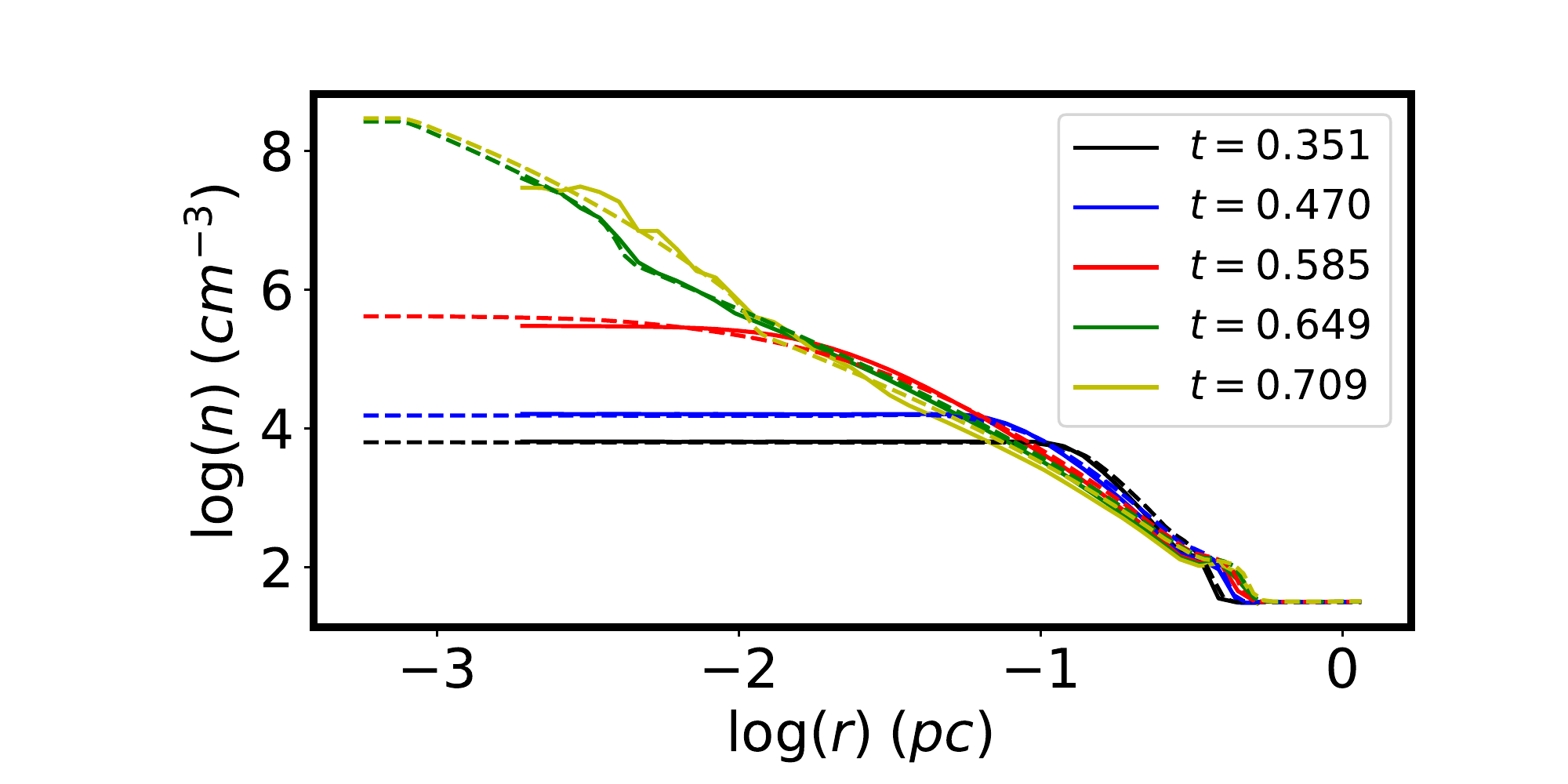}}
\put(1.5,7.7){$A0.5M0.1$}
\put(7,0){\includegraphics[width=8cm]{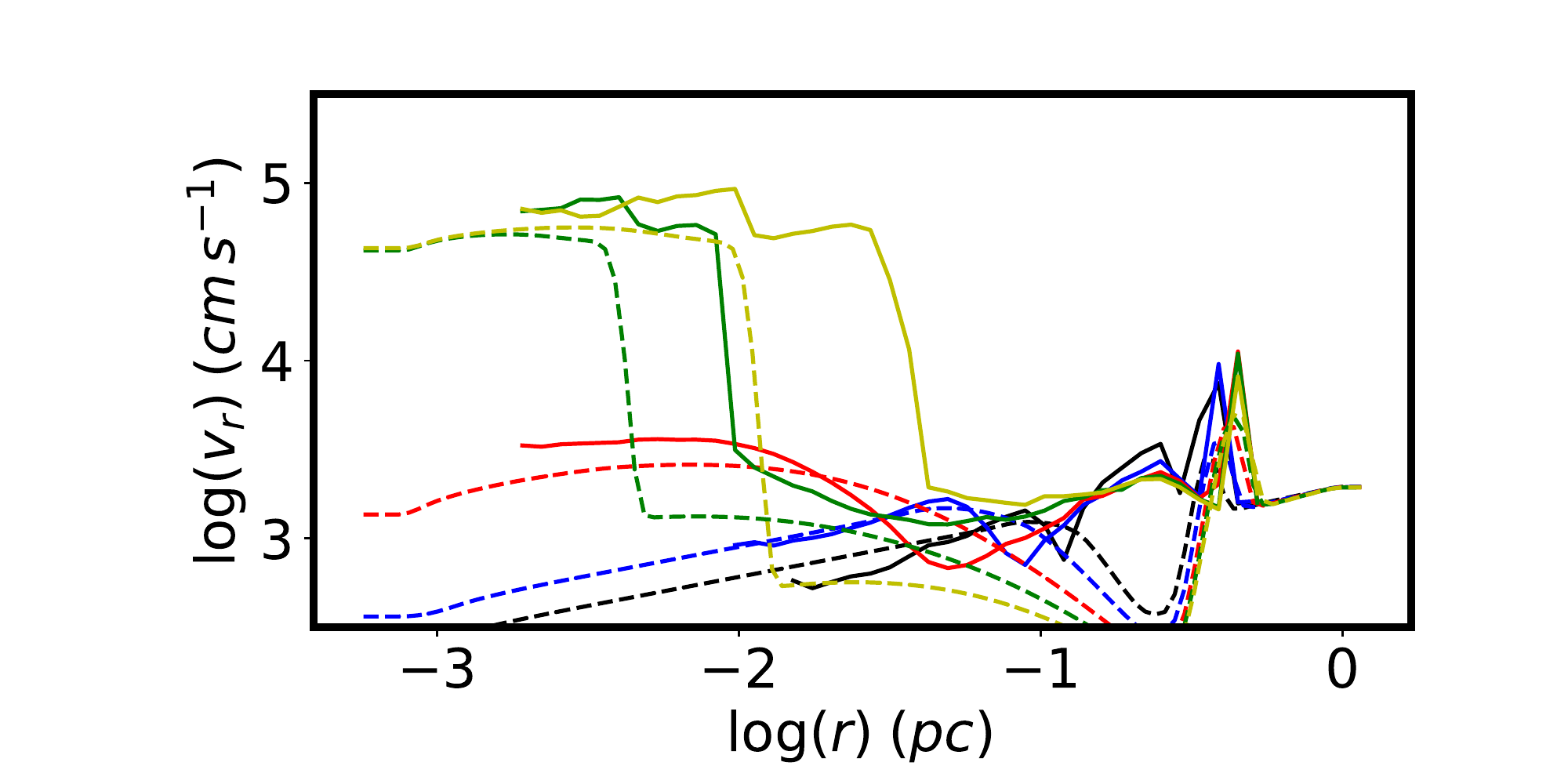}}
\put(0,0){\includegraphics[width=8cm]{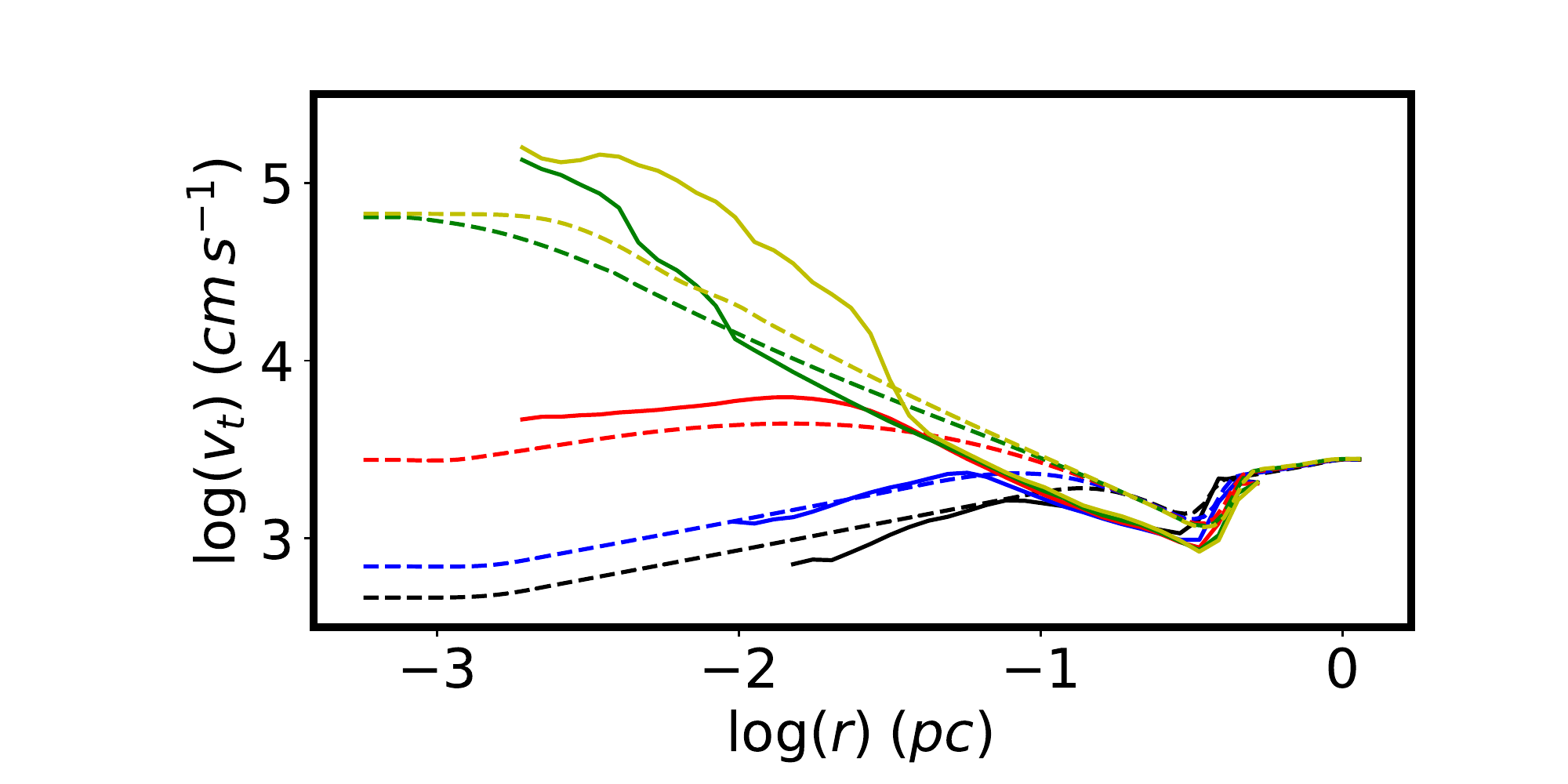}}
\end{picture}
\caption{Run $A0.5M0.1$ ($\alpha=0.5$, $\mathcal{M}=0.1$) at 5 five timesteps (3 before and 2 after 
the sink formation). Full lines: 3D simulations, dashed ones: 1D simulations. The timesteps have been adjusted using 
the value of the central density or the mass of the sink particle. 
The agreement is generally 
quite good for the density and radial velocity (except in the cloud inner part after sink formation). 
Before the sink formation, the agreement with the transverse velocity component is also good though less tight than for 
$n$ and $V_r$. After the sink formation, we observe major deviations for $v_t$ and $v_{r}$. 
 For the radial velocity fluctuation, $v_{r}$, the agreement 
remains comparable than for $v_t$ except at late time in the cloud inner part and generally outside the cloud
($r > 0.1$ pc.) }
\label{gam1_alpha0.5}
\end{figure*}

\setlength{\unitlength}{1cm}
\begin{figure*}
\begin{picture} (0,8)
\put(7,4){\includegraphics[width=8cm]{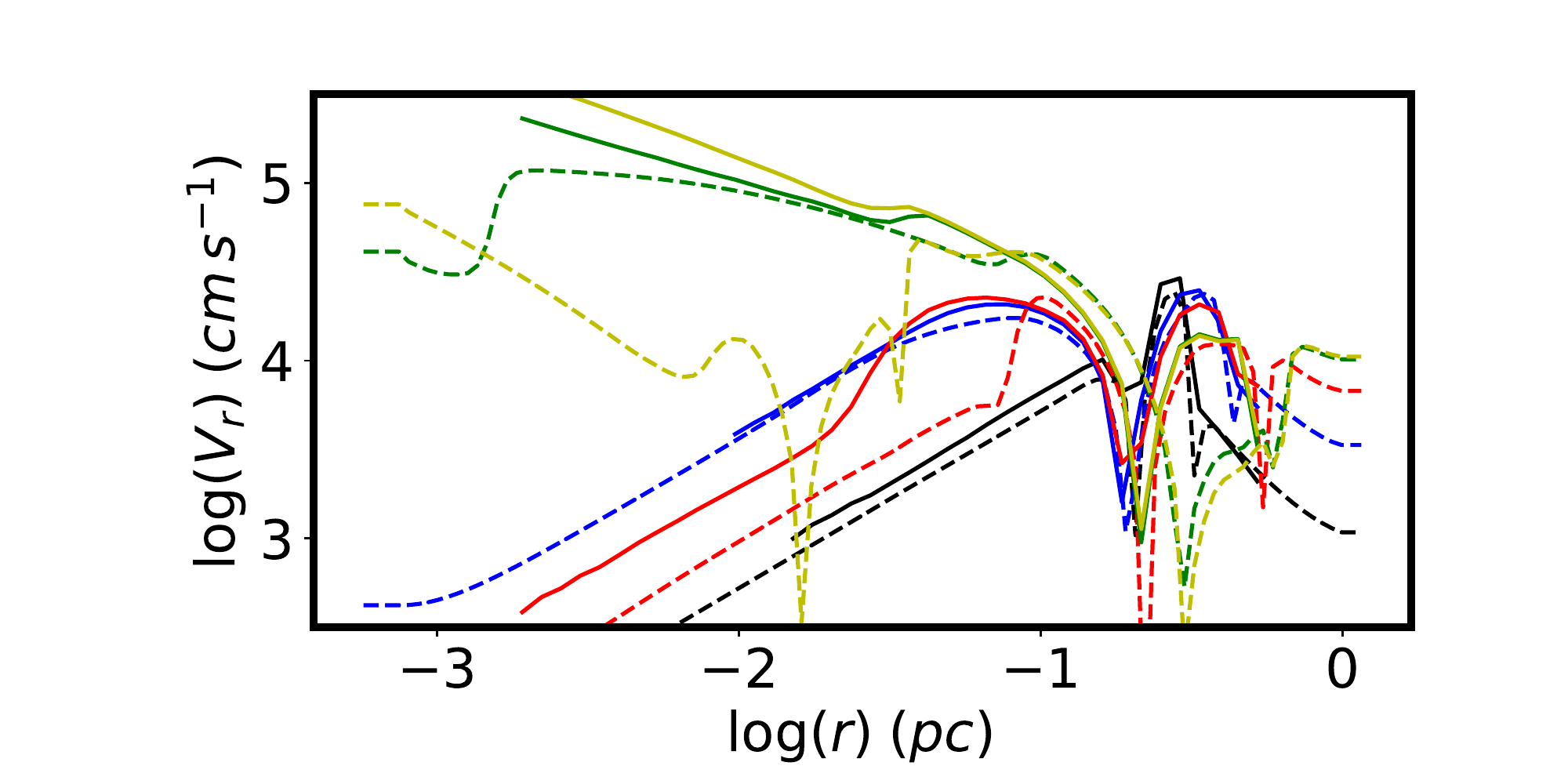}}
\put(0,4){\includegraphics[width=8cm]{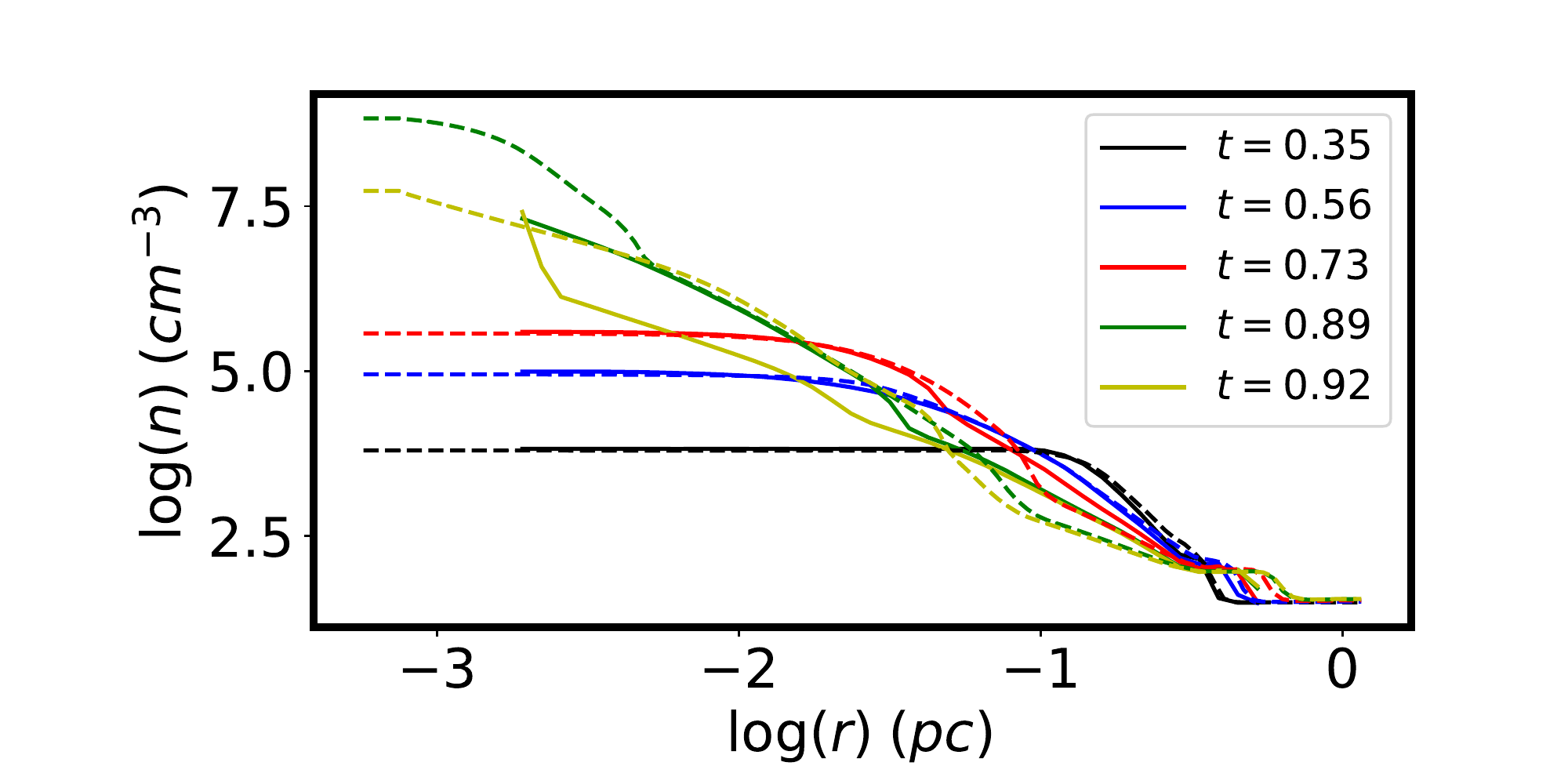}}
\put(1.5,7.7){$A0.5gam1.25M0.1$}
\put(7,0){\includegraphics[width=8cm]{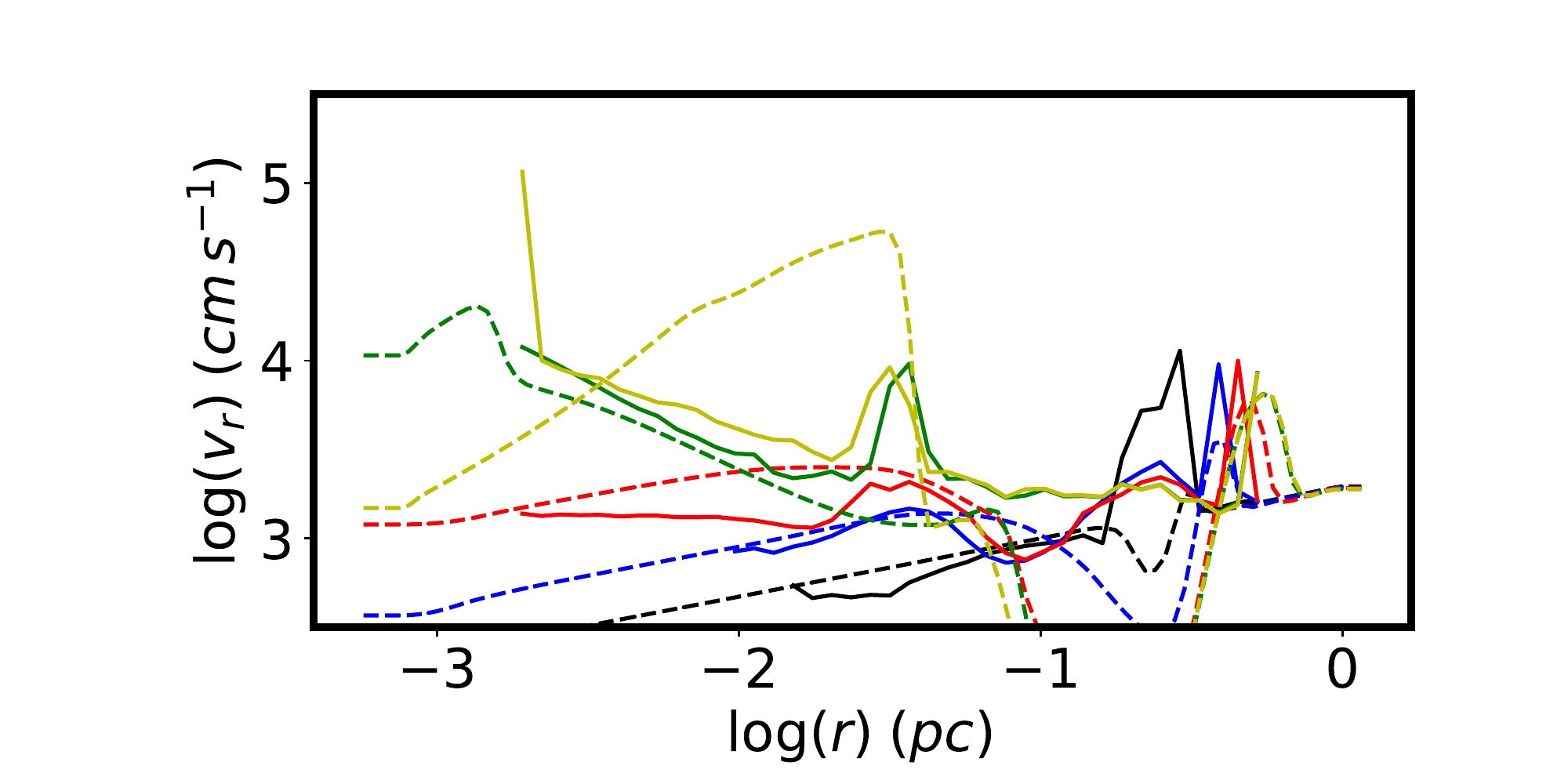}}
\put(0,0){\includegraphics[width=8cm]{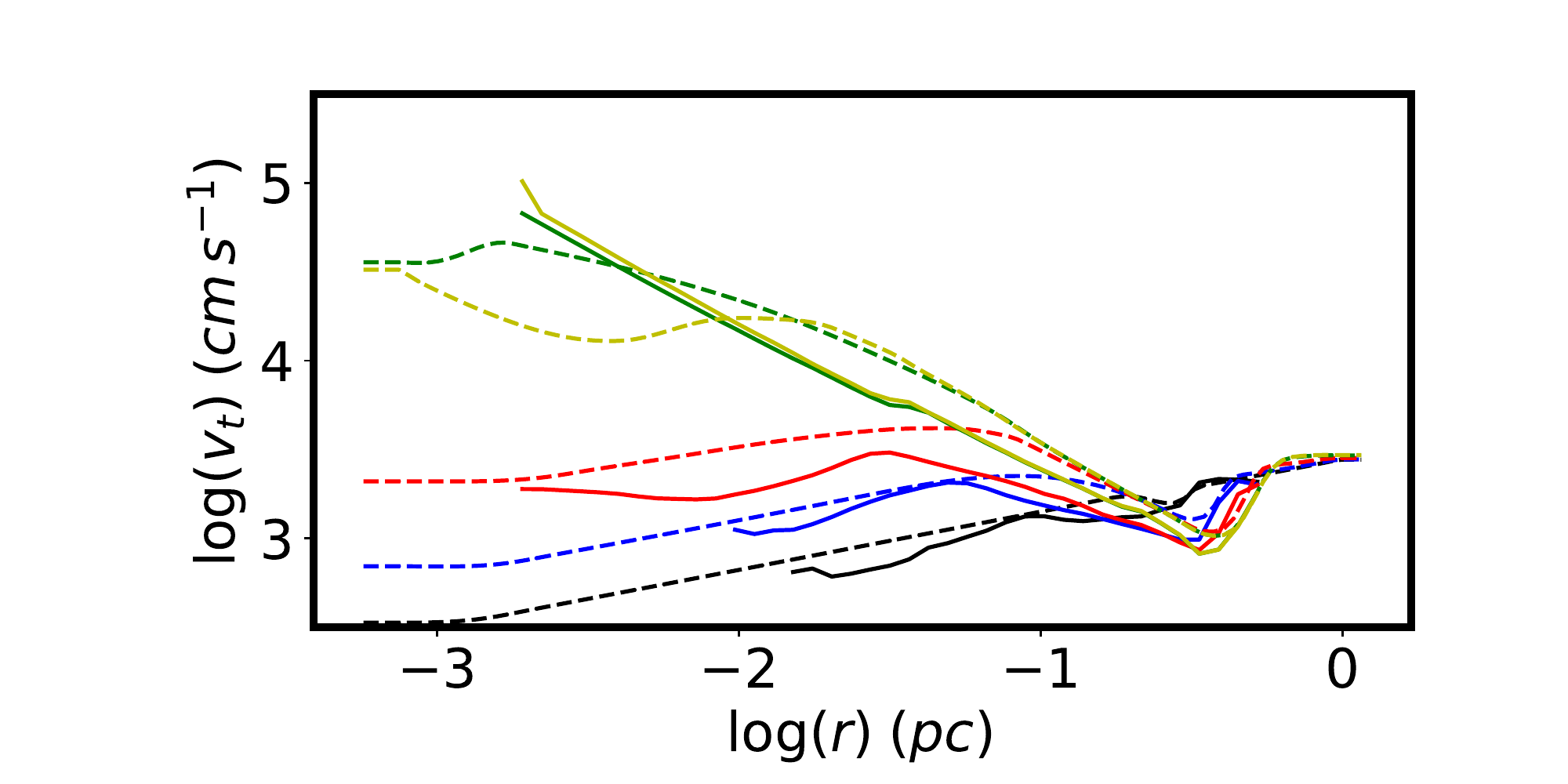}}
\end{picture}
\caption{Run $A0.5gam1.25M0.1$ ($\alpha=0.5$, $\Gamma=1.25$, $\mathcal{M}=0.1$) at 5 five timesteps (3 before and 2 after 
the sink formation). Full lines: 3D simulations, dashed ones: 1D simulations. The timesteps have been adjusted using 
the value of the central density or the mass of the sink particle. As can be seen the agreement is generally 
quite good for the density and radial velocity (except in the cloud inner part after sink formation). 
The agreement with the transverse velocity component is also good though less tight than for 
$n$ and $V_r$. For the radial velocity fluctuation, $v_{r}$, the agreement 
remains comparable than for $v_t$ except at late time in the cloud inner part and generally outside the cloud
($r > 0.1$ pc.)
}
\label{gam125}
\end{figure*}

\setlength{\unitlength}{1cm}
\begin{figure*}
\begin{picture} (0,13.5)
\put(0,9){\includegraphics[width=6.cm]{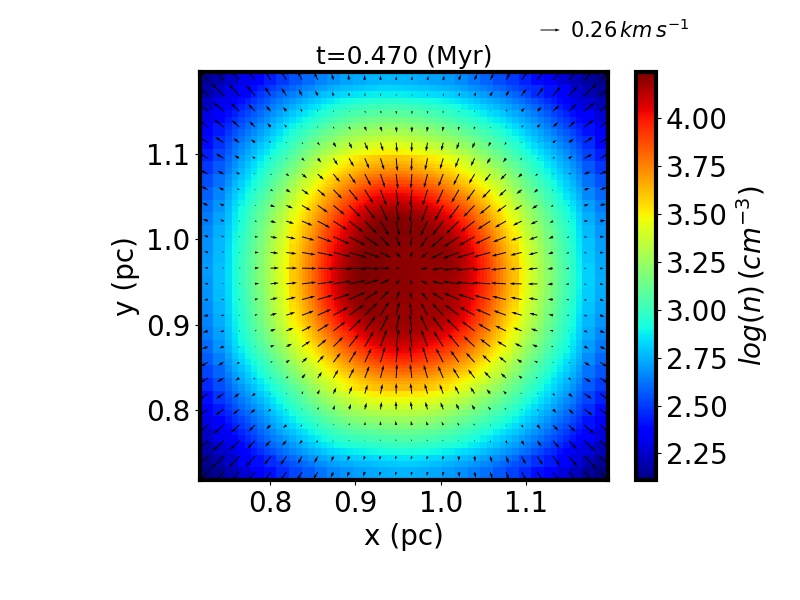}}
\put(6,9){\includegraphics[width=6.cm]{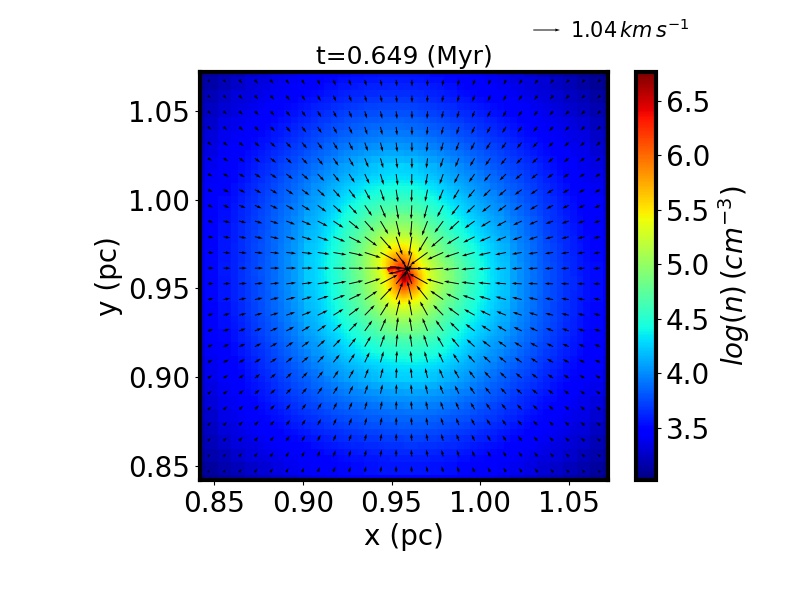}}
\put(12,9){\includegraphics[width=6.cm]{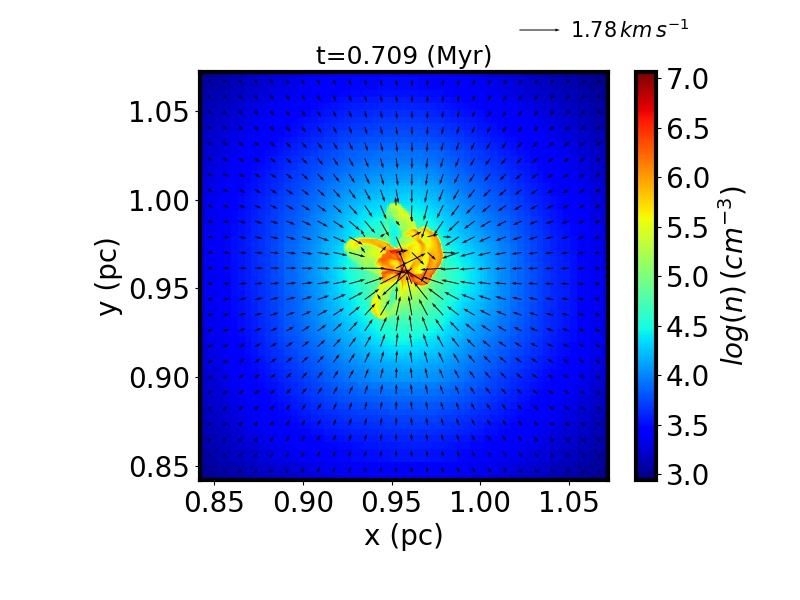}}
\put(1,13.3){$A0.5M0.1$}
\put(0,4.5){\includegraphics[width=6.cm]{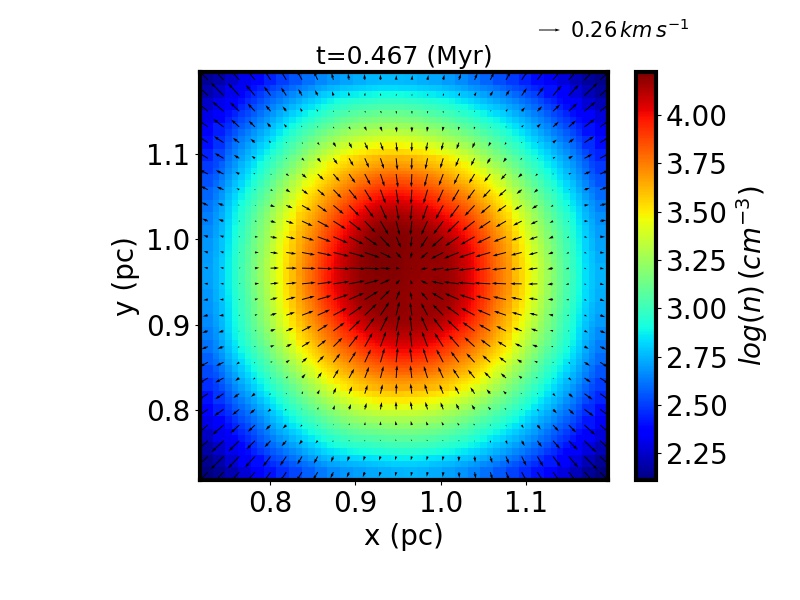}}
\put(6,4.5){\includegraphics[width=6.cm]{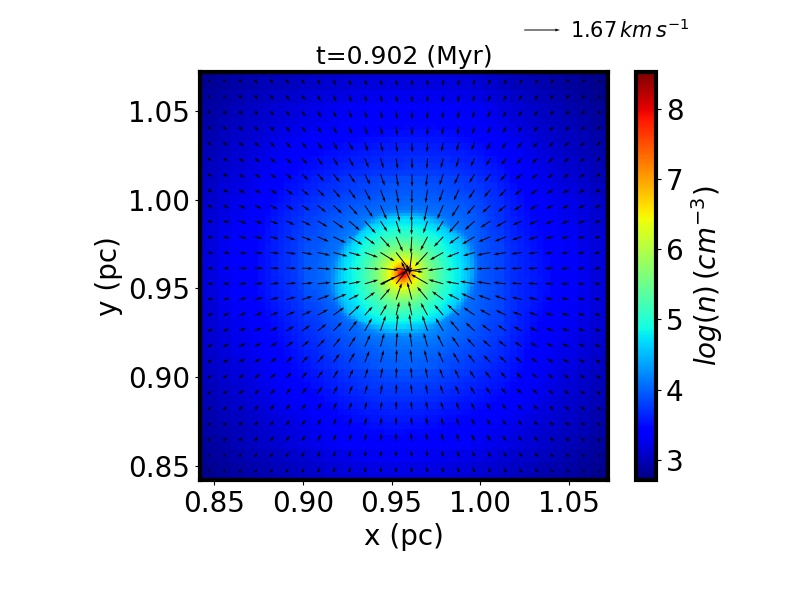}}
\put(12,4.5){\includegraphics[width=6.cm]{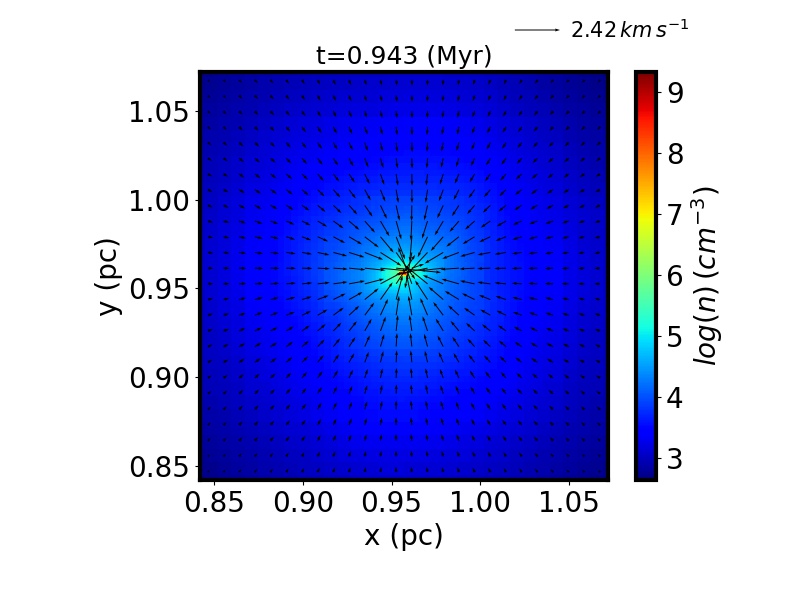}}
\put(1,8.8){$A0.5gam1.25M0.1$}
\put(0,0){\includegraphics[width=6.cm]{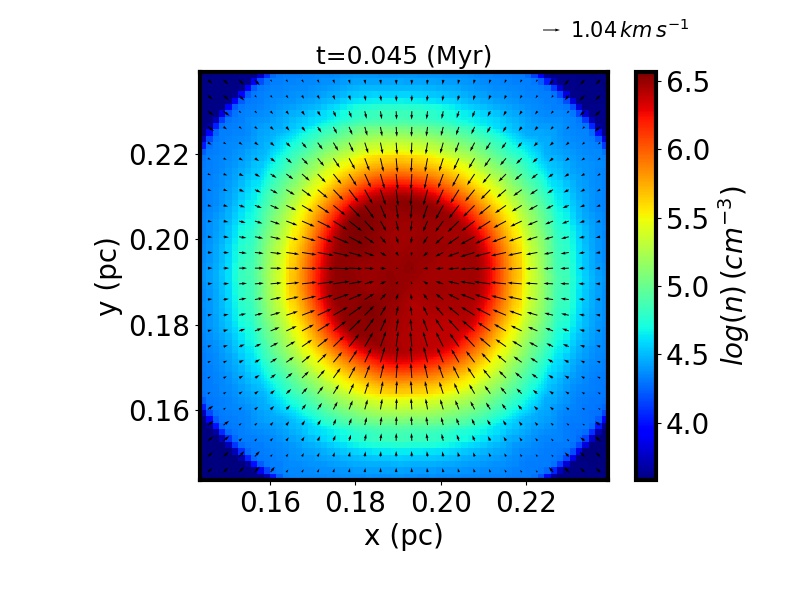}}
\put(6,0){\includegraphics[width=6.cm]{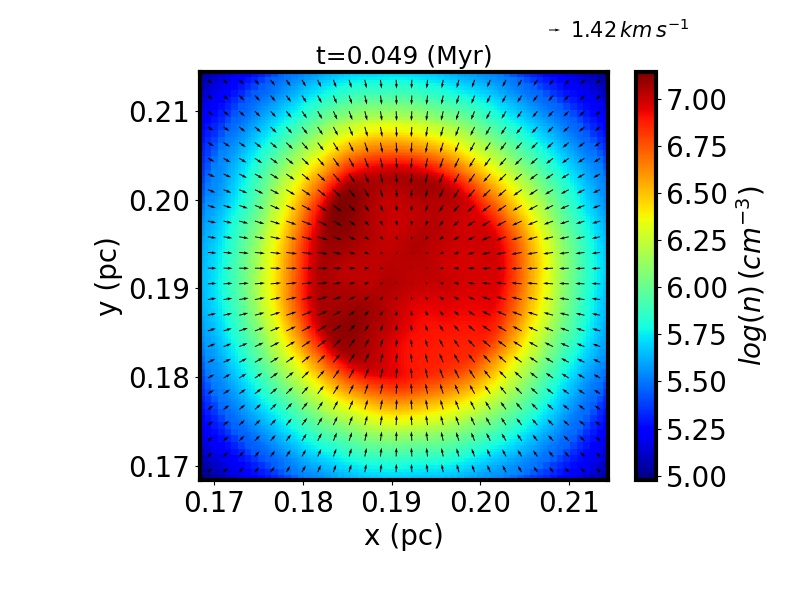}}
\put(12,0){\includegraphics[width=6.cm]{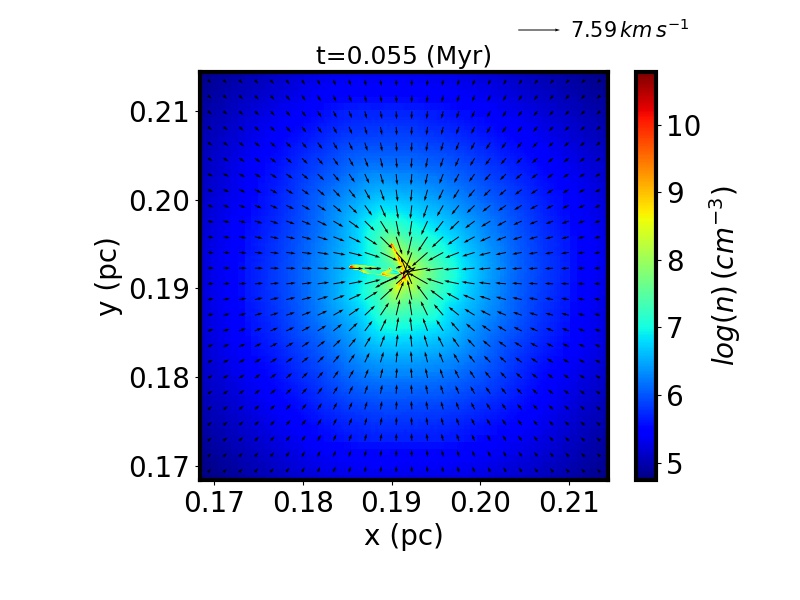}}
\put(1,4.3){$A0.1M0.1$}
\end{picture}
\caption{Density and velocity cut at three timesteps for runs $A0.5M01$, $A0.5gam1.25M01$, $A0.1M01$.
When thermal energy is high (runs $A0.5M01$ and $A0.5gam1.25M01$) the collapse 
remains well symmetric, except after the pivotal stage (top-right panel).
When thermal energy is low (runs $A0.1M01$), the collapsing cloud is extremely unstable and non-axisymmetric
motions develop even before the pivotal stage. 
}
\label{belleimage}
\end{figure*}

\section{Estimating turbulent  dissipation through global comparison between 1D and 3D simulations}
As mentioned above, we first need to estimate the parameter $\eta _{diss}$ which remains undetermined. 

\subsection{Previous estimates}
Performing non self-gravitating driven turbulence simulations, \citet{maclow1999} inferred
values of the turbulent dissipation. Writing
\begin{eqnarray}
\dot{E}_{turb} = - {2 \pi \over L_d} \eta M \sigma^3,
\label{maclow_eta}
\end{eqnarray}
he inferred $\eta=0.067$, where $L_d$ is the turbulent driving scale.

More centently, \citet{guerrero2020} discussed turbulent dissipation in a collapsing cloud.
By assuming a local energy balance, namely $d/dt (E_g / E_{turb}) =0$ and further assuming that the radial 
and infall speeds are comparable, as confirmed 
from their simulations, they infer that $L_d / R \simeq 8.6 \eta$.
Thus the value of $\eta$ depends on the driving scale. Assuming that $L_d=R$, they 
get a value of $\eta \simeq 0.12$. 

Estimating the value of $\eta_{diss}$ in a collapsing cloud is not an easy task. In particular we note 
that from Eqs.~(\ref{turb_rad2_jeans})-(\ref{turb_trans2_jeans}), there are several 
contributions to the turbulence variations, which is advected, amplified in two different ways along 
radial and orthoradial directions and possibly locally generated if the core is locally gravitationally 
unstable. Indeed as we show later, there are cases where turbulence is amplified but not generated. 

We note that from Eq.~(\ref{maclow_eta}) and Eqs.~(\ref{turb_rad2_jeans})-(\ref{turb_trans2_jeans}), 
we have $2 \pi  \eta / L_d \simeq \eta_{diss} / r$. Thus assuming that $L _d \simeq 2 r$, i.e. that the  driving scale 
is a local diameter, we get $\eta_{diss} \simeq \pi \eta = 0.21$.

\subsection{Estimating the dissipation parameter}
To estimate the value of $\eta_{diss}$, we  perform a series of 1D simulations that we compare with the 3D ones. 
More precisely, for each 3D simulations listed in Table~\ref{table_param_num}, we run seven  1D simulations
for $\eta_{diss}=0.1, 0.15, 0.2, 0.25, 0.5, 0.75$ and 1. These values have been chosen because they 
fall in the relevant expected range of values from previous studies but also because they allow 
 to explore the effects of large changes of turbulent dissipation.
A good and important indicator for collapsing 
clouds is certainly the amount of mass, that has been accreted as a function of time. Indeed the accreted 
mass depends on the complete collapse history.
Figure~\ref{accret_comp} portrays the results. The rows and lines respectively 
correspond to constant initial Mach numbers and constant $\alpha$ values. There is one 
exception however, run $A01M3$ is placed for space reason at the bottom and right panel. 

 Obviously the difficulty 
in estimating $\eta_{diss}$ is that in 3D, the collapse may not remain spherically symmetric particularly when 
the turbulent energy is initially high and this is why several Mach numbers are explored.
We start with discussing run $A0.5M0.1$ (which we remind has $\alpha=0.5$ and $\mathcal{M}=0.1$).
The black solid line represents the accreted mass of the 3D simulation, the dotted one is the log of the 
number of formed sink particles. The dashed and colored lines represent 1D models with a 
value of $\eta_{diss}$ as indicated in the legend. 
As can be seen the turbulent dissipation parameter, $\eta_{diss}$, indeed appears to play an 
important role, once the collapse has started i.e. after time $t=0.65$ Myr. A remarkable agreement 
is obtained between the 3D run and the 1D ones with $\eta_{diss}=0.2-0.25$ at least up to time
$t \simeq 0.8$ Myr, where $M \simeq 2$ M$_\odot$. Interestingly, the behaviour for 
smaller and larger values of $\eta_{diss}$ are quite different. For  $\eta_{diss} < 0.15$ 
the accretion rate is significantly reduced while for $\eta_{diss} > 0.25$, it 
is substantially higher. This clearly is a consequence of the turbulence that is respectively too strongly dissipated or 
 amplified. While the collapse behaviour stiffly depends on $\eta_{diss}$ when the value of this parameter 
is low, the collapse is completely unchanged when $\eta_{diss}$ is further increased from 0.5 to 1. This is because the turbulence 
is so efficiently dissipated that it does not contribute significantly to the collapse. 

This conclusion is supported 
by  run $A0.5M0.044$, which has an initial turbulent energy roughly 4 times lower than run $A0.5M0.1$. 
The collapse with high  $\eta_{diss}$ is nearly indistinguishable from runs $A0.5M0.1$ having high $\eta_{diss}$. 
The dependence on $\eta_{diss}$ is generally also similar although good agreement between 1D and 3D runs 
are obtained for $\eta_{diss} \simeq 0.15-0.2$, suggesting a slightly lower values than for runs $A0.5M0.1$. 

The behaviour of run $A0.3M0.1$, which has  $\alpha=0.3$ and $\mathcal{M}=0.1$ is very similar to
run $A0.5M0.1$. The agreement between the 3D simulation and the 1D one with $\eta_{diss}=0.25$
appears to be very good. 

For run $A0.1M0.1$, the agreement is also good up to $M \simeq 1.5$ M$_\odot$ after 
which the 3D simulation accretes much faster. The best value is also clearly
$\eta_{diss}=0.25$. We note that in this run, because of the low thermal 
energy ($\alpha=0.1$) several sink particles have formed, which probably helps to accrete 
gas more rapidly in the 3D simulation compared to the 1D ones. 

Run  $A0.02M0.1$ presents a slightly different behaviour. Generally speaking 
the trends are similar but the effective value of $\eta_{diss}$ for which 
the agreement at first sight appears to be the best is between 0.1 and 0.15. However, we see
that for $\eta_{diss}=0.15$ the shape of the curve is quite different from the 3D simulation
which  appears more compatible with $\eta_{diss}=0.2-0.25$. 

The runs with relatively high Mach numbers, namely $A0.5M1$, $A0.1M1$ and $A0.1M3$ present 
a significantly different behaviour. They all have in common to require much higher 
values of $\eta_{diss}>0.5-1$ for the 3D and 1D simulations to match reasonably well. 
Moreover for small values of $\eta_{diss}$, no central mass forms in the 1D runs. Indeed 
the gas would simply bounce back after some contraction. Clearly a large 
dissipation is required to avoid unrealistic turbulent support. The physical 
interpretation is however not straighforward and as discussed in the next section,
it is likely that, at least in part, the high value of $\eta_{diss}$ that is needed, 
is a consequence of several sink particles being formed in the 3D runs.

\subsection{Turbulent support}
It is interesting to compare the timescale for the various runs with same thermal energy 
but different levels of turbulence as $A0.5M0.1$ with $A0.5M1$ or 
$A0.1M0.1$ and $A0.1M1$ with $A0.1M3$.
Typically turbulence delays the collapse and slows down accretion but this delay
remains modest. Indeed while run $A0.1M3$ has a turbulent support that is about 30 times larger 
than run  $A0.1M0.1$, the total accreted mass reaches 2 M$_\odot$ with a delay 
of about 10$\%$ only. As can be seen from Fig.~\ref{accret_comp}, much longer 
delays are obtained for 1D simulations with smaller $\eta_{diss}$ while 
for large $\eta_{diss}$ the behaviour and the timescale are reasonably reproduced. 

A possible major difference between 1D and 3D simulations, is the number of sink particles 
that the latter are generating. This certainly leads to faster accretion. More generally 
this also illustrates the dual role that turbulence is playing by generating 
density fluctuations that favor local collapses. It is likely that these effects 
promote the early formation of sink particles and therefore limit the large delay that 
turbulence would have introduced otherwise. 

Similarly, the high value of $\eta_{diss}$ seemingly suggested by the comparisons
between 1D and 3D $A0.1M3$ simulations is likely, at least in part, a consequence of the 
turbulently induced collapse. In principle, a more advanced model could take into account 
these effects for instance by computing the probability of finding self-gravitating 
density fluctuations \citep[e.g.][]{HC08}. Moreover, once material has been accreted into a star/sink
particles, thermal and turbulent support are retrieved from the gas phase and do not 
contribute to support the gas against gravity.

\setlength{\unitlength}{1cm}
\begin{figure*}
\begin{picture} (0,4)
\put(1.5,3.7){$A0.1M0.1$ - no Jeans source term}
\put(7,0){\includegraphics[width=8cm]{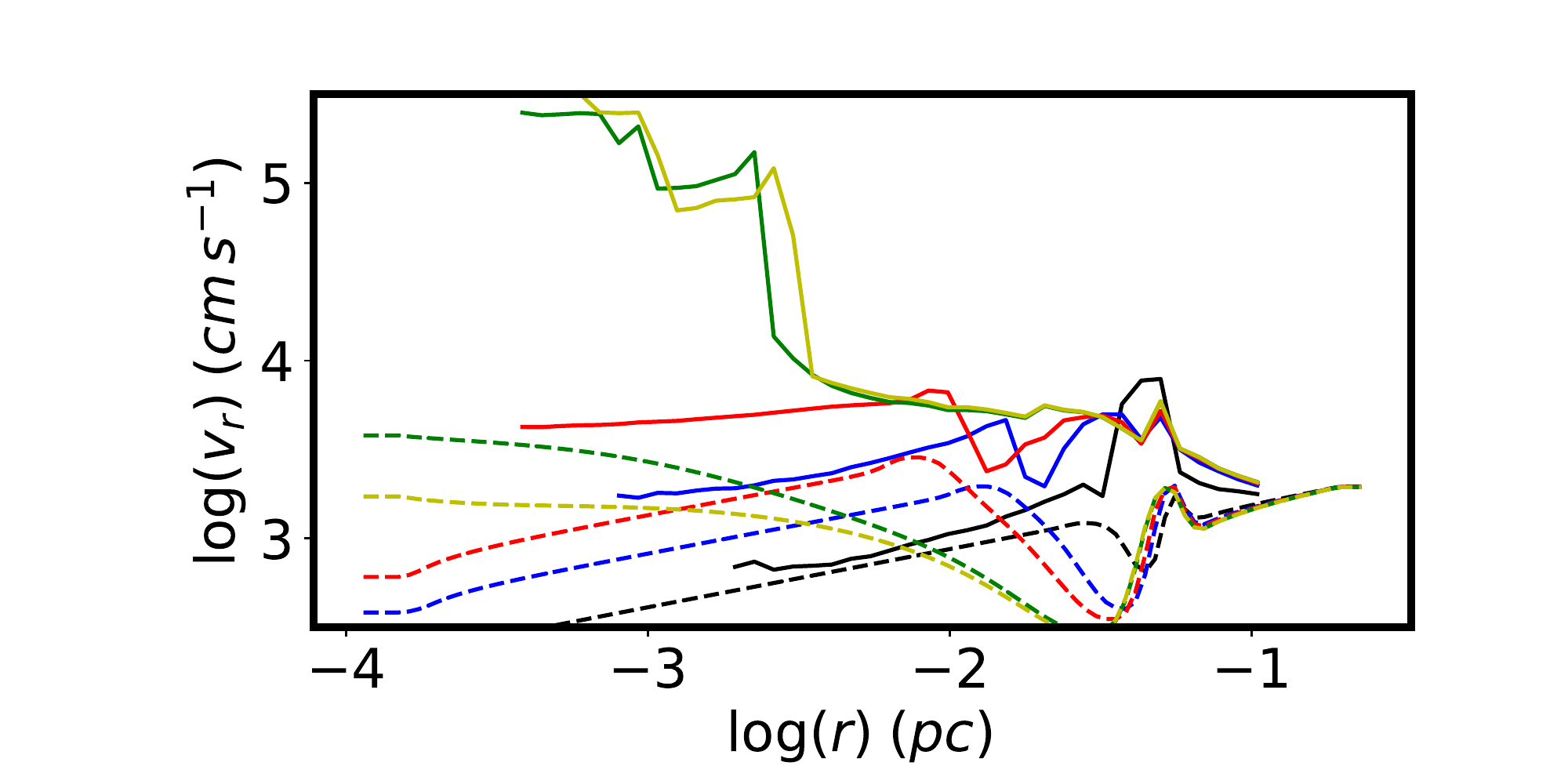}}
\put(0,0){\includegraphics[width=8cm]{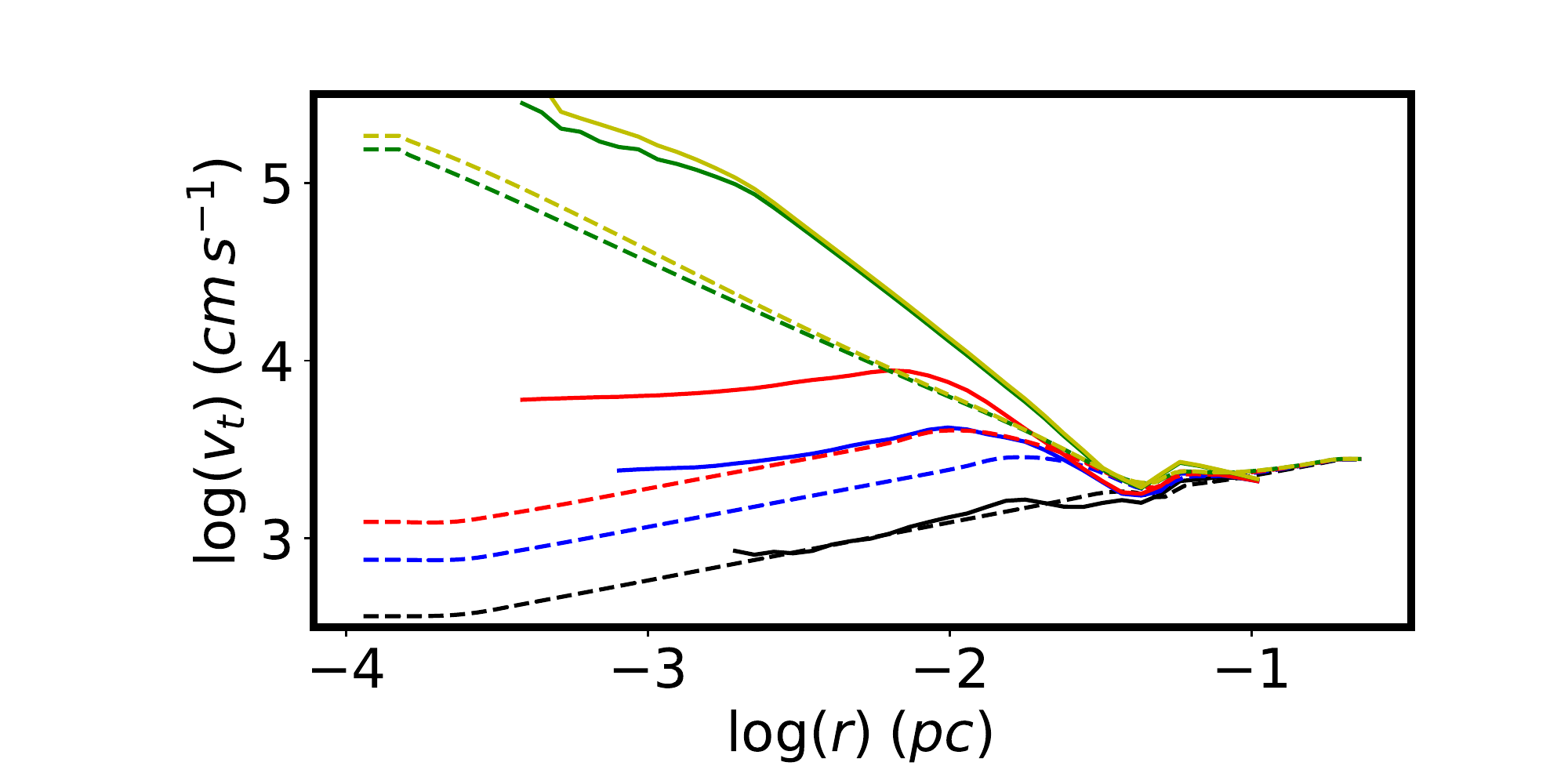}}
\end{picture}
\caption{Turbulent component $v_r$ and $v_t$ for 
run $A0.1M0.1$ ($\alpha=0.1$, $\mathcal{M}=0.1$).
The 1D simulations have been performed with Eqs.~(\ref{turb_rad2}) and (\ref{turb_trans2})
that is to say without the contribution of the Jeans instability 
(Eqs.~(\ref{turb_rad2_jeans}) and (\ref{turb_trans2_jeans}).
The transverse component $v_t$ is significantly 
larger in the 3D calculations than in the 1D ones implying that another source of turbulence must be accounted for
in the 1D simulations.
}
\label{alpha0.1_mach0.1_nojeans}
\end{figure*}

\setlength{\unitlength}{1cm}
\begin{figure*}
\begin{picture} (0,8)
\put(7,4){\includegraphics[width=8cm]{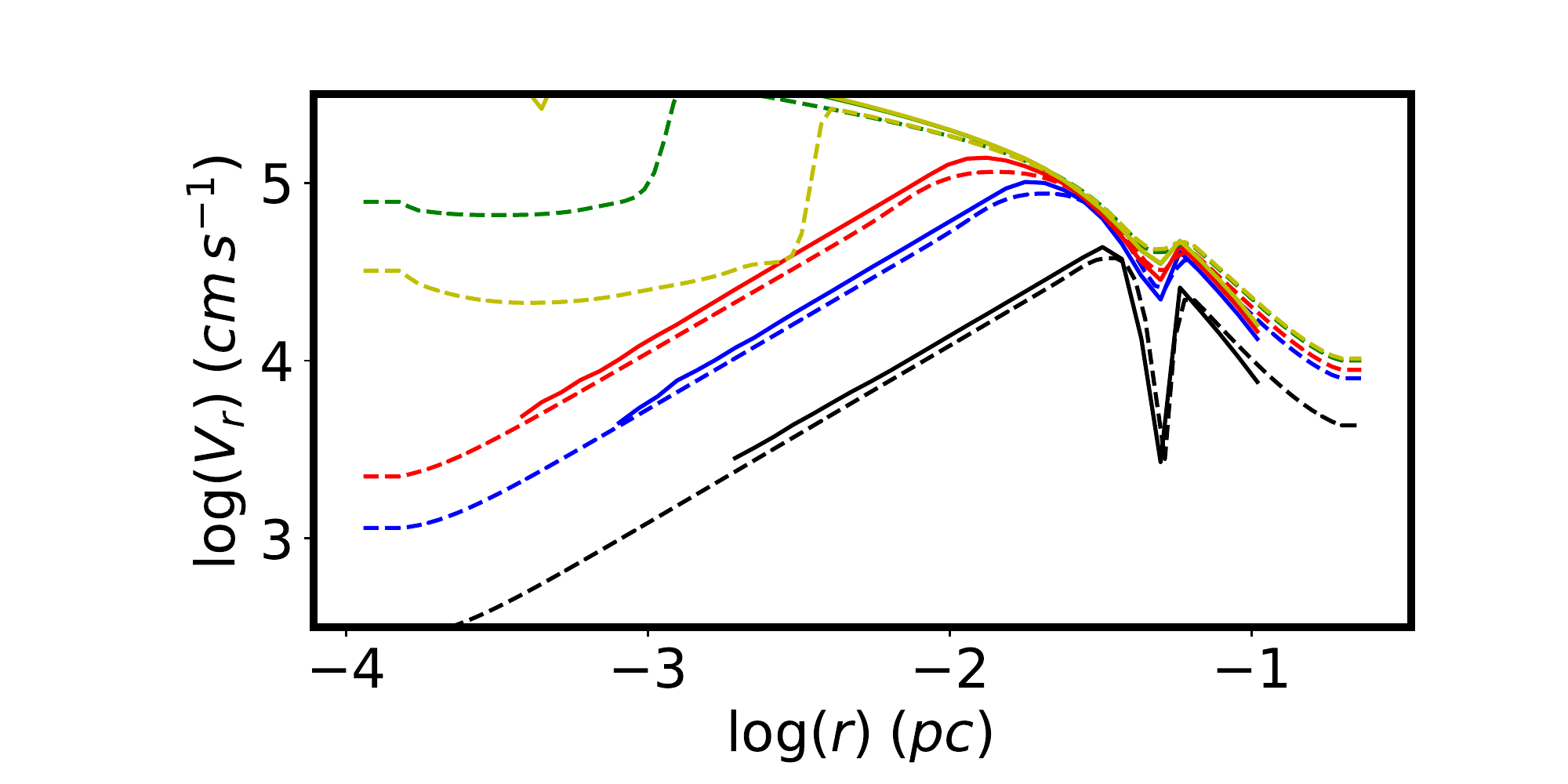}}
\put(0,4){\includegraphics[width=8cm]{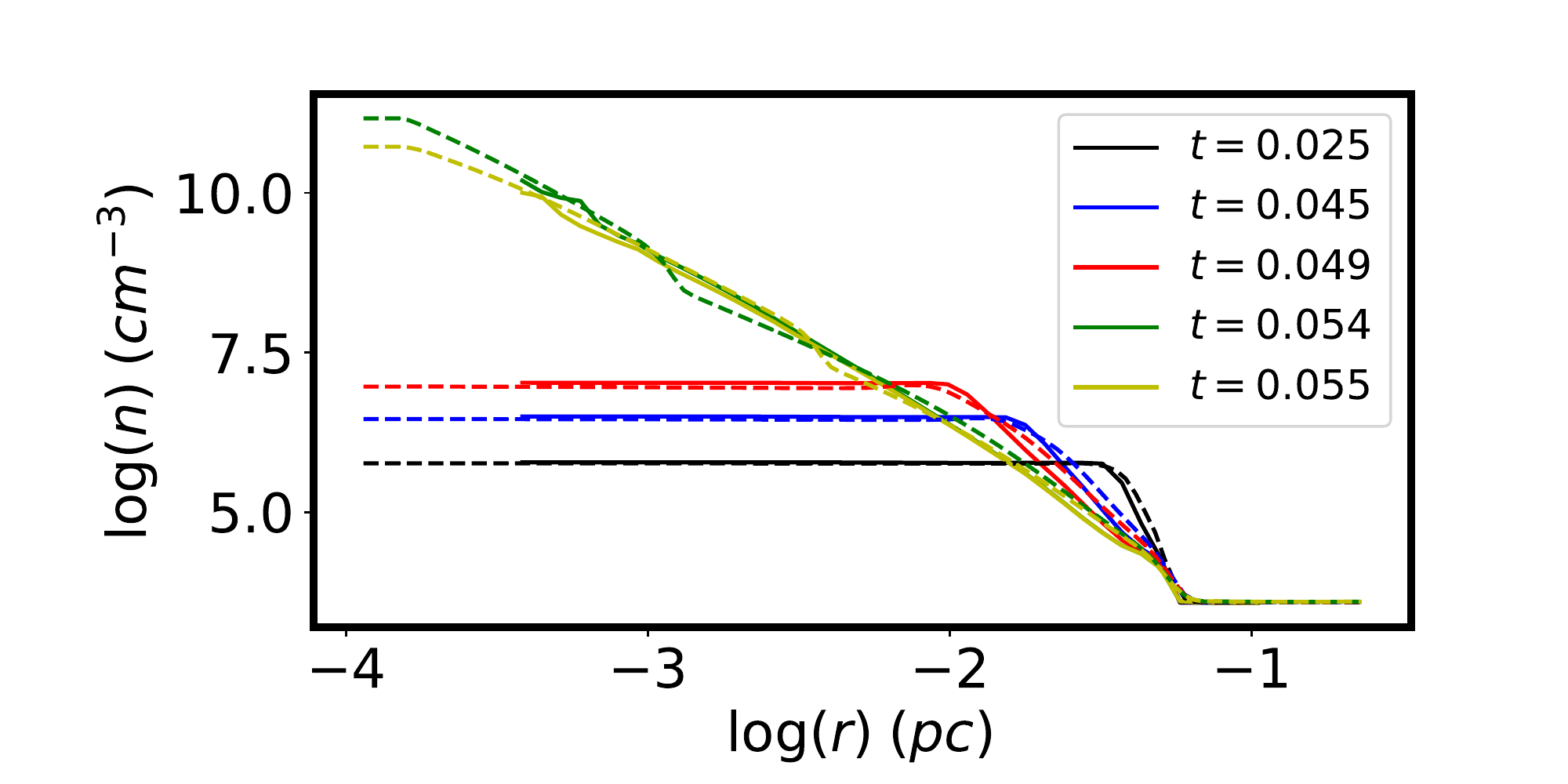}}
\put(1.5,7.7){$A0.1M0.1$}
\put(7,0){\includegraphics[width=8cm]{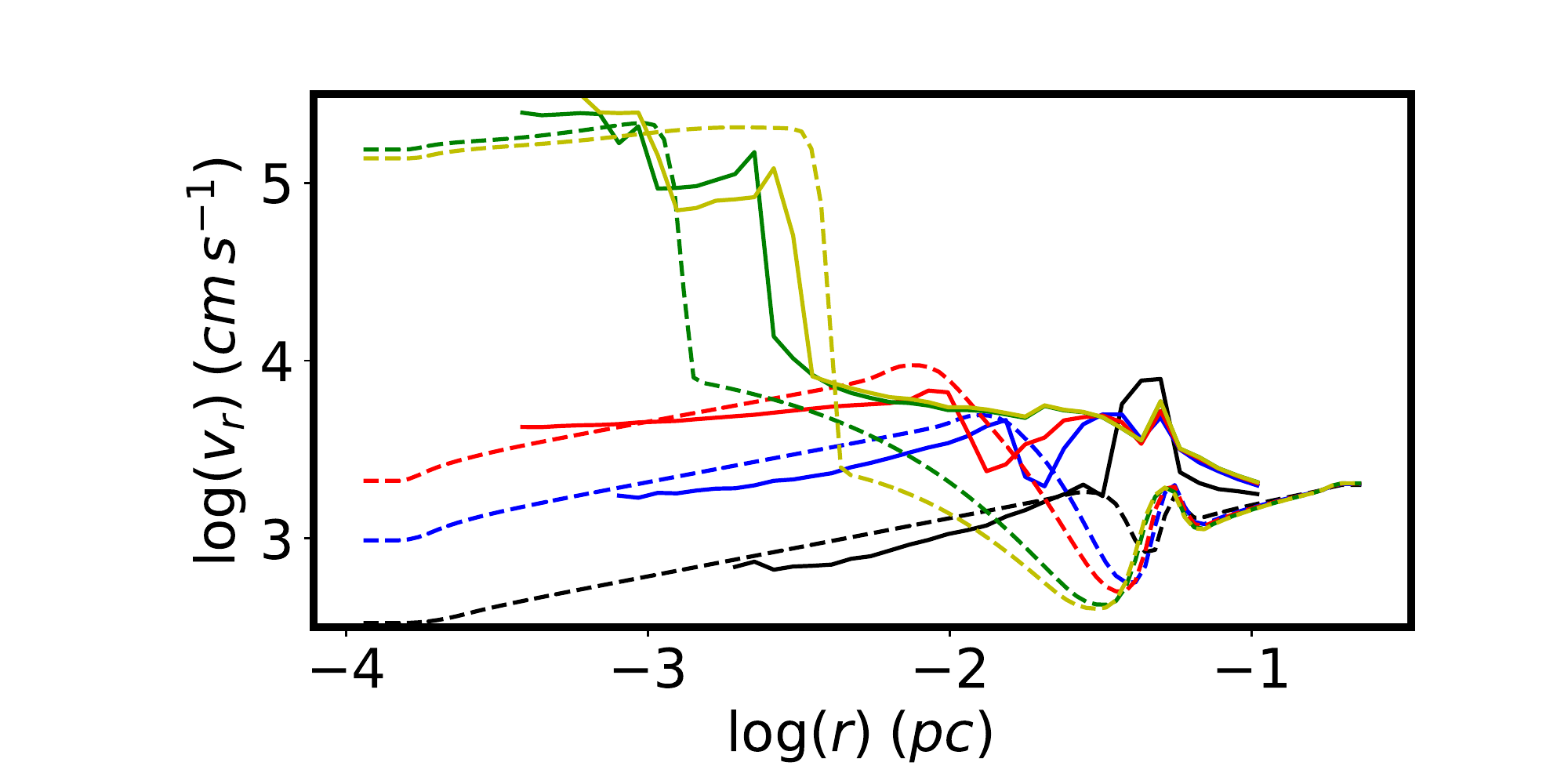}}
\put(0,0){\includegraphics[width=8cm]{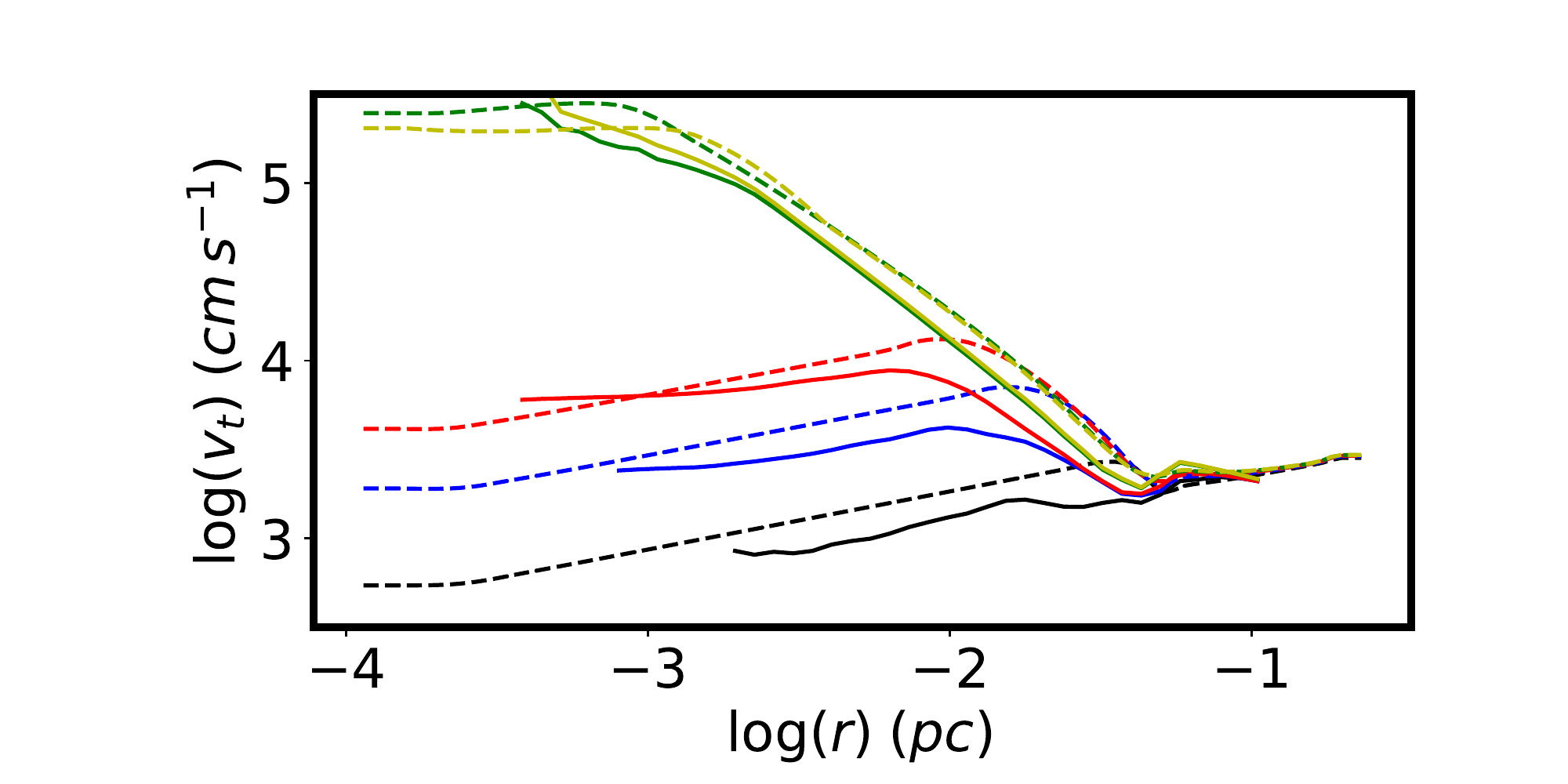}}
\end{picture}
\caption{Same as Fig.~\ref{gam1_alpha0.5} for
run $A0.1M0.1$ ($\alpha=0.1$, $\mathcal{M}=0.1$).  
Unlike in Fig.~\ref{alpha0.1_mach0.1_nojeans}, Eqs.~(\ref{turb_rad2_jeans}) and (\ref{turb_trans2_jeans}) have 
been used to perform the 1D simulations.
The agreement between the 1D (dashed lines) and 3D (solid lines) 
simulation results  is overall very good and  much better than when the local Jeans instability  is not accounted for
as in Fig.~\ref{alpha0.1_mach0.1_nojeans}.}
\label{alpha0.1_mach0.1}
\end{figure*}

\setlength{\unitlength}{1cm}
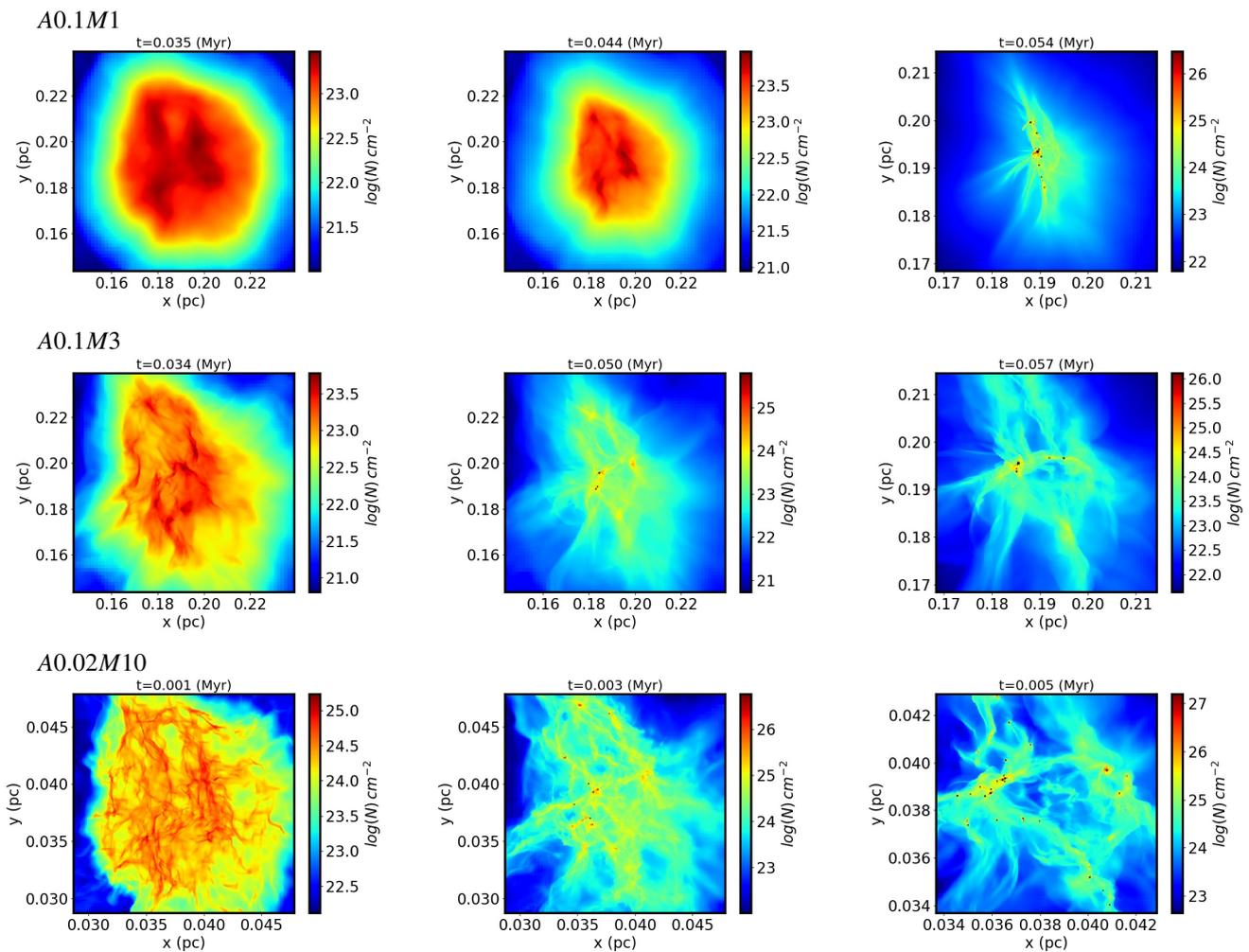
\begin{figure*}
\begin{picture} (0,13.5)
\put(0,9){\includegraphics[width=6.cm]{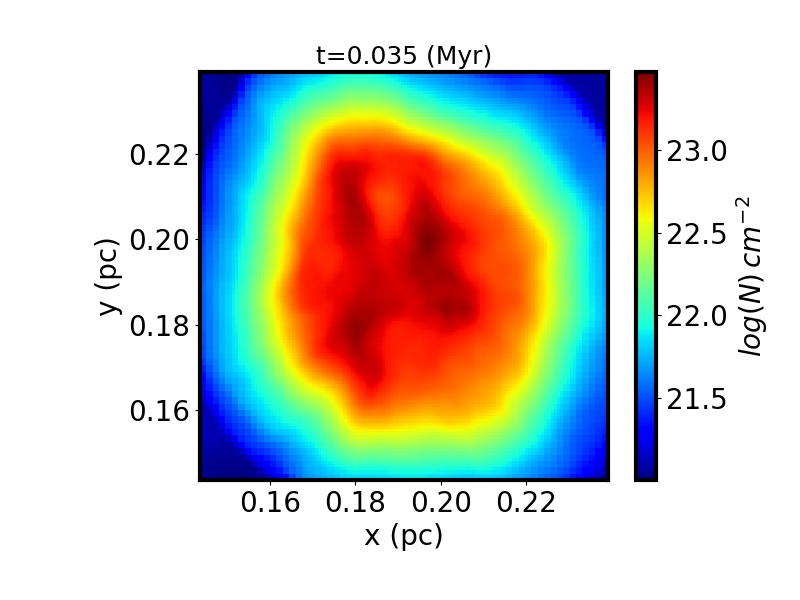}}
\put(6,9){\includegraphics[width=6.cm]{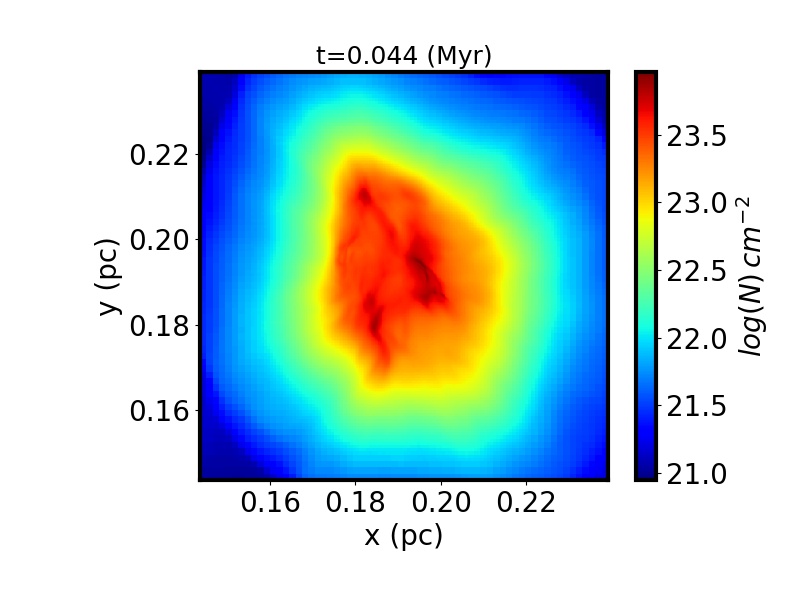}}
\put(12,9){\includegraphics[width=6.cm]{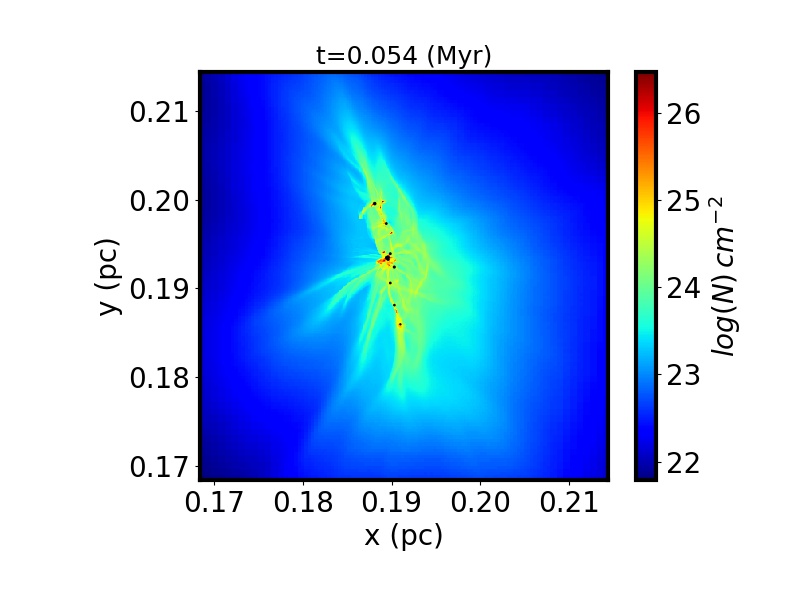}}
\put(1,13.3){$A0.1M1$}
\put(0,4.5){\includegraphics[width=6.cm]{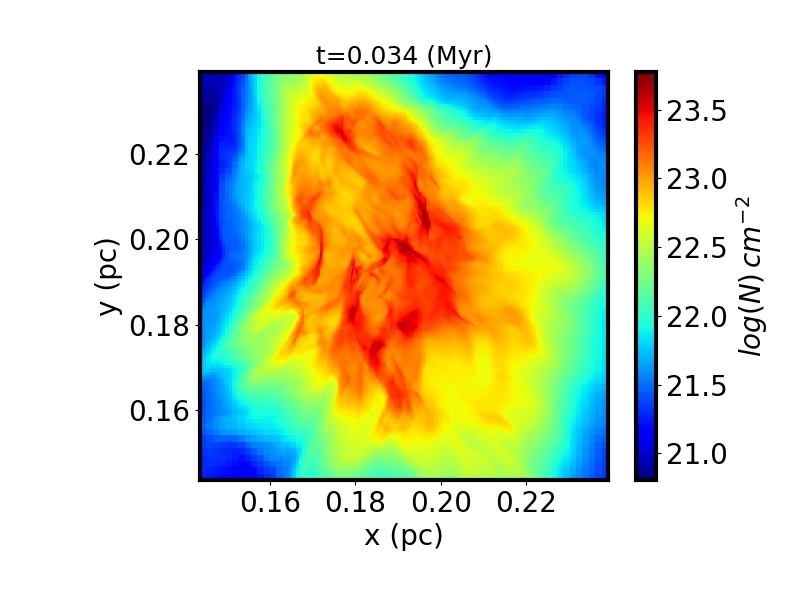}}
\put(6,4.5){\includegraphics[width=6.cm]{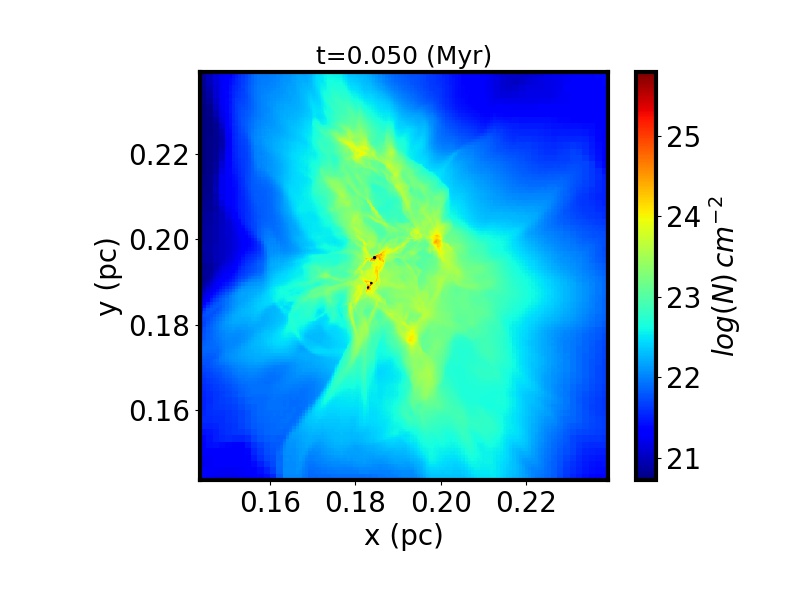}}
\put(12,4.5){\includegraphics[width=6.cm]{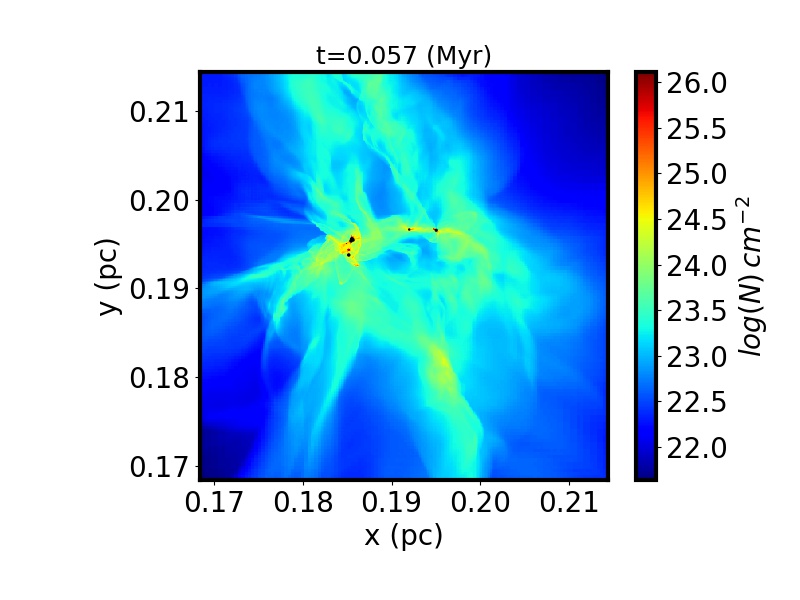}}
\put(1,8.8){$A0.1M3$}
\put(0,0){\includegraphics[width=6.cm]{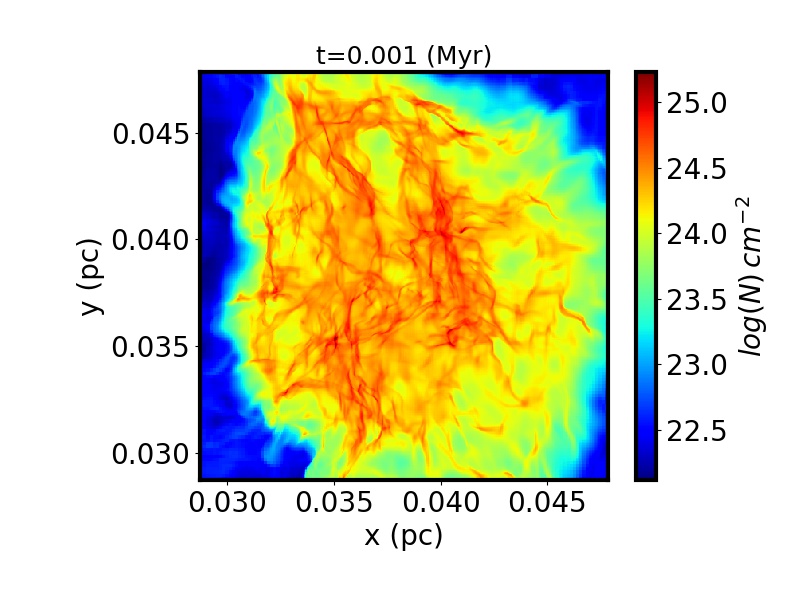}}
\put(6,0){\includegraphics[width=6.cm]{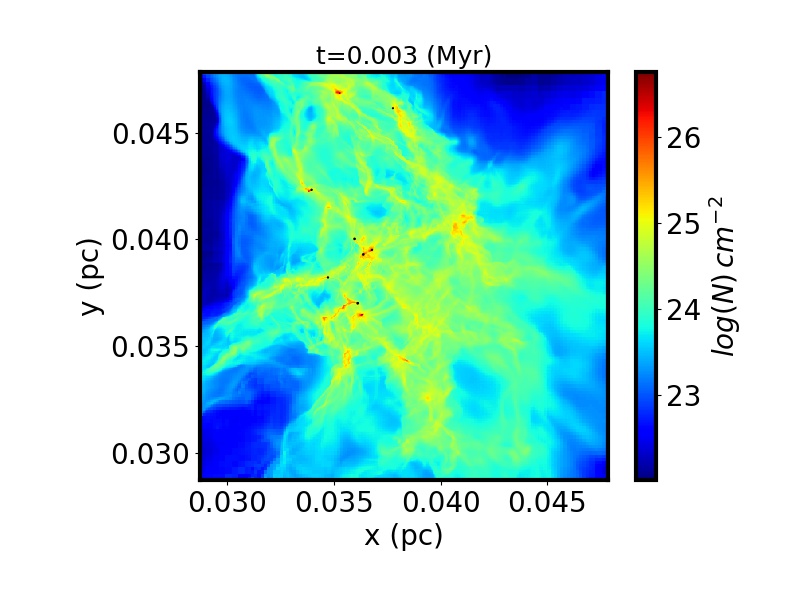}}
\put(12,0){\includegraphics[width=6.cm]{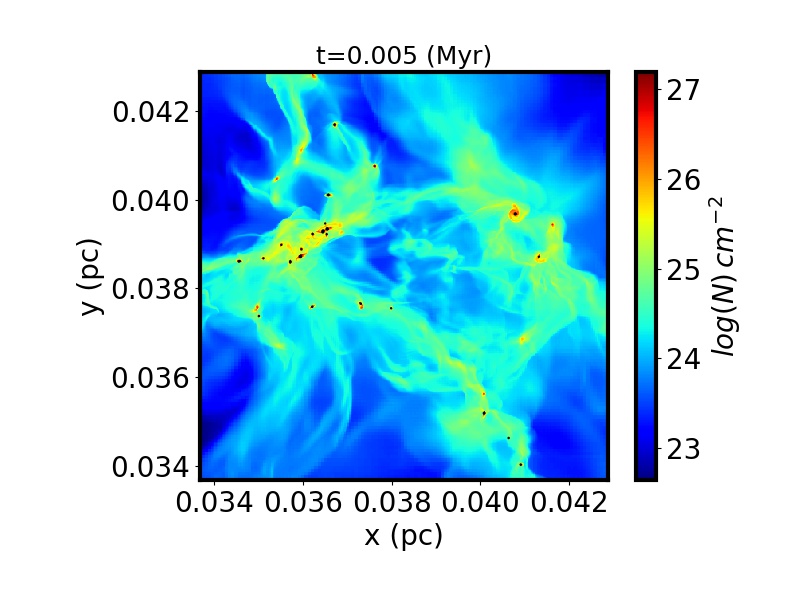}}
\put(1,4.3){$A0.02M10$}
\end{picture}
\caption{Column density at three timesteps of runs $A0.1M1$, $A0.1M3$ and  $A0.02M10$.
As these runs have high turbulence initially, the collapsing clouds are extremely non-axisymmetric.
}
\label{belleimage_cd}
\end{figure*}

\setlength{\unitlength}{1cm}
\begin{figure*}
\begin{picture} (0,8)
\put(7,4){\includegraphics[width=8cm]{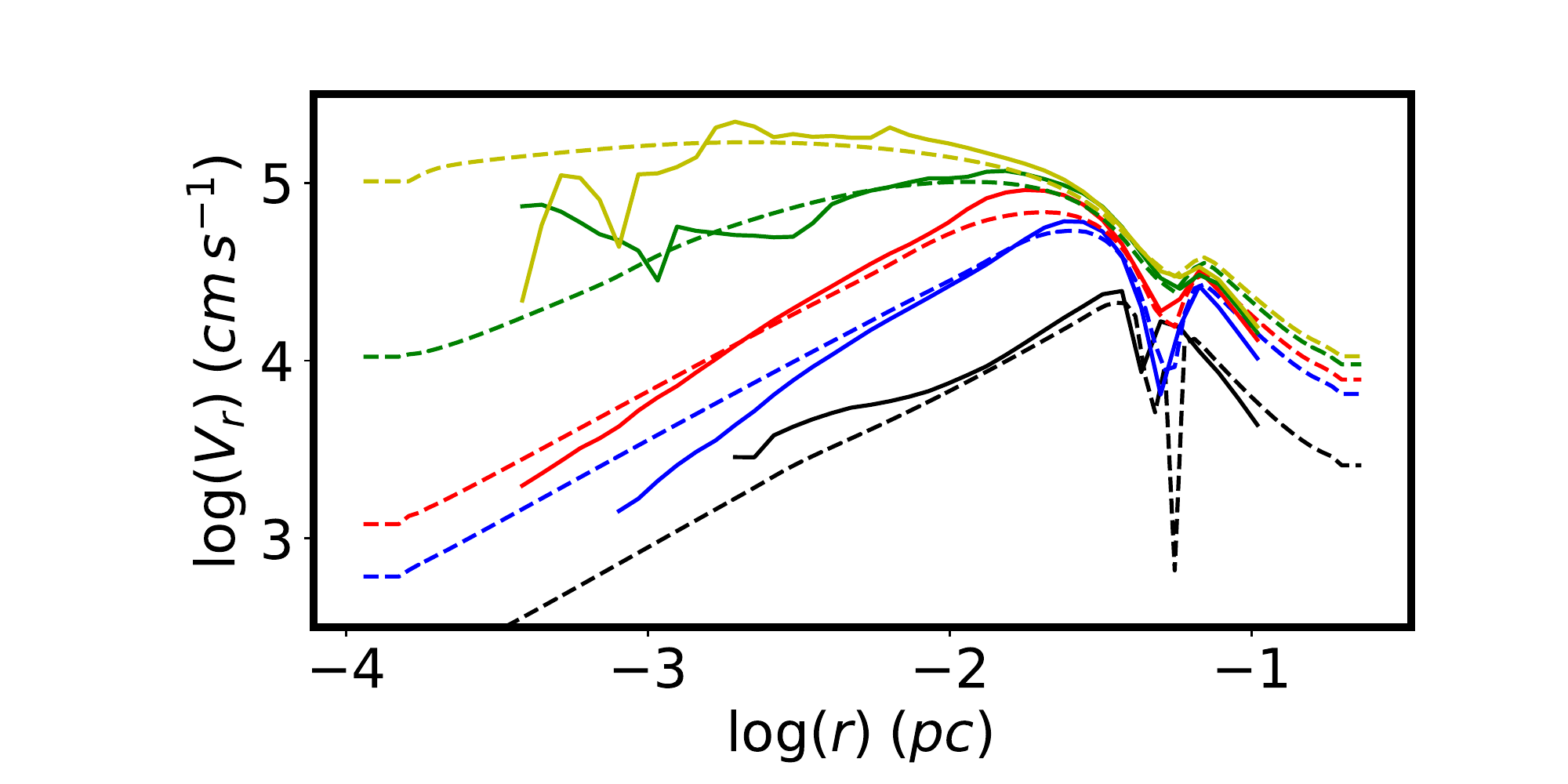}}
\put(0,4){\includegraphics[width=8cm]{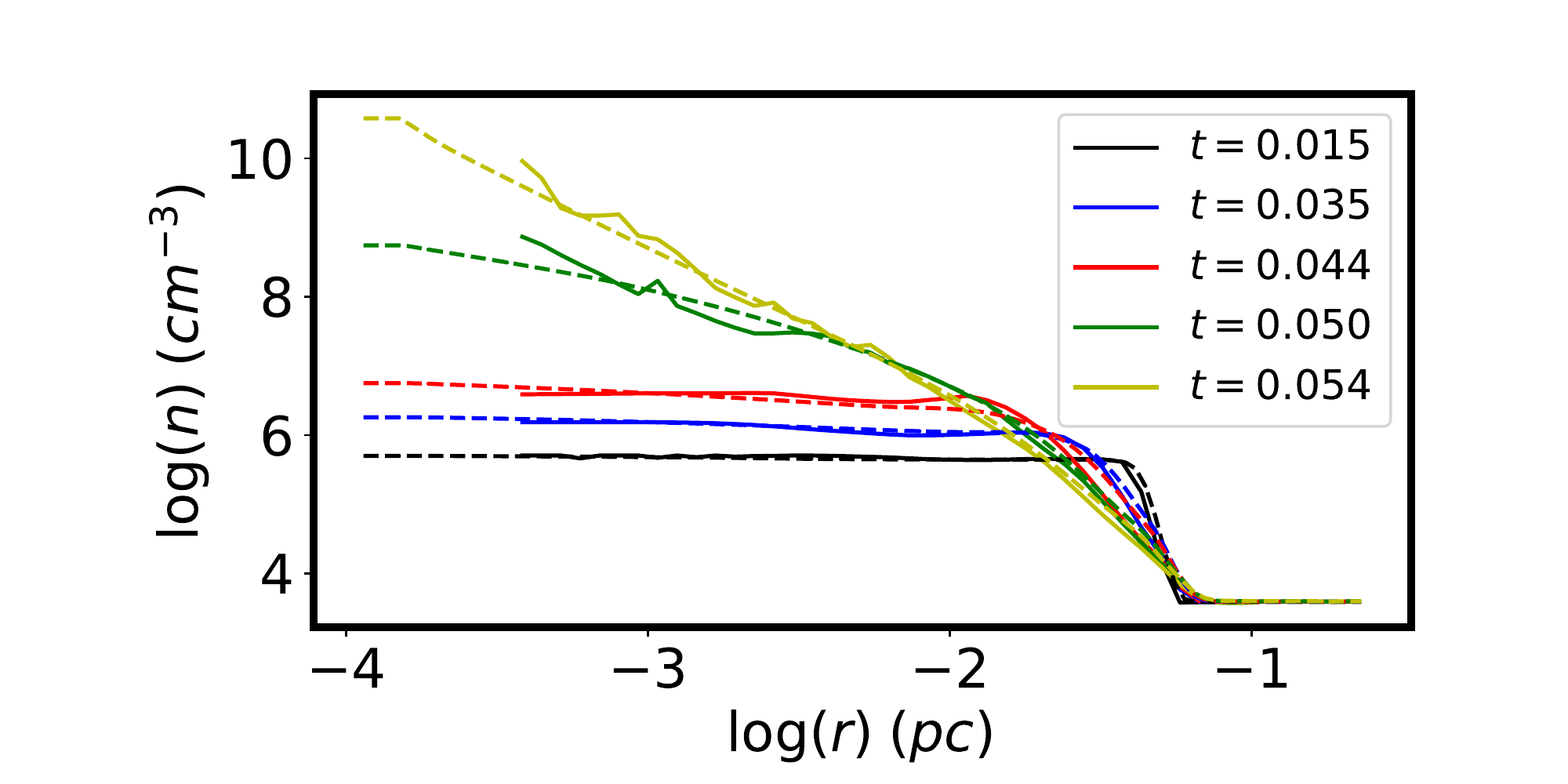}}
\put(1.5,7.7){$A0.1M1$}
\put(7,0){\includegraphics[width=8cm]{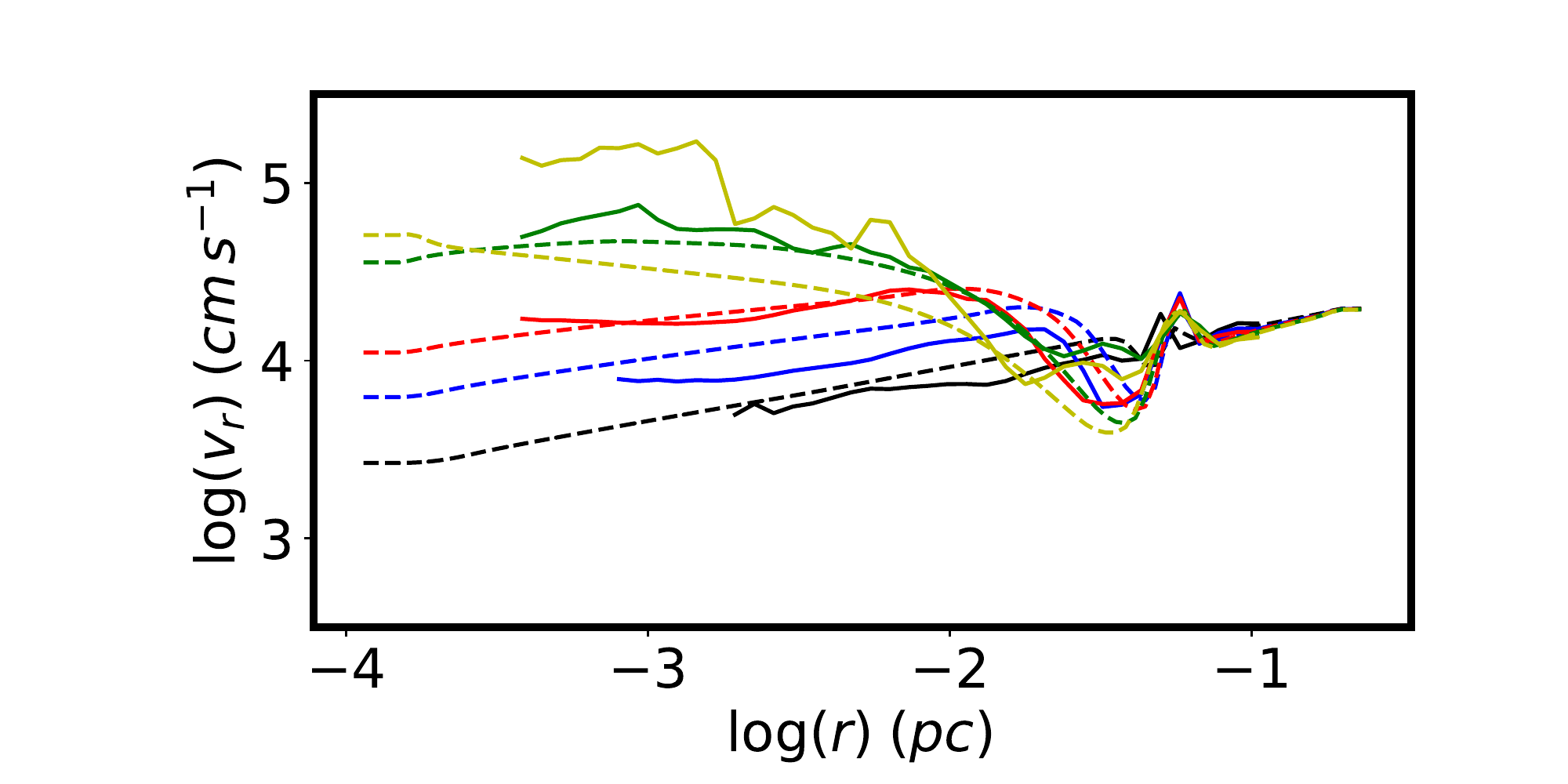}}
\put(0,0){\includegraphics[width=8cm]{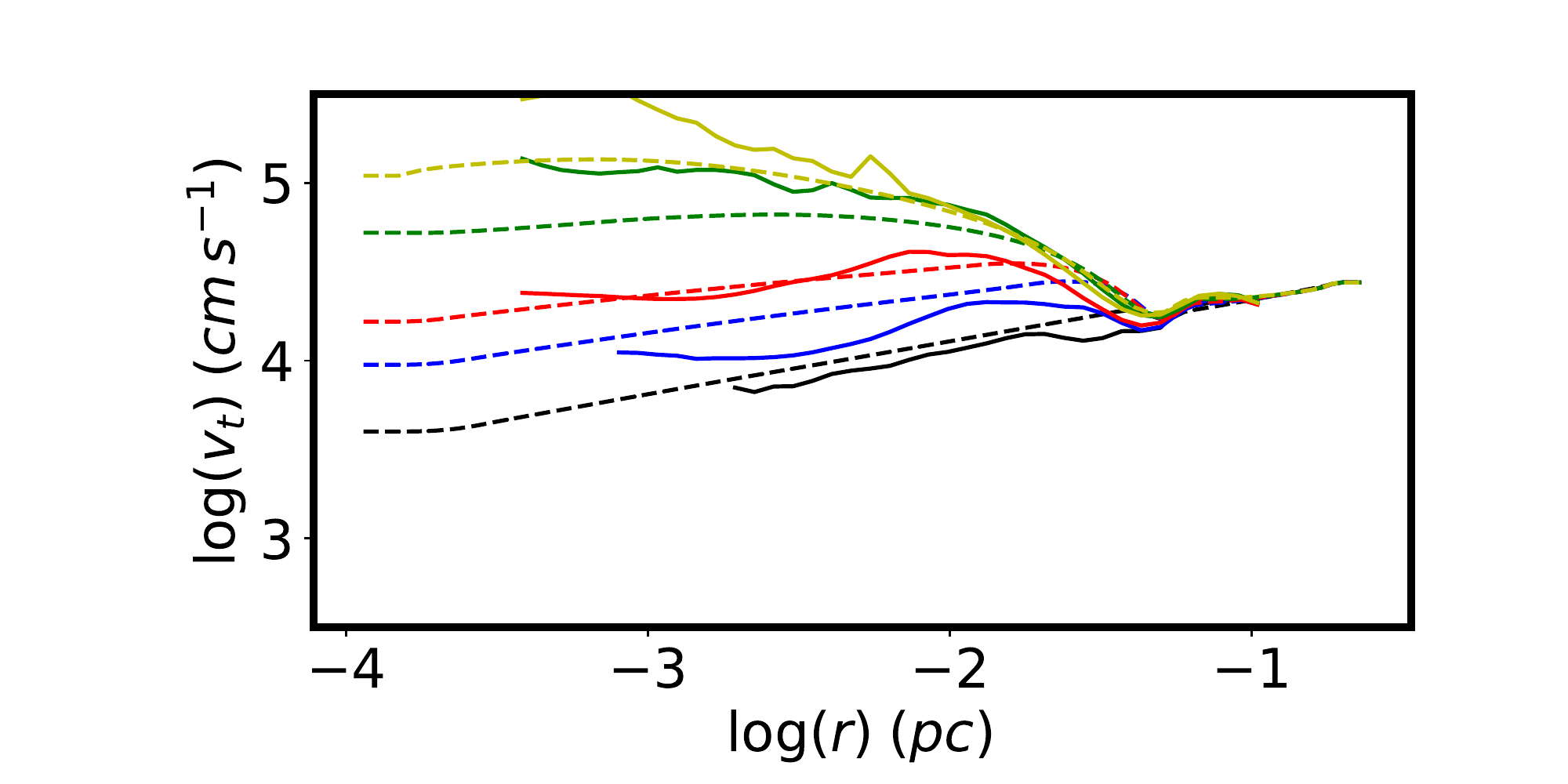}}
\end{picture}
\caption{Same as Fig.~\ref{gam1_alpha0.5} for
run $A0.1M1$  ($\alpha=0.1$, $\mathcal{M}=1$). Also turbulent energy is initially much larger 
than in run  $A0.1M0.11$
}
\label{alpha0.1_mach1}
\end{figure*}

\setlength{\unitlength}{1cm}
\begin{figure*}
\begin{picture} (0,8)
\put(7,4){\includegraphics[width=8cm]{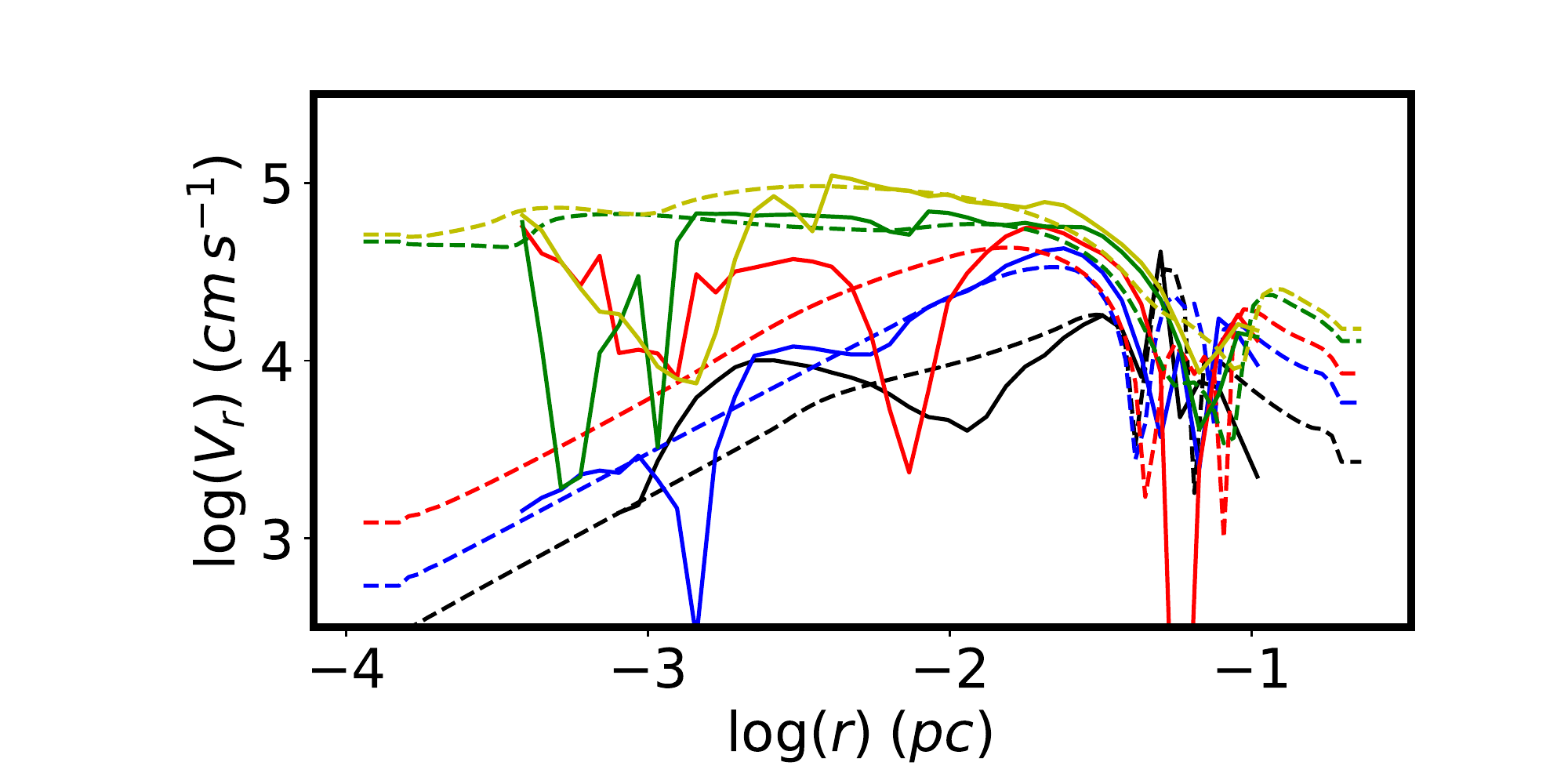}}
\put(0,4){\includegraphics[width=8cm]{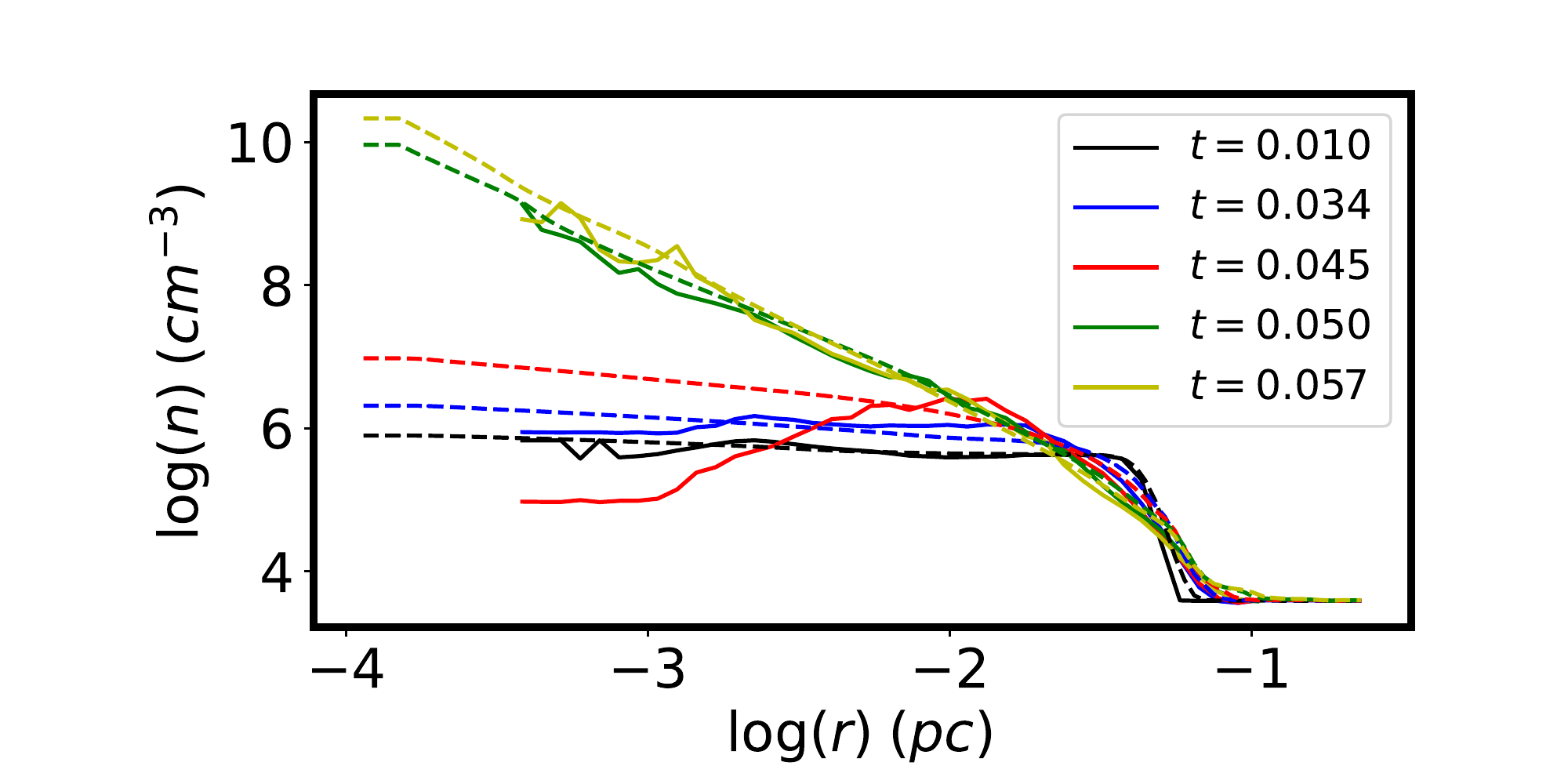}}
\put(1.5,7.7){$A0.1M3$}
\put(7,0){\includegraphics[width=8cm]{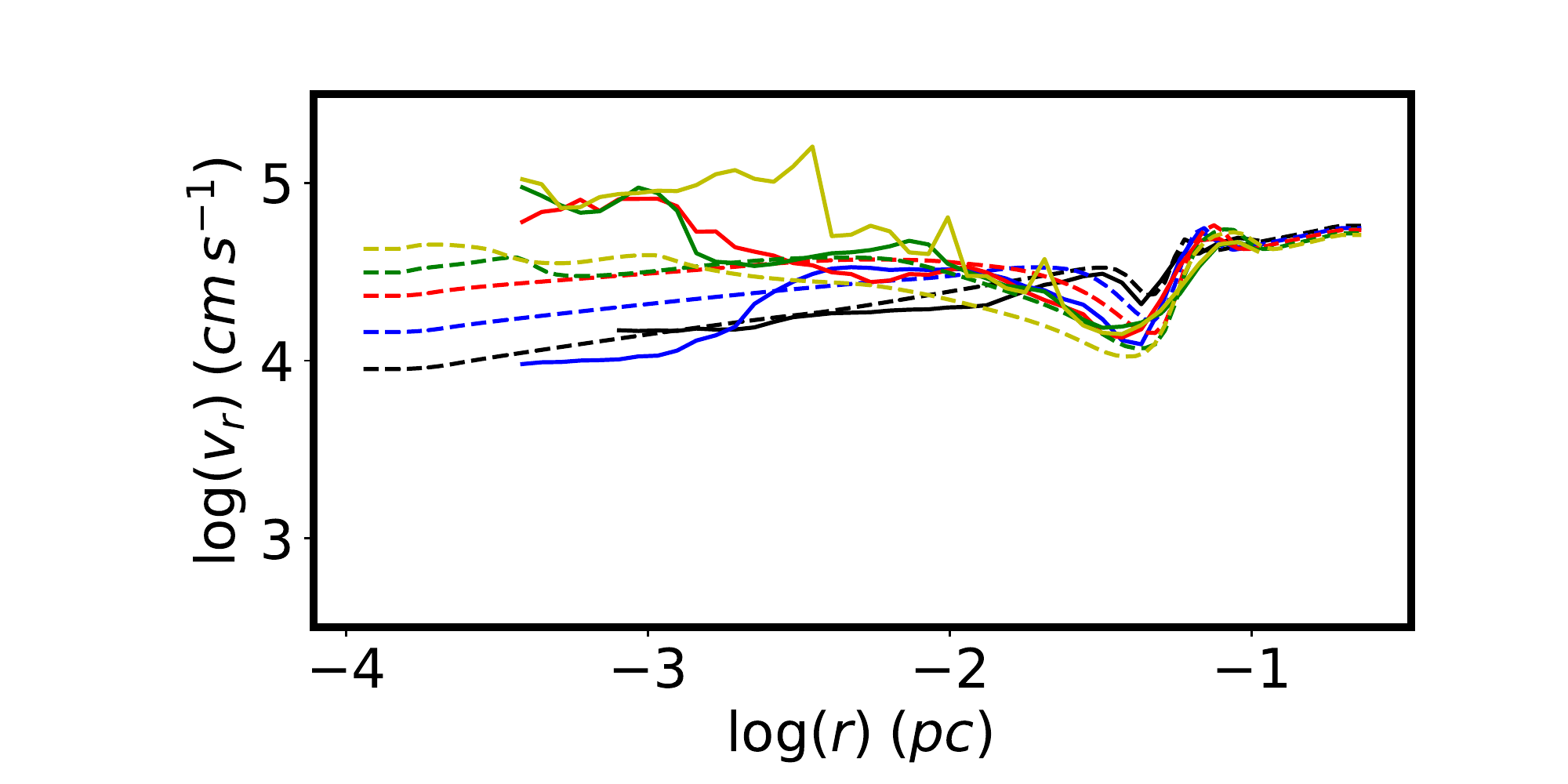}}
\put(0,0){\includegraphics[width=8cm]{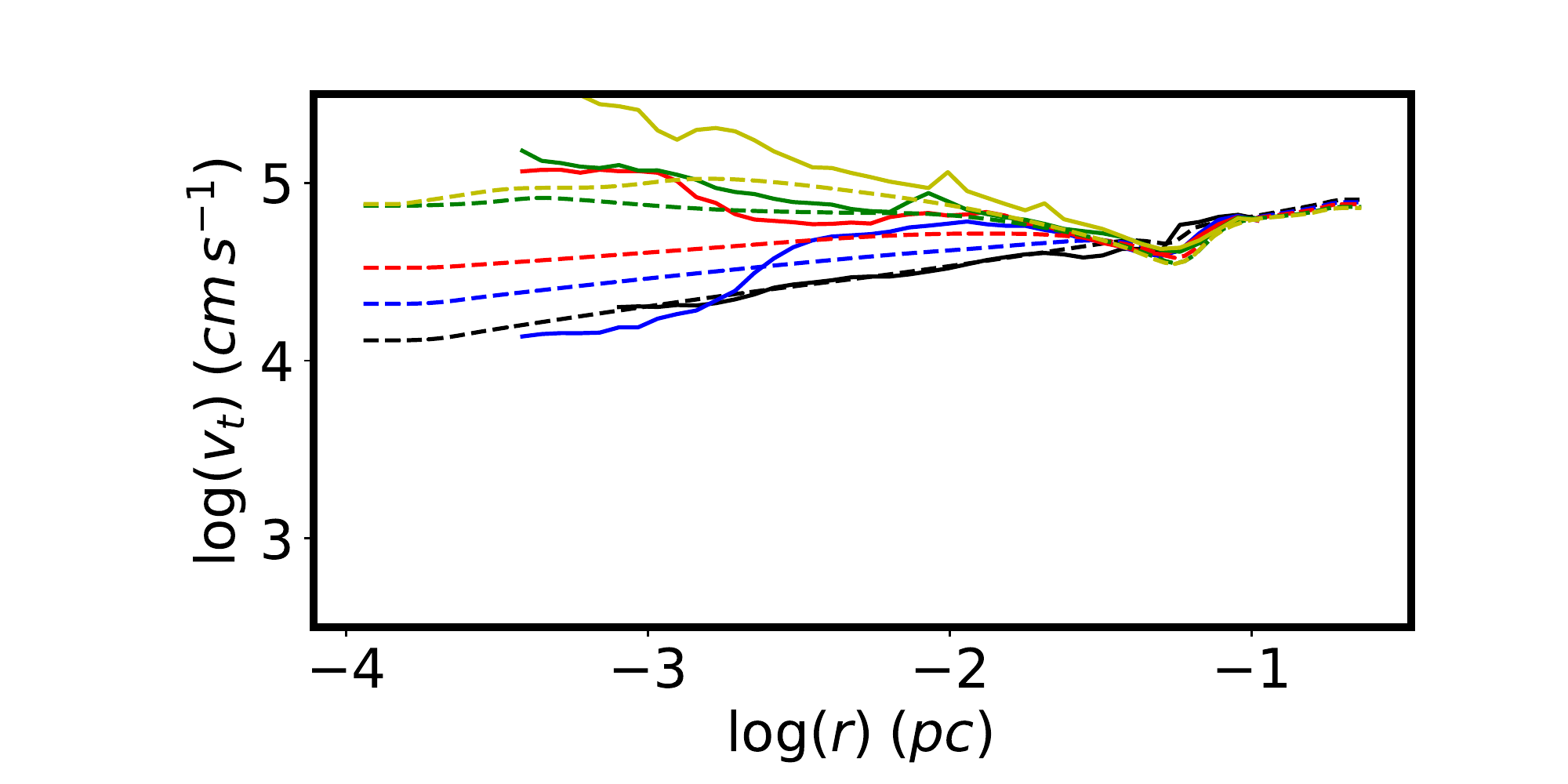}}
\end{picture}
\caption{Same as Fig.~\ref{gam1_alpha0.5} for
run $A0.1M3$ ($\alpha=0.1$, $\mathcal{M}=3$). Since the turbulence is strong initially, the 
collapse proceeds a non-symmetric way and therefore the agreement cannot be quantitative.
 Qualitatively the 1D and 3D solutions remain similar. 
}
\label{alpha0.1_mach3}
\end{figure*}

\setlength{\unitlength}{1cm}
\begin{figure*}
\begin{picture} (0,8)
\put(7,4){\includegraphics[width=8cm]{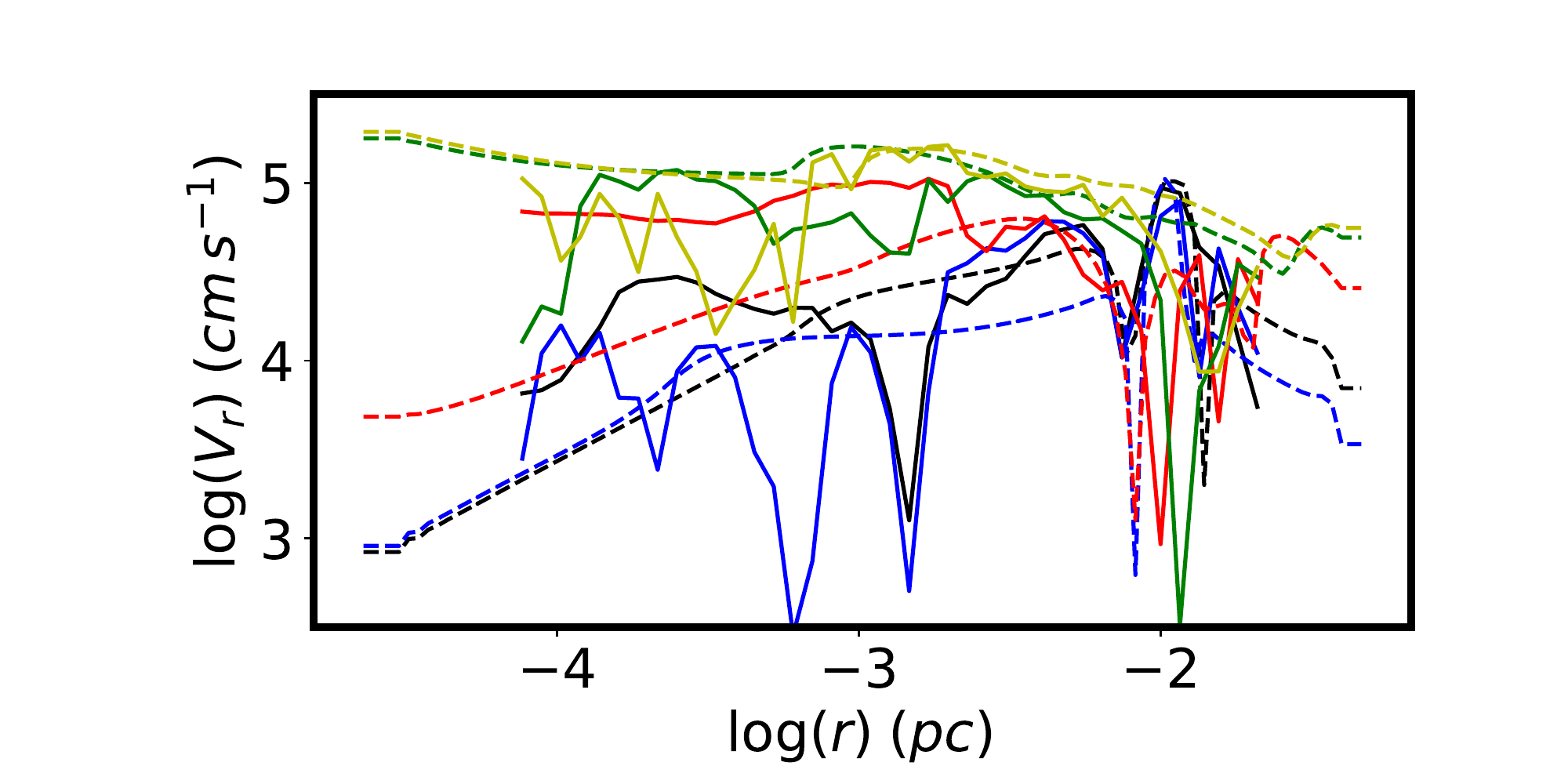}}
\put(0,4){\includegraphics[width=8cm]{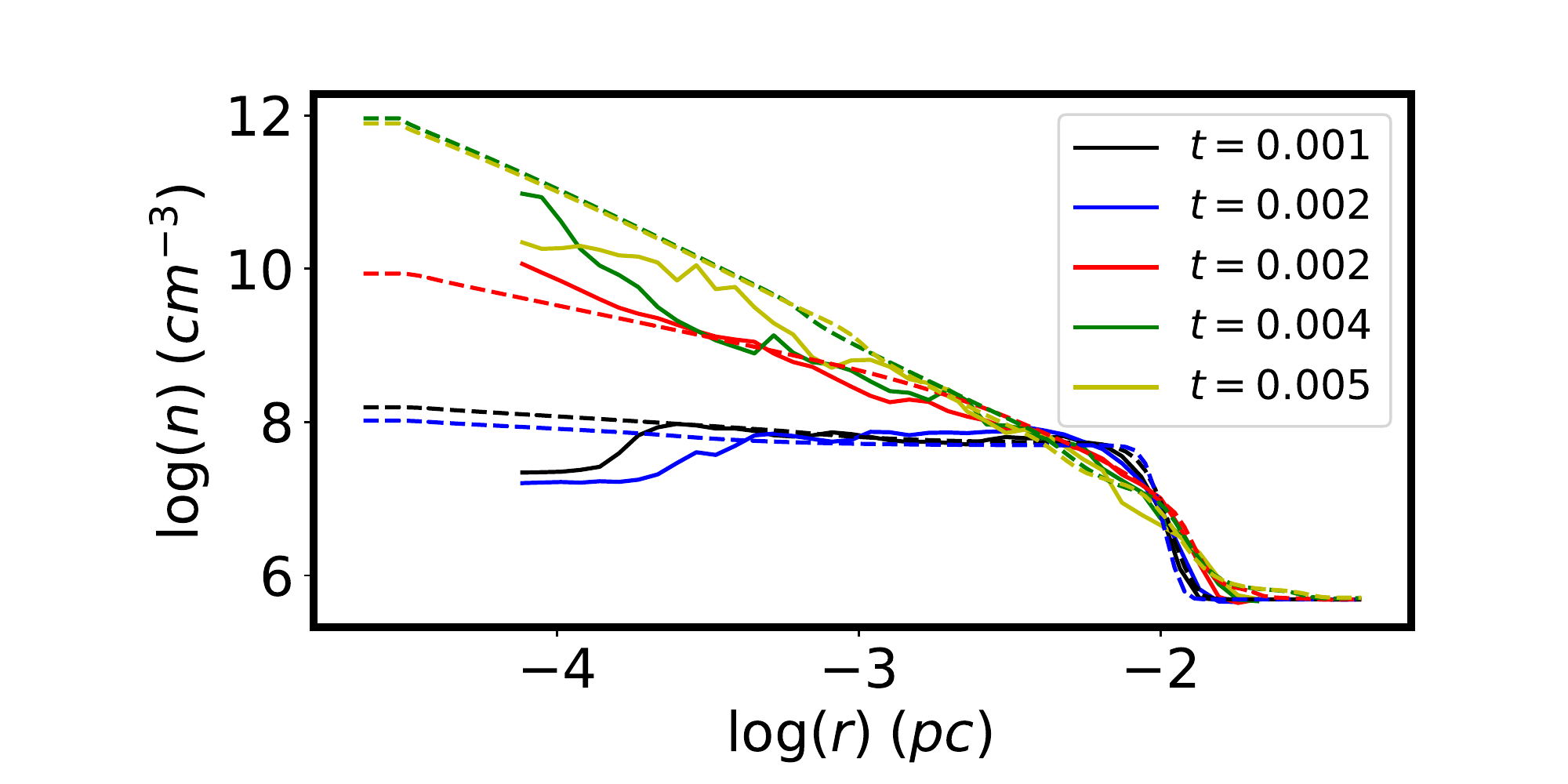}}
\put(1.5,7.7){$A0.02M10$}
\put(7,0){\includegraphics[width=8cm]{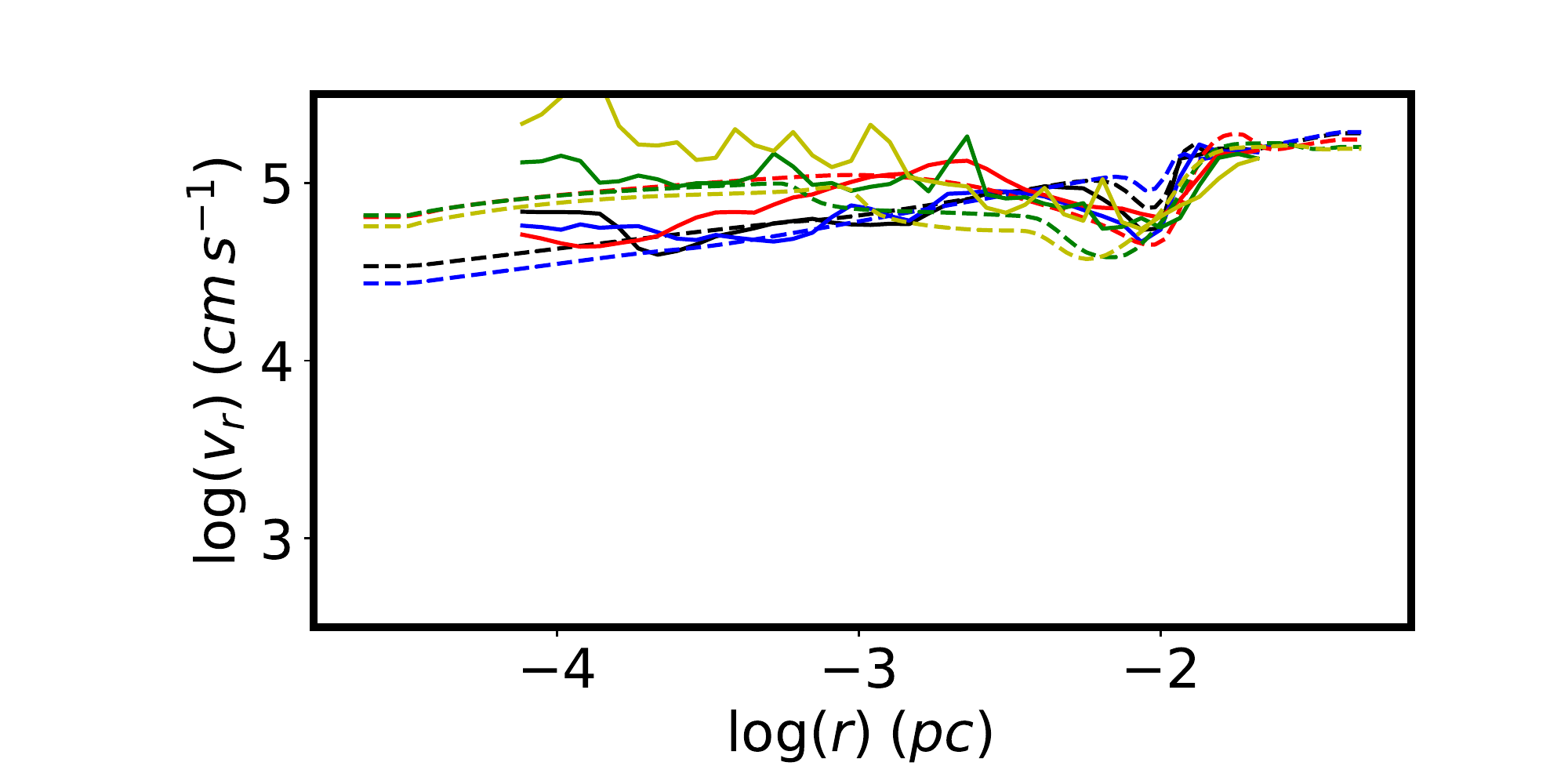}}
\put(0,0){\includegraphics[width=8cm]{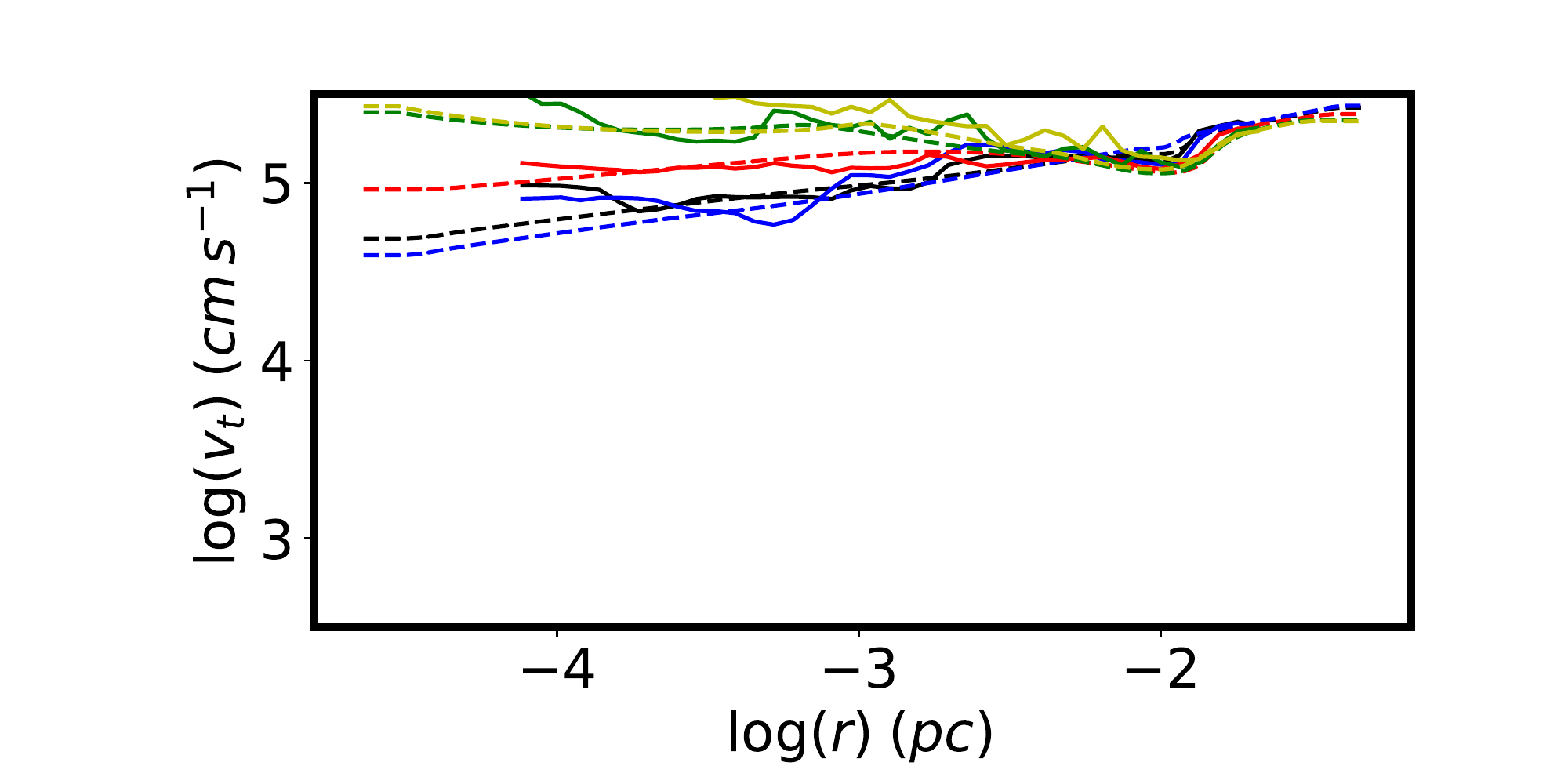}}
\end{picture}
\caption{Same as Fig.~\ref{gam1_alpha0.5} for
run $A0.02M10$ ($\alpha=0.02$, $\mathcal{M}=10$). 
}
\label{alpha0.02_mach10}
\end{figure*}

\section{Detailed comparison between 1D and 3D numerical simulations}
\label{results}

In this section, we present the detailed results of some of the runs performed. We start with run 
$A0.5M0.1$, which is the more thermally supported isothermal run and 
thus more prone to maintain the spherical symmetry. Then we present results for 
 $A0.1gam1.25M0.1$ which has the same initial thermal support than run $A0.5M0.1$
but since it has $\Gamma=1.25$ the thermal energy during the collapse stays larger.
We then study in more details run $A0.1M0.1$ where thermal support is weak. Finally we end
with a discussion on runs having high Mach numbers initially, namely 
$A0.1M1$, $A0.1M3$ and  $A0.02M10$.

\subsection{Weak turbulence and high thermal support}

\subsubsection{Isothermal case}

Figure~\ref{gam1_alpha0.5} presents the results for run $A0.5M0.1$ (which has $\alpha=0.5$, $\mathcal{M}=0.1$). 
Before to investigate the agreement between 1D and 3D values, it is worth 
discussing the results from the 3D calculations (solid lines).
The general behaviour displayed by the density, $n$, and radial velocity, $V_r$, is typical of collapsing motions. 
The 3 first timesteps (dark, blue and red) reveal that in the inner part of the cloud, the  density is uniform 
while the radial velocity  remains homologous. Both increases with time. In the outer part of the cloud, 
both fields connect to the outer values. 
As already explained, the two last timesteps (green and yellow)  arise after
the pivotal stage \citep{Shu77}. At these 2 timesteps, the sink particle, 
which typically represents the central object (say the star or the compact object), 
has a mass of about 0.2 and 1 $M_\odot$. 
Regarding the turbulent components, $v_t$ and $v_r$, 
 we see that both 
components are indeed amplified during the collapse and in the very inner part, reach values
comparable to $V_r$. Also the two components do not present the same behaviour 
as anticipated from the different nature of the source terms in 
Eqs.~(\ref{turb_rad2}) and (\ref{turb_trans2}).

In Fig.~\ref{gam1_alpha0.5} the dashed lines represent the 1D calculations which have been performed 
with $\eta_{diss}=0.25$ (this value is used for all simulations except those from Sect.~\ref{high_turb}). 
The densities, $n$, and radial velocities, $V_r$,  of  the 1D and 3D calculations agree remarkably well at all  
timesteps. 
The comparison between the 1D and 3D values of $v_t$ is also remarkably good though the 3D field presents
more fluctuations due to the stochastic nature of turbulence.
This is particularly the case before sink formation (black, blue and red lines) 
 Indeed, after sink formation and at small radii,  $v_t$ is 
  higher in the 3D simulations than in the 1D ones. For instance at time $t=0.709$ Myr (yellow line), 
$v_t$ is up to 3  times larger in 3D than in 1D simulations. We also see that there is a relatively well 
defined  transition between the outer region where both values are close and the inner region,
 where the 3D values are much larger. This transition propagates from  inside-out. 
 Apart for the two last timesteps for which again the 
values at $r < 0.01$ pc present major deviations, the agreement is usually better than few tens of percents while 
overall $v_t$ varies by more than one order of magnitude.
This demonstrates that Eq.~(\ref{turb_trans2_jeans})  accurately captures the evolution of $v_t$.
Note that for this run, there is actually no difference between 
Eq.~(\ref{turb_trans2}) and Eq.~(\ref{turb_trans2_jeans}) because due to the high value 
of $\alpha=0.5$, the source term due to the development of gravitational instability in Eq.~(\ref{turb_trans2_jeans}) 
is vanishing for all $k$ and all radius $r$.

The situation for $v_{r}$ is a little more complex. We see that  within the cloud, 
i.e for $r< 0.1$ pc, the agreement is still good for the first three timesteps 
 (except on few places) though less accurate than for $v_t$.
Major differences are seen in the outer part of the cloud and almost everywhere at the last timesteps.
In particular, strong disagreements seem to appear when the radial velocity gradient $\partial _r  V_r$ reverses. 
This suggests that there may be a missing source term that should be added to $v_r$, perhaps the growth
of an instability or the production of sound waves. Indeed, in the context of collapsing stars, 
it is now well established that accoustic waves can be generated 
\citep{abdikamalov2020} as a consequence of vorticity conservation.

\subsubsection{$\Gamma=1.25$}

Figure~\ref{gam125} presents the results for run $A0.5gam1.25M0.1$ (which has $\alpha=0.5$, $\Gamma=1.25$, $\mathcal{M}=0.1$)
both for the 3D simulation (solid lines) and 1D one (dashed one). 
First of all a major difference with run $A0.5M0.1$ needs to be stressed. 
This regards the duration of the 3D and 1D runs
as the latter is  about 40$\%$ longer than the former. This seems to be due to 
the fact that run $A0.5gam1.25M0.1$ is marginally unstable as it has an exponent close 
to $4/3$ and a high thermal over gravitational energy ratio. Indeed small differences for 
instance on the gas outside the cloud have been found to make substantial differences. 
However, by using the synchronisation procedure based on density, we find 
as for run $A0.5M0.1$, that the overall agreement between 1D and 3D calculations is very good 
for $n$ and $V_r$ except in the inner part ($r<0.03$ pc) well after the sink formation (yellow line).
This again is likely because small differences in the inner boundaries (the choice made
regarding the sink particle and accretion) are important since the cloud is marginally unstable.

The agreement for $v_t$ and $v_r$ is also quite good (except at time $t=0.92$ 
in the cloud inner part for $v_t$ and nearly everywhere for $v_r$).
Indeed  an important differences with run $A0.5M0.1$ is that
in the 3D simulations,  $v_t$ and $v_r$ 
reach smaller values (up to a factor of 3 below $r=0.01$). By contrast 
the 1D runs tend to predict similar values for $A0.5M0.1$ and $A0.5gam1.25M0.1$.

This clearly indicates that thermal support is making an important difference 
regarding turbulent generation, 
at the advanced stage of the collapse. To understand the origin of these 
differences, Fig.~\ref{belleimage} portrays the  density cuts as well as the 
projected velocity field in 3 snapshots (2 before and 1 after the pivotal stage). 
As can be seen for the 2 first snapshots, the clouds remain fairly spherical. However, 
for the third snapshot the situation is quite different. While for run 
$A0.5gam1.25M0.1$ the density field remains spherical, it is not the case in run $A0.5M0.1$, 
where prominent spiral patterns develop. Their origin is very likely due to 
angular momentum as they are reminiscent of what has been found in many simulations \citep[e.g.][]{matsumoto2003,brucy2021}. 
In particular \citet{verliat2020} show that the formation of centrifugally supported disks is favored by 
symmetry breaking as in such circumstances the collapse center is not the center of mass and therefore
angular momentum is not conserved with respect to the collapse center. For run $A0.5gam1.25M0.1$
the thermal support maintain spherical symmetry preventing prominent axisymmetry breaking to occur.

\subsection{Weak turbulence and low thermal support: turbulent generation}

\subsubsection{Evidence for another source of turbulence}

We first start by performing 1D runs with  Eqs.~(\ref{turb_rad2}) and (\ref{turb_trans2}) which we remind 
do not include the generation of turbulence through gravitational instability. 
Figure~\ref{alpha0.1_mach0.1_nojeans} presents results for run $A0.1M0.1$ (which has $\alpha=0.1$, $\Gamma=1$, $\mathcal{M}=0.1$)
restricting to the turbulent variables $v_r$ and $v_t$.
Clearly the 3D values are significantly  larger than the 1D ones even well before reaching the pivotal 
stage, for instance it is the case at time $t=0.049$ Myr. The disagreement typically increases with time and unlike what is 
observed for  $A0.5M0.1$, the amplification does not proceed from the inside-out but appears
to be global.

These results clearly suggest that Eqs.~(\ref{turb_rad2}) and (\ref{turb_trans2}) are missing a
source of turbulence.  
To get a hint on what may be happening, Fig.~\ref{belleimage} portrays 
density cuts of run $A0.1M0.1$. As can be seen, strong non-axisymmetric 
perturbations develop therefore generating further turbulence. 
Note that in run $A0.1M0.1$ and at time $0.045$ and $0.049$ Myr, the densest 
density fluctuations, are clearly located in a shell of radius $\simeq$0.01 pc which 
is reminiscent of the shell instability known to develop in collapsing clouds with 
low thermal support \citep[e.g.][]{ntormousi2015}.

\subsubsection{Evidence for turbulent generation by local gravitational instability}

Figure~\ref{alpha0.1_mach0.1} portrays the results of 1D calculations performed 
with Eqs.~(\ref{turb_rad2_jeans}) and~(\ref{turb_trans2_jeans}) and a value of $\eta_{diss}=0.25$.
Unlike for the 1D simulations displayed in Fig.~\ref{alpha0.1_mach0.1_nojeans}, 
the contribution of local Jeans instabilities in the development of turbulence
is taken into account.
The agreement is overall very good. In most locations, and except towards the cloud center the 3D and 1D results 
barely differ to more than few tens of percents. 
This clearly shows that turbulence is not only amplified by the contraction, it is also {\it generated} by the 
local development of gravitational instabilities. 

Note that $v_t$ is possibly a little too high at time $t=0.045$ and $0.049$ Myr and around 
$r \simeq 0.01$ pc. This may indicate that the gravitational instability growth rate used in 
Eqs.~(\ref{turb_rad2_jeans}) and~(\ref{turb_trans2_jeans}) is a little too high there and that a more
accurate spatial dependent analysis should be considered \citep[e.g.][]{nagai1998,fiege2000,ntormousi2015}.

\subsection{High initial turbulence}
\label{high_turb}

In many astrophysical situations of interest, the turbulence is not
initially  small when the collapse starts and it is worth investigating to what 
extend the 1D approach is  nevertheless still relevant. The obvious difficulty is that when turbulence
is strong, it induces major geometrical deviations from spherical geometry as can be seen 
from Fig.~\ref{belleimage_cd} that displays at  3 timesteps, the column density for runs
$A.01M1$, $A0.1M3$  and $A0.02M10$.

Figure~\ref{alpha0.1_mach1} shows the detailed profiles for run $A0.1M1$ (which has $\alpha=0.1$, $\Gamma=1$, $\mathcal{M}=1$).
In this run, the turbulent energy is  initially hundred times larger than in run $A0.1M0.1$ and 
 represents 10$\%$ of the gravitational energy initially. 
In this section,  we use $\eta_{diss}=1$ as this value is suggested by 
Fig.~\ref{accret_comp}. 
In spite of a  relatively large initial turbulence, 
 which induces significant departures from spherical geometry (see Fig.~\ref{belleimage_cd}), 
the agreement between the 3D and 1D simulations is still remarkably good. 
In particular, the level of turbulence is close in 3D and 1D runs 
with a global amplification on the order of a factor 10 at a few $r=0.001$ pc.
There is possible discrepancy and the order 
of a factor of $\simeq 2$ in the inner part and 
at time $t=0.05$ Myr where the 3D turbulence appears to be slightly larger than the 1D ones. 
Interestingly the two components $v_r$ and $v_t$ appears to present  profiles
much more similar than in runs with lower Mach numbers, where prominent differences between 
$v_r$ and $v_t$ appear.  
Interestingly we see that the turbulent velocities have amplitude which are comparable to 
the mean radial one, $V_r$.

Figure~\ref{alpha0.1_mach3} portrays results for run $A0.1M3$ (which has $\alpha=0.1$, $\Gamma=1$, $\mathcal{M}=3$).
For this run, the turbulent energy is  9 times larger than the thermal ones and is therefore 
comparable to the gravitational energy. 
As expected we see from Fig.~\ref{belleimage_cd} that the collapse proceeds in a highly non-symmetric way.  
In spite of this major departure from spherical symmetry, the agreement between 1D and 3D fields 
is still reasonably good in spite of  the high level of fluctuations present in the 3D run. 

 Figure~\ref{alpha0.02_mach10} portrays results for run $A0.02M10$ (which has $\alpha=0.02$, $\Gamma=1$, $\mathcal{M}=10$).
This run has a thermal energy which is only one percent 
of the turbulent one initially while this latter is comparable to the gravitational energy. 
As for Fig.~\ref{alpha0.1_mach3}, we see that the agreement between 1D and 3D runs remains entirely reasonable.

\section{Discussion}
We have presented quantitative evidences that the velocity fluctuations are greatly amplified 
within a collapsing cloud. However the exact nature of these velocity fluctuations requires
a better description. In particular while, given our general understanding 
of fluid behaviour, it sounds likely that turbulence is developing, the exact way that the  cascade 
may proceed and what is the driving scale are interesting avenues for future investigations. While these 
questions are beyond the scope of the present paper and clearly require detailed investigations, several aspects can  
already be discussed.

First of all, the modeling of the turbulent dissipation used in this work appears to play 
a decisive role. In the absence of dissipation for instance ($\eta_{diss}=0$), we observe
that the collapse is quickly halted and the 1D models do not ressemble the 3D ones. 
On the other-hand, when a sufficiently high value of $\eta_{diss}$ is used, the agreement 
between 1D and 3D models become really good. Since the 3D simulations do not have 
explicit viscosity, numerical dissipation that occurs at the mesh scale, is the only 
dissipation chanel. This is a strong indication that indeed a turbulent cascade 
is taking place.
Moreover, at least for low Mach values, 
the value of $\eta _{diss}$ for which the best agreement between 1D and 3D simulations is obtained, 
appears to be close to the value that has been inferred from 3D compressible turbulence simulations.  

Second, we can refer to the study of \citet{higashi2021} where powerspectra of kinetic energy 
have meen measured. For instance from their figure~9 they inferred that both the compressible
and solenoidal modes are amplified and present powerspectra close to $k^{-2}$ which has been 
inferred in compressible simulations \citep[e.g.][]{kritsuk2007}. Clearly, there is however 
more complexity. For instance starting from velocity fluctuations which present 
an energy powerspectra proportional to $k^{-3}$ instead of $k^{-2}$, \citet{higashi2021}
found that it remains proportional  to $k^{-3}$ at low density but get flatter at higher 
density where it is closer to $k^{-2}$.

As a matter of fact, there are likely several regimes of turbulence taking place in collapsing clouds. 
When turbulence is initally weak, there is a transition between the outer and inner parts of the cloud as
in the latter one, turbulence likely is amplified up to values comparable to gravitational energy. 
When turbulence is weak, as we saw above, it is likely highly anisotropic, the radial and transverse
components having different source terms. Even when turbulence is initially strong and likely 
more homogeneous and less spatially dependent (see for instance Figs.~\ref{alpha0.1_mach3} 
and~\ref{alpha0.02_mach10}), turbulence likely remains anisotropic because the mean radial 
velocity never vanishes and is of comparable amplitude than the other velocity components.

\section{Conclusion}
We have inferred a new set of 1D equations which describe the evolution of the spherically averaged 
variables during the collapse of a turbulent cloud. These equations, while similar, present 
significant differences with the ones used in the literature. We developed a 1D code 
which solve these equations and we performed a series of 1D but also 3D simulations. To carry out
the latter, the Ramses code has been used. The simulation sample covers a wide range of Mach 
numbers and thermal  support expressed by the parameter $\alpha$, the ratio between thermal and gravitational energy.
For each set of initial conditions we have performed several 1D runs with different values of $\eta_{diss}$
which controls the turbulent dissipation. By comparing the central mass as a function of time in 1D and 3D 
simulations, we can determine that the value of $\eta_{diss}$ requested to get good agreement between the 
1D and 3D runs is about 0.2-0.25 for low initial Mach numbers. This value is in good agreement with previous 
estimates inferred for turbulent non-self-gravitating gas. For larger initial Mach numbers, we found that 
values of $\eta_{diss}$ up to 5 times larger are requested to obtain good match between 1D and 3D simulations. 
For several set of initial conditions, we have then performed detailed comparisons between the 1D and 3D simulations using 
the previously inferred values of $\eta_{diss}$. Generally speaking we obtain remarkable agreement between 
the 1D and 3D runs. From these detailed comparisons we show that when the thermal support is large, 
initial turbulence is being amplified by the collapsing motions. However, when thermal support is 
low, it is shown that amplification is not sufficient to reproduce the 3D simulations. When turbulent 
generation through the development  of local gravitational instabilities is accounted for, very good 
agreement between the 1D and 3D runs is obtained. 
Finally, we show that even when turbulence is initially strong, the spherically averaged equations still 
predict behaviours that remain quantitatively similar to the 3D simulations.  
The spherically averaged 1D equations can be used in various context to predict the amplification and 
generation of turbulence within collapse. 


\begin{acknowledgements}
I thank the referee for their useful report as well as
Thierry Foglizzo for enlighting discussions. 
This work was granted access to HPC
   resources of CINES and CCRT under the allocation x2014047023 made by GENCI (Grand
   Equipement National de Calcul Intensif). 
   This research has received funding from the European Research Council
synergy grant ECOGAL (Grant : 855130).
\end{acknowledgements}

\bibliography{lars}{}

\begin{thebibliography}{47}
\expandafter\ifx\csname natexlab\endcsname\relax\def\natexlab#1{#1}\fi

\bibitem[{{Abdikamalov} \& {Foglizzo}(2020)}]{abdikamalov2020}
{Abdikamalov}, E. \& {Foglizzo}, T. 2020, \mnras, 493, 3496

\bibitem[{{Balbus} \& {Papaloizou}(1999)}]{balbus1999}
{Balbus}, S.~A. \& {Papaloizou}, J. C.~B. 1999, \apj, 521, 650

\bibitem[{{Bate} {et~al.}(2003){Bate}, {Bonnell}, \& {Bromm}}]{bate2003}
{Bate}, M.~R., {Bonnell}, I.~A., \& {Bromm}, V. 2003, \mnras, 339, 577

\bibitem[{{Bleuler} \& {Teyssier}(2014)}]{Bleuler14}
{Bleuler}, A. \& {Teyssier}, R. 2014, \mnras, 445, 4015

\bibitem[{{Brucy} \& {Hennebelle}(2021)}]{brucy2021}
{Brucy}, N. \& {Hennebelle}, P. 2021, \mnras, 503, 4192

\bibitem[{{Couch} \& {Ott}(2013)}]{couch2013}
{Couch}, S.~M. \& {Ott}, C.~D. 2013, \apjl, 778, L7

\bibitem[{{Davidovits} \& {Fisch}(2016)}]{davidovits2016}
{Davidovits}, S. \& {Fisch}, N.~J. 2016, \prl, 116, 105004

\bibitem[{{Elia} {et~al.}(2017){Elia}, {Molinari}, {Schisano}, {Pestalozzi},
  {Pezzuto}, {Merello}, {Noriega-Crespo}, {Moore}, {Russeil}, {Mottram},
  {Paladini}, {Strafella}, {Benedettini}, {Bernard}, {Di Giorgio}, {Eden},
  {Fukui}, {Plume}, {Bally}, {Martin}, {Ragan}, {Jaffa}, {Motte}, {Olmi},
  {Schneider}, {Testi}, {Wyrowski}, {Zavagno}, {Calzoletti}, {Faustini},
  {Natoli}, {Palmeirim}, {Piacentini}, {Piazzo}, {Pilbratt}, {Polychroni},
  {Baldeschi}, {Beltr{\'a}n}, {Billot}, {Cambr{\'e}sy}, {Cesaroni},
  {Garc{\'\i}a-Lario}, {Hoare}, {Huang}, {Joncas}, {Liu}, {Maiolo}, {Marsh},
  {Maruccia}, {M{\`e}ge}, {Peretto}, {Rygl}, {Schilke}, {Thompson},
  {Traficante}, {Umana}, {Veneziani}, {Ward-Thompson}, {Whitworth}, {Arab},
  {Band ieramonte}, {Becciani}, {Brescia}, {Buemi}, {Bufano}, {Butora},
  {Cavuoti}, {Costa}, {Fiorellino}, {Hajnal}, {Hayakawa}, {Kacsuk}, {Leto}, {Li
  Causi}, {Marchili}, {Martinavarro-Armengol}, {Mercurio}, {Molinaro},
  {Riccio}, {Sano}, {Sciacca}, {Tachihara}, {Torii}, {Trigilio}, {Vitello}, \&
  {Yamamoto}}]{elia2017}
{Elia}, D., {Molinari}, S., {Schisano}, E., {et~al.} 2017, \mnras, 471, 100

\bibitem[{{Federrath} {et~al.}(2011){Federrath}, {Chabrier}, {Schober},
  {Banerjee}, {Klessen}, \& {Schleicher}}]{federrath2011}
{Federrath}, C., {Chabrier}, G., {Schober}, J., {et~al.} 2011, \prl, 107,
  114504

\bibitem[{{Fiege} \& {Pudritz}(2000)}]{fiege2000}
{Fiege}, J.~D. \& {Pudritz}, R.~E. 2000, \mnras, 311, 105

\bibitem[{{Foglizzo}(2001)}]{foglizzo2001}
{Foglizzo}, T. 2001, \aap, 368, 311

\bibitem[{{Fromang} {et~al.}(2006){Fromang}, {Hennebelle}, \&
  {Teyssier}}]{Fromang06}
{Fromang}, S., {Hennebelle}, P., \& {Teyssier}, R. 2006, \aap, 457, 371

\bibitem[{{Goodwin} {et~al.}(2004{\natexlab{a}}){Goodwin}, {Whitworth}, \&
  {Ward-Thompson}}]{goodwin04a}
{Goodwin}, S.~P., {Whitworth}, A.~P., \& {Ward-Thompson}, D.
  2004{\natexlab{a}}, \aap, 414, 633

\bibitem[{{Goodwin} {et~al.}(2004{\natexlab{b}}){Goodwin}, {Whitworth}, \&
  {Ward-Thompson}}]{goodwin2004}
{Goodwin}, S.~P., {Whitworth}, A.~P., \& {Ward-Thompson}, D.
  2004{\natexlab{b}}, \aap, 423, 169

\bibitem[{{Guerrero-Gamboa} \& {V{\'a}zquez-Semadeni}(2020)}]{guerrero2020}
{Guerrero-Gamboa}, R. \& {V{\'a}zquez-Semadeni}, E. 2020, \apj, 903, 136

\bibitem[{{Hennebelle} \& {Chabrier}(2008)}]{HC08}
{Hennebelle}, P. \& {Chabrier}, G. 2008, \apj, 684, 395

\bibitem[{{Hennebelle} {et~al.}(2019){Hennebelle}, {Lee}, \&
  {Chabrier}}]{h2019}
{Hennebelle}, P., {Lee}, Y.-N., \& {Chabrier}, G. 2019, \apj, 883, 140

\bibitem[{{Higashi} {et~al.}(2021){Higashi}, {Susa}, \& {Chiaki}}]{higashi2021}
{Higashi}, S., {Susa}, H., \& {Chiaki}, G. 2021, arXiv e-prints,
  arXiv:2105.07701

\bibitem[{{Janka} {et~al.}(2007){Janka}, {Langanke}, {Marek},
  {Mart{\'\i}nez-Pinedo}, \& {M{\"u}ller}}]{janka2007}
{Janka}, H.~T., {Langanke}, K., {Marek}, A., {Mart{\'\i}nez-Pinedo}, G., \&
  {M{\"u}ller}, B. 2007, \physrep, 442, 38

\bibitem[{{Jeans}(1902)}]{jeans1902}
{Jeans}, J.~H. 1902, Philosophical Transactions of the Royal Society of London
  Series A, 199, 1

\bibitem[{{Joos} {et~al.}(2013){Joos}, {Hennebelle}, {Ciardi}, \&
  {Fromang}}]{joos2013}
{Joos}, M., {Hennebelle}, P., {Ciardi}, A., \& {Fromang}, S. 2013, \aap, 554,
  A17

\bibitem[{{Kritsuk} {et~al.}(2007){Kritsuk}, {Norman}, {Padoan}, \&
  {Wagner}}]{kritsuk2007}
{Kritsuk}, A.~G., {Norman}, M.~L., {Padoan}, P., \& {Wagner}, R. 2007, \apj,
  665, 416

\bibitem[{{Lee} \& {Hennebelle}(2018{\natexlab{a}})}]{leeh2018a}
{Lee}, Y.-N. \& {Hennebelle}, P. 2018{\natexlab{a}}, \aap, 611, A88

\bibitem[{{Lee} \& {Hennebelle}(2018{\natexlab{b}})}]{leeh2018b}
{Lee}, Y.-N. \& {Hennebelle}, P. 2018{\natexlab{b}}, \aap, 611, A89

\bibitem[{{Mac Low}(1999)}]{maclow1999}
{Mac Low}, M.-M. 1999, \apj, 524, 169

\bibitem[{{Mandal} {et~al.}(2020){Mandal}, {Federrath}, \&
  {K{\"o}rtgen}}]{mandal2020}
{Mandal}, A., {Federrath}, C., \& {K{\"o}rtgen}, B. 2020, \mnras, 493, 3098

\bibitem[{{Matsumoto} \& {Hanawa}(2003)}]{matsumoto2003}
{Matsumoto}, T. \& {Hanawa}, T. 2003, \apj, 595, 913

\bibitem[{{Misugi} {et~al.}(2019){Misugi}, {Inutsuka}, \&
  {Arzoumanian}}]{misugi2019}
{Misugi}, Y., {Inutsuka}, S.-i., \& {Arzoumanian}, D. 2019, \apj, 881, 11

\bibitem[{{Mocz} {et~al.}(2017){Mocz}, {Burkhart}, {Hernquist}, {McKee}, \&
  {Springel}}]{mocz2017}
{Mocz}, P., {Burkhart}, B., {Hernquist}, L., {McKee}, C.~F., \& {Springel}, V.
  2017, \apj, 838, 40

\bibitem[{{M{\"u}ller} \& {Janka}(2015)}]{muller2015}
{M{\"u}ller}, B. \& {Janka}, H.~T. 2015, \mnras, 448, 2141

\bibitem[{{Murray} \& {Chang}(2015)}]{murray2015}
{Murray}, N. \& {Chang}, P. 2015, \apj, 804, 44

\bibitem[{{Nagai} {et~al.}(1998){Nagai}, {Inutsuka}, \& {Miyama}}]{nagai1998}
{Nagai}, T., {Inutsuka}, S.-i., \& {Miyama}, S.~M. 1998, \apj, 506, 306

\bibitem[{{Ntormousi} \& {Hennebelle}(2015)}]{ntormousi2015}
{Ntormousi}, E. \& {Hennebelle}, P. 2015, \aap, 574, A130

\bibitem[{{Pope}(2000)}]{pope2000}
{Pope}, S.~B. 2000, {Turbulent Flows}

\bibitem[{{Pringle}(1981)}]{pringle1981}
{Pringle}, J.~E. 1981, \araa, 19, 137

\bibitem[{{Robertson} \& {Goldreich}(2012)}]{robertson2012}
{Robertson}, B. \& {Goldreich}, P. 2012, \apjl, 750, L31

\bibitem[{{Santos-Lima} {et~al.}(2012){Santos-Lima}, {de Gouveia Dal Pino}, \&
  {Lazarian}}]{santos2012}
{Santos-Lima}, R., {de Gouveia Dal Pino}, E.~M., \& {Lazarian}, A. 2012, \apj,
  747, 21

\bibitem[{{Schleicher} {et~al.}(2010){Schleicher}, {Banerjee}, {Sur},
  {Arshakian}, {Klessen}, {Beck}, \& {Spaans}}]{schleicher2010}
{Schleicher}, D.~R.~G., {Banerjee}, R., {Sur}, S., {et~al.} 2010, \aap, 522,
  A115

\bibitem[{{Schmidt} {et~al.}(2006){Schmidt}, {Niemeyer}, \&
  {Hillebrandt}}]{schmidt2006}
{Schmidt}, W., {Niemeyer}, J.~C., \& {Hillebrandt}, W. 2006, \aap, 450, 265

\bibitem[{{Schober} {et~al.}(2012){Schober}, {Schleicher}, {Federrath},
  {Glover}, {Klessen}, \& {Banerjee}}]{schober2012}
{Schober}, J., {Schleicher}, D., {Federrath}, C., {et~al.} 2012, \apj, 754, 99

\bibitem[{{Seifried} {et~al.}(2012){Seifried}, {Banerjee}, {Pudritz}, \&
  {Klessen}}]{seifried2012}
{Seifried}, D., {Banerjee}, R., {Pudritz}, R.~E., \& {Klessen}, R.~S. 2012,
  \mnras, 423, L40

\bibitem[{{Shu}(1977)}]{Shu77}
{Shu}, F.~H. 1977, \apj, 214, 488

\bibitem[{{Teyssier}(2002)}]{Teyssier02}
{Teyssier}, R. 2002, \aap, 385, 337

\bibitem[{{Verliat} {et~al.}(2020){Verliat}, {Hennebelle}, {Maury}, \&
  {Gaudel}}]{verliat2020}
{Verliat}, A., {Hennebelle}, P., {Maury}, A.~J., \& {Gaudel}, M. 2020, \aap,
  635, A130

\bibitem[{{Viciconte} {et~al.}(2018){Viciconte}, {Gr{\'e}a}, \&
  {Godeferd}}]{viciconte2018}
{Viciconte}, G., {Gr{\'e}a}, B.-J., \& {Godeferd}, F.~S. 2018, \pre, 97, 023201

\bibitem[{{Ward-Thompson} {et~al.}(2007){Ward-Thompson}, {Andr{\'e}},
  {Crutcher}, {Johnstone}, {Onishi}, \& {Wilson}}]{ward2007}
{Ward-Thompson}, D., {Andr{\'e}}, P., {Crutcher}, R., {et~al.} 2007, Protostars
  and Planets V, 33

\bibitem[{{Xu} \& {Lazarian}(2020)}]{xu2020}
{Xu}, S. \& {Lazarian}, A. 2020, \apj, 890, 157

\end{thebibliography}
\bibliographystyle{aa} 

\appendix

\end{document}